\documentclass[12pt,twoside]{article}

\newcommand{\mytitle}{The supersymmetric Higgs sector 
                    and \bbm\ for large $\tan \beta$}

\usepackage[normalem]{ulem}
\usepackage{amsmath}
\usepackage{cite}
\usepackage{graphicx}
\usepackage{amsfonts}
\usepackage{amssymb}%
\usepackage{epsf}
\usepackage{times}
\usepackage{a4wide}
\usepackage{epsfig}
\usepackage{pstricks}
\newcmykcolor{darkgreen}{1 0 0.6 0.2}  
\newcmykcolor{darkblue}{1 0 0 0.4}
\newcmykcolor{ST}{1 0.5 0 0}
\newcmykcolor{darkred}{0 1 0 0.3}

\usepackage{fancyheadings}
\advance \headheight by 3.0truept       

\newlength{\nseparation}
\newenvironment{nfigure}[1]
      {\begin{figure}[#1]\hrule\vspace{\nseparation}\par}
      {\vspace{\nseparation}\par \hrule \end{figure}}
\newenvironment{ntable}[1]
      {\begin{table}[#1]\hrule\vspace{\nseparation}\par}
      {\vspace{\nseparation}\par \hrule \end{table}}

\def\openone{\leavevmode\hbox{\small1\kern-3.8pt\normalsize1}}%

\newcommand{\Zbmu}{\bar Z}
\newcommand{\Qbmu}{\bar Q}
\newcommand{\Ebmu}{\bar E}
\newcommand{\gammabmu}{\bar \gamma}
\newcommand{\ds}{\displaystyle}
\newcommand{\lt}{\left}
\newcommand{\rt}{\right}
\newcommand{\no}{\nonumber}
\newcommand{\nn}{\nonumber \\}
\newcommand{\ov}[1]{\overline{#1}}
\newcommand{\eq}[1]{Eq.~(\ref{#1})}
\newcommand{\eqsand}[2]{Eqs.~(\ref{#1}) and (\ref{#2})}
\newcommand{\eqsto}[2]{Eqs.~(\ref{#1}--\ref{#2})}

\newcommand{\gev}{\,\mbox{GeV}}

\newcommand{\Bbar}{\,\overline{\!B}}

\newcommand{\bbd}{\ensuremath{B_d\!-\!\Bbar{}_d\,}}
\newcommand{\bbs}{\ensuremath{B_s\!-\!\Bbar{}_s\,}}
\newcommand{\bbq}{\ensuremath{B_q\!-\!\Bbar{}_q\,}}
\newcommand{\bb}{\ensuremath{B\!-\!\Bbar{}\,}}
\newcommand{\bbms}{\bbs\ mixing}

\newcommand{\bbmq}{\bbq\ mixing}
\newcommand{\bbm}{\bb\ mixing}

\newcommand{\bra}[1]{\ensuremath{\langle #1 |}}
\newcommand{\ket}[1]{\ensuremath{| #1 \rangle }}
\newcommand{\fig}[1]{Fig.~\ref{#1}}
\newcommand{\lqcd}{\Lambda_{\textit{\scriptsize{QCD}}}}

\newcommand{\dm}{\ensuremath{\Delta M}}
\newcommand{\dg}{\ensuremath{\Delta \Gamma}}

\newcommand{\BsMM}{B_s \to \mu^+ \mu^-}
\newcommand{\BTN}{B^+ \to \tau^+ \nu_\tau}
\newcommand{\bsg}{b \to s \gamma}

\begin{document}

\thispagestyle{empty}
\noindent
TUM-HEP-707/09  \hfill January 2009 \\ 
TTP09-01
\\
SFB/CPP-09-03 
\\
CERN-PH-TH/2008-204
\vspace*{0.8cm}

\boldmath
\centerline{{\large\bf \mytitle} }
\unboldmath
\vspace*{1.5cm}
\centerline{{\sc Martin Gorbahn${}^1$,~~~Sebastian J\"ager${}^2$
,~~Ulrich Nierste${}^3$~~and~~St\'ephanie Trine${}^3$}}
\bigskip \bigskip
\begin{center}
\sl ${}^1$
\small
Technische Universit\"at M\"unchen, 
Institute for Advanced Study, \\
Arcisstra\ss e 21, D-80333 M\"unchen, Germany
\end{center}
\vspace*{-5mm}
\begin{center}
\sl ${}^2$
\small
Technische Universit\"at M\"unchen,
Excellence Cluster ``Universe'', \\
Boltzmannstra\ss e 2,
D-85748 Garching, Germany
\end{center}
\vspace*{-5mm}
\begin{center}
\sl ${}^3$ 
\small
Institut f{\"u}r Theoretische Teilchenphysik,
Universit{\"a}t Karlsruhe,
Karlsruhe Institute of Technology, \\
Engesserstra\ss e 7,
D-76128 Karlsruhe, Germany
\end{center}

\vspace*{1cm}
\centerline{\bf Abstract}
\vspace*{0.3cm}
\noindent
We match the Higgs sector of the most general flavour breaking and CP
violating minimal supersymmetric standard model (MSSM) onto a generic
two-Higgs-doublet model, paying special attention to the definition of
$\tan\beta$ in the effective theory. In particular no
$\tan\beta$-enhanced loop corrections appear in the relation to $\tan\beta$
defined in the $\overline{\textrm{DR}}$ scheme in the MSSM. The
corrections to the Higgs-mediated flavour-changing amplitudes which
result from this matching are especially relevant for the $B_d$ and
$B_s$ mass differences $\dm_{d,s}$ for minimal flavour violation,
where the superficially leading contribution vanishes. We give a
symmetry argument to explain this cancellation and perform a
systematic study of all Higgs-mediated effects, including Higgs
loops. The corrections to $\dm_s$ are at most 7\%  for
$\mu>0$ and $M_A < 600\gev$ if constraints from other observables are
taken into account. For $\mu<0$ they can be larger,
but are always less than about 20\%. Contrary to
recent claims we do not find numerically large contributions here,
nor do we find any $\tan\beta$-enhanced contributions
from loop corrections to the Higgs potential in
$B^+ \to \tau^+ \nu$ or $B \to X_s \gamma$. 
We further update
supersymmetric loop corrections to the Yukawa couplings, where we
include all possible CP-violating phases and correct errors in the
literature.  The possible presence of CP-violating phases generated by
Higgs exchange diagrams is briefly discussed as well. Finally we
provide improved values for the bag factors $P^{\rm VLL}_1$, $P^{\rm
  LR}_2$, and $P^{\rm SLL}_1$ at the electroweak scale.
\vspace*{.8cm}

\noindent
PACS numbers: 11.30.Pb~~~12.60.Fr~~~12.15.Ff~~~14.40.Nd

\vfill

\newpage
\setcounter{page}{2}

\tableofcontents

\section{Introduction}
\label{sect:intro}

Supersymmetry constrains the structure of the Yukawa
couplings of the minimal supersymmetric standard model (MSSM) to those of a
special two-Higgs-doublet model (2HDM).  In this 2HDM of type II one Higgs
doublet, $H_u$, only couples to up-type fermions, while the other one, $H_d$,
only couples to down-type fermions. As a consequence, there are no dangerous
tree-level flavour-changing neutral current (FCNC) couplings of the neutral
Higgs bosons. However, the presence of supersymmetry-breaking terms destroys
this pattern at the one-loop level, permitting couplings of both Higgs
doublets to all fermions.  Thus the resulting Higgs sector is that of a general
2HDM, often called 2HDM of type III. As pointed out first by Hall, Rattazzi
and Sarid, the loop-induced Yukawa couplings can compete with the tree-level
ones in the limit of a large $\tan \beta=v_u/v_d$, which is the
ratio of the vacuum expectation values (vevs) of $H_u$ and $H_d$ \cite{hrs}:
in the relationship between $H_{u,d}$-couplings and observed masses of
the down-type
fermions the loop suppression factor $\sim 0.02$ is offset by a factor of
$\tan\beta$, so that ${\cal O}(1)$ corrections to the type-II 2HDM are
possible for $\tan\beta\sim 50$. In such scenarios also ${\cal O}(1)$
loop-induced FCNC couplings of neutral Higgs bosons appear \cite{hpt}, which
allow the branching fractions of (yet unobserved) leptonic $B$ decays to
exceed their standard-model values by more than two orders of magnitude
\cite{bk}.  This observation has stimulated a large activity in flavour
physics and powerful constraints on the MSSM Higgs sector in scenarios with
large $\tan\beta$ have been derived from $B$ factory data 
\cite{bk,ir,bcrs,ltb}.
These Higgs-induced effects in flavour physics are very transparent in the~limit
\begin{eqnarray}
M_{\rm SUSY}\gg M_{A} \sim v \, , \label{hier}
\end{eqnarray}
where $M_{\rm SUSY}$ denotes the generic mass scale of the superpartners and
the masses $M_{A}$, $M_{H}$, $M_{h}$ and $M_{H^\pm}$ of the five
physical Higgs bosons are taken to be of the order of the electroweak scale
$v\equiv \sqrt{v_u^2+v_d^2}=246 \,\gev$.  All low-energy observables can be
computed in the type-III 2HDM, which emerges as the effective theory in the
limit of \eq{hier}.  The new couplings can be calculated from finite one-loop
diagrams with supersymmetric particles and thus become functions of the MSSM
parameters, so that the desired constraints on the supersymmetric parameter
space can be derived. The effective 2HDM Lagrangian
efficiently incorporates all large-$\tan\beta$ effects, equivalent to a
perturbative all-order resummation of those radiative corrections which are
enhanced by a factor of $\tan\beta$ \cite{cgnw}.

\bbmq\ (with $q=d$ or $s$) plays a special role among the FCNC transitions of
$B$ mesons. Here the leading new effect stems from effective tree-level diagrams with
neutral Higgs bosons (see \fig{fig:bbtree}).
\begin{nfigure}{tb}
\centerline{\includegraphics[width=6cm]{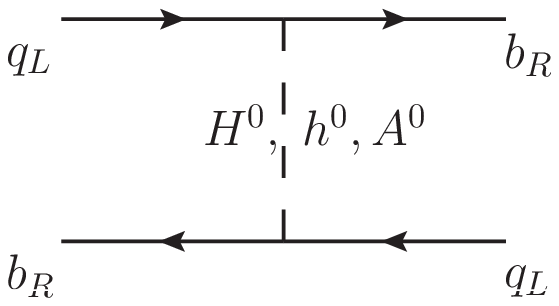}}
\caption{Leading contributions to \bbmq\ from supersymmetric Higgs bosons.
 The FCNC couplings are induced by supersymmetric loops.  The
 coefficient of $Q_1^{\rm SLL}=(\ov b_R q_L) (\ov b_R q_L)$
 vanishes, if the tree-level relations between Higgs masses and
 mixing angles are used.}\label{fig:bbtree}
\end{nfigure}
A priori the dominant contribution is expected from Yukawa couplings to
right-handed $b$ quarks, generating the effective $\Delta B=2$ operator
\begin{eqnarray}
Q_1^{\rm SLL} & \equiv & 
\left( \ov b_R q_L \right) \left( \ov b_R q_L \right) \, .\label{defqll}
\end{eqnarray}
However, the corresponding coefficient $C_1^{\rm SLL}$ vanishes exactly, if one
employs the tree-level relations between the Higgs masses and mixing
angles \cite{hpt}.  Nevertheless, sizeable effects in \bbms\ are possible
even in scenarios with minimal flavour violation (MFV)
\cite{Hall:1990ac,Ciuchini:1998xy,Ali:1999we,Buras:2000qz,Buras:2000dm,D'Ambrosio:2002ex,Isidori:2006qy,Altmannshofer:2007cs,Colangelo:2008qp},
in which the Cabibbo-Kobayashi-Maskawa (CKM) matrix \cite{ckm} is the only
source of flavour violation: keeping the strange Yukawa
coupling non-zero one finds a non-vanishing contribution to the
coefficient~of
\begin{eqnarray}
Q_2^{\rm LR} \equiv \left( \ov b_R q_L \right) 
\left( \ov b_L q_R \right) \, , \label{defqlr}
\end{eqnarray}
which depletes the \bbs\ mass difference $\dm_s$ \cite{bcrs}.  The
tree-level vanishing of $C_1^{\rm SLL}$
calls for a systematic analysis of all subleading effects.
In particular, the contribution that stems from $Q^{\rm SLL}_1$ can a
priori compete with the contribution of the operator $Q^{\rm LR}_2$ above
if the one-loop corrections to the MSSM Higgs potential
\cite{Okada:1990vk,Haber:1990aw,Ellis:1990nz,Brignole:1992uf,Dabelstein:1994hb,Chankowski:1992er,cepw}
are taken into account.
While a lot of work has been
devoted to the analysis of the Yukawa sector
\cite{hpt,bk,cgnw,bcrs,ltb}, little attention has been given to effects
from the Higgs potential.
An exception is Ref.~\cite{fgh}, which finds large contributions.
We revisit these effects
in the present paper and perform a systematic matching of the MSSM Higgs sector
onto the type-III 2HDM. The result is not only relevant for the
calculation of $C_1^{\rm SLL}$, it also clarifies the relationship
between the definitions of $\tan\beta$ in the MSSM and the effective
2HDM.
This is important to link the constraints from flavour physics to other
fields of MSSM phenomenology, in particular Higgs physics. Our paper
is organized as follows.
We derive the corrected \bbm\ amplitude in Sect.~\ref{sect:bbm},
including all relevant subleading contributions. The renormalization
of $\tan\beta$ and som further technial issues are the subject of
Sect.~\ref{sect:higgs}.
In Sect.~\ref{sect:phen} we apply our new formulae to the phenomenology of
\bbm, analysing the mass differences $\dm_d$ and $\dm_s$ as well as
CP-violation. Our results are summarized in Sect.~\ref{sect:ccl}.
We list our notation and our technical results in four appendices.
Parts of our results were previously presented by one of us at
a conference~\cite{Trine:2007ma}.

\section{\boldmath Higgs-mediated effects in $\bb$ mixing \unboldmath}
\label{sect:bbm}

The quantity governing the $\bbq$ mass difference is the
off-diagonal element of the $\bbq$ meson mass matrix:
$\Delta M_{q}=2\left\vert M_{21}^{q}\right\vert $, with
\begin{equation}
M_{21}^{q}
= \frac{\left\langle \overline{B}_{q} 
     \left \vert \mathcal{H}_{\rm eff}^{\Delta B=2} \right\vert B_{q} 
     \right\rangle }{2M_{B_{q}}} \, .
 \label{eq:heff}
\end{equation}
The $\Delta B=2$ effective weak
Hamiltonian $\mathcal{H}_{\rm eff}^{\Delta B=2}$ consists
in general of eight dimension-six operators: 
\begin{equation}
\mathcal{H}_{\rm eff}^{\Delta B=2}=\frac{G_{F}^{2}M_{W}^{2}\lambda_{qb}^{2}}%
{16\pi^{2}}\sum_{i=1}^{8}C_{i}(\mu_{h})Q_{i}(\mu_{h})\, , \label{3.2}%
\end{equation}
with $\lambda_{qb}\equiv V_{tq}V_{tb}^{\ast}$.
The set of operators in \eq{3.2} comprises the 
standard-model operator,
\begin{equation}
Q_1^{\rm VLL}
=\left(  \overline{b}_{L}\gamma_{\mu}q_{L}\right)  \left(  \overline
{b}_{L}\gamma^{\mu}q_{L}\right) \, ,\label{3.3}%
\end{equation}
the two scalar operators defined in 
\eqsand{defqll}{defqlr},
the operator
\begin{eqnarray}
Q_1^{\rm SRR} & \equiv & 
\left( \ov b_L q_R \right) \left( \ov b_L q_R \right) \, ,\label{defqrr}
\end{eqnarray}
and four other operators. The complete list of operators plus the
relevant evanescent operators is given in Eq.~(\ref{eq:33}) and
Eq.~(\ref{eq:31}) of Appendix \ref{sect:rghp}.  We express our results in
terms of matrix elements at the high scale $\mu_{h}$ which we choose equal to
the top mass $\ov{m}_t(\ov{m}_t)=164\gev$. In this way the other four
operators do not appear in our formulae. However, some of them are needed to
connect $Q_i(\mu_{h})$ with $Q_i(\mu_b)$ at the low scale $\mu_b\sim m_b$ at
which their matrix elements are computed, because they mix with $Q_1^{\rm
SLL}$, $Q_1^{\rm SRR}$ or $Q_2^{\rm LR}$ under renormalization.
We follow the conventions of Refs.~\cite{bmu} and
\cite{Buras:2001ra} for operators and matrix elements. In particular
we parametrize the hadronic matrix elements as
\begin{equation}
\left\langle \overline{B}_{q}\left\vert Q_{i} (\mu_{h}) 
   \right\vert B_{q}\right\rangle
=\frac{2}{3}M_{B_{q}}^{2}f_{B_{q}}^{2}P_{i} \, . \label{3.6}
\end{equation}
The $P_i$'s are obtained \cite{Buras:2001ra} by renormalization-group
evolution from the conventional bag factors $B_{i}$
computed at the low scale $\mu_b$.  We calculate the $P_i$'s
from up-to-date lattice QCD results in Appendix~\ref{sect:rghp}, where
we fully exploit constraints from heavy-quark relations.  This is a
new feature of our analysis compared to previous studies of
new-physics effects in \bbm.

\subsection{Effective tree-level Higgs exchange}
\label{sec:fcnc-higgs-tree}

The Higgs sector of the MSSM contains two $SU(2)$ doublets $H_u$ and
$H_d$,
\begin{equation}
H_{u}=\binom{h_{u}^{+}}{h_{u}^{0}} \, ,\qquad
H_{d}=\epsilon\binom{h_{d}^{+}}{h_{d}^{0}}^{\ast} \, ,
\qquad 
\epsilon=\left( \begin{array}[c]{cc}
 0 & 1\\
-1 & 0
\end{array} \right) \, , \label{2.3}
\end{equation}
of hypercharge $+1/2$ and $-1/2$, respectively, with
vacuum expectation values (vevs)
$\langle h_{u,d}^0\rangle  = v_{u,d}/\sqrt{2}$
of relative size $\tan\beta = v_u/v_d$.
Integrating out supersymmetric particles, the Lagrangian of the resulting
effective 2HDM is no longer restricted to be of type II, and is
constrained only by the electroweak symmetry. Neither will it be
renormalizable, with operators of dimension greater than four encoding
effects that decouple at least as $v/M_{\rm SUSY}$ for heavy superpartners.
We begin  with a short review of some pertinent aspects of the general 2HDM.

Defining
\begin{equation}   \label{eq:vevbasis}
 \left( \begin{array}{c} \Phi \\ \Phi' \end{array} \right) =
    \left( \begin{array}{cc}
           \cos\beta & \sin\beta \\
           -\sin\beta & \cos\beta
    \end{array} \right) \,
\left( \begin{array}{c} - \epsilon H_d^* \\ H_u \end{array} \right) \, ,
\end{equation}
the most general fermion-Higgs interactions up to dimension four read
\begin{equation}   \label{eq:LY}
\begin{split}
  {\cal L}_Y =&
   - \frac{\sqrt{2}}{v} \bar d_{Ri}\, M_{d_{ij}}\, \Phi^\dagger Q_{Lj}
   - \bar d_{Ri}\, \kappa_{ij}\, \Phi'^\dagger Q_{Lj}\\
  &- \frac{\sqrt{2}}{v} \bar u_{Ri}\, M_{u_{ij}}\, Q_{Lj} \cdot \! \Phi
   - \bar u_{Ri}\, \tilde \kappa_{ij}  Q_{Lj} \cdot \! \Phi'
  + \textrm{h.c.} \, ,
\end{split}
\end{equation}
where we have employed the notation $a \cdot b \equiv a^T \epsilon\, b$.
By construction, the vev of $\Phi'$ vanishes,
whereas $\Phi$ has $\langle \Phi \rangle = (0, v/\sqrt{2})^T$ 
and contains all three Goldstone bosons. Hence only $\Phi$ can contribute
to the fermion masses and only $\Phi'$ can have flavour-violating
neutral couplings.
The flavour basis is defined such that the down-quark mass
matrix $M_d$ is diagonal.
In this basis the FCNC Higgs couplings to
$b$-quarks are governed by $\kappa_{bq}$ or $\kappa_{qb}$ ($q=d$ or $s$).

The renormalizable Higgs self-interactions are comprised in the most general
gauge-invariant dimension-four two-Higgs-doublet potential \cite{Haber:1997dt},
\begin{align}
&  V\underset{}{=}m_{11}^{2}H_{d}^{\dagger}H_{d}+m_{22}^{2}H_{u}^{\dagger
}H_{u}+\left\{  m_{12}^{2}H_{u}\cdot H_{d}+h.c.\right\} \nonumber\\
&  +\frac{\lambda_{1}}{2}(H_{d}^{\dagger}H_{d})^{2}+\frac{\lambda_{2}}%
{2}(H_{u}^{\dagger}H_{u})^{2}+\lambda_{3}(H_{u}^{\dagger}H_{u})(H_{d}%
^{\dagger}H_{d})+\lambda_{4}(H_{u}^{\dagger}H_{d})(H_{d}^{\dagger}%
H_{u})\nonumber\\
&  +\left\{  \frac{\lambda_{5}}{2}\left(  H_{u}\cdot H_{d}\right)
^{2}-\lambda_{6}(H_{d}^{\dagger}H_{d})\left(  H_{u}\cdot H_{d}\right)
-\lambda_{7}(H_{u}^{\dagger}H_{u})\left(  H_{u}\cdot H_{d}\right)
+\mbox{h.c.}\right\}  .\label{2.1}%
\end{align}
The couplings $m_{12}^{2}$, $\lambda_{5}$, $\lambda_{6}$, and $\lambda_{7}$ are
in general complex, yet the vevs $v_{u,d}$
can be made real by a $U(1)$ transformation on the Higgs fields. The
definitions of $m^2_{ij}$ and $\lambda_{i}$ in \eq{2.1}
coincide with Ref.~\cite{Haber:1997dt} except for
$\lambda_{3}$ and $\lambda_{4}$: we associate a different operator
with $\lambda_{4}$ to eliminate it from tree-level
neutral-Higgs phenomenology and have instead
$\lambda_{3}=\lambda_{3}^{\text{\textrm{\cite{Haber:1997dt}}}}+\lambda_{4}
^{\text{\textrm{\cite{Haber:1997dt}}}}$ and
$\lambda_{4}=-\lambda_{4}^{\text{\textrm{\cite{Haber:1997dt}}}}$.

Shifting the fields in Eq.~(\ref{2.1}) by their vevs,
which minimize $V$ at tree level,\footnote{
``Tree level'' here refers to the 2HDM.
We defer a discussion of quantum corrections to $v_u$ and $v_d$ to
Sect.~\ref{sect:higgs}.
}
\begin{equation}
h_{u,d}^0 = \frac{1}{\sqrt{2}}(v_{u,d} + \phi_{u,d} + i \chi_{u,d})\, ,
\end{equation}
determines the physical Higgs-boson mass matrices and interactions.
We write the neutral-Higgs mass matrix
in the basis $(\phi_d,\phi_u,\chi_d,\chi_u)$
in terms of $2\times2$ blocks,
\begin{equation}     \label{eq:higgsmassneut}
M^2_0 = \left( \begin{array}{cc}
           M^2_{R} & M^2_{RI} \\[2mm]
           M^{2\,T}_{RI} & M^2_{I}
    \end{array} \right)\, , 
\end{equation}
with $M^2_{R}$, $M^2_{RI}$, and $M^2_{I}$ given in
Eqs.~(\ref{Delta}-\ref{CPVMassMat}) below. In the CP-conserving case,
$M^2_{RI}=0$, and $M^2_{R}$ and $M^2_{I}$ are diagonalized by rotating
the CP-even and CP-odd Higgs fields through angles $\alpha$ and
$\beta$, respectively:
\begin{equation}
\left(
\begin{array}[c]{c}
\phi_{d}\\
\phi_{u}%
\end{array}
\right)  = \left(
\begin{array}[c]{cc}
\cos\alpha & -\sin\alpha\\
\sin\alpha & \cos\alpha
\end{array}\right)  \left(
\begin{array}[c]{c}
H^{0}\\
h^{0}
\end{array} \right) , \quad
 \left(
\begin{array}[c]{c}
\chi_{d}\\
\chi_{u}
\end{array} \right)  =
 \left(
\begin{array}[c]{cc}
\cos\beta & -\sin\beta\\
\sin\beta & \cos\beta
\end{array}
\right)  \left(
\begin{array}[c]{c}
G^{0}\\
A^{0}
\end{array}
\right)  .
\label{eq:alphabeta}
\end{equation}%
The same angle $\beta= \arctan v_u/v_d$ as defined above appears because
(and only when) $v_u$, $v_d$ minimize $V$. If CP violation is present,
four physical mixing angles $\alpha_{1,2,3}$ and $\beta$ are required to
diagonalize $M_0^2$.
The charged-Higgs mass matrix $M_+^2$ is always diagonalized by $\beta$,
\begin{equation}
\left(
\begin{array}[c]{c}
h_{d}^{+}\\
h_{u}^{+}
\end{array}
\right)  =\left(
\begin{array}[c]{cc}
\cos\beta & -\sin\beta\\
\sin\beta & \cos\beta
\end{array}
\right)  \left(
\begin{array}[c]{c}
G^{+}\\
H^{+}
\end{array}
\right)  .
\end{equation}

The non-standard effective operators $Q_2^{\rm LR}$, $Q_1^{\rm SLL}$, and
$Q_{1}^{\rm SRR}$  are generated at tree level via the exchange of
neutral Higgs bosons (see \fig{fig:bbtree}) with the Wilson
coefficients
\begin{equation}
C_2^{\rm LR} =
-\frac{8\pi^{2}}{G_{F}^{2}M_{W}^{2} \lambda_{qb}^2}
\left(  \kappa_{qb}^{\ast}\,\kappa_{bq}\right)  \mathcal{F}^{+},   \qquad
C_1^{\rm SLL}=
-\frac{4\pi^{2}}{G_{F}^{2}M_{W}^{2} \lambda_{qb}^2}
\left(  \kappa_{bq}\right)^{2} \mathcal{F}^{-}\label{3.9},
\end{equation}
and $C^{\rm SRR}_1$ obtained from $C^{\rm SLL}_1$ through the replacement
$\left(\kappa_{bq}\right)^{2} \mathcal{F}^{-}\to \left(\kappa_{qb}^\ast\right)^{2} \mathcal{F}^{-\ast}$.
We find that, in the general case, the Higgs propagation factors can be expressed as follows:
\begin{align}
\mathcal{F}^{+} &
= \frac{\det\left( M_{R}^{2}+M_{I}^{2}+iM_{RI}^{2}-iM_{RI}^{2\,T}\right)}
       {m_{1}^{2}m_{2}^{2}m_{3}^{2}} \equiv \frac{\det B}{m_1^2 m_2^2 m_3^2},
                                              \label{3.13bis}  \\[2mm]
\mathcal{F}^{-} & 
= -\frac{\det\left( M_{R}^{2}-M_{I}^{2}-iM_{RI}^{2}-iM_{RI}^{2\,T}\right)}
        {m_{1}^{2}m_{2}^{2}m_{3}^{2}} \equiv -\frac{\det A}{m_1^2 m_2^2 m_3^2},  
                                              \label{3.14}
\end{align}
where the denominators contain the product of the three nonzero
eigenvalues of $M^2_0$. In the CP-conserving case,
\eqsand{3.13bis}{3.14} reduce to the well-known expressions
\begin{equation}
\label{3.10}
\mathcal{F}^{\pm}=\frac{\sin^{2}(\alpha-\beta)}{M_{H}^{2}}+\frac{\cos
^{2}(\alpha-\beta)}{M_{h}^{2}}\pm\frac{1}{M_{A}^{2}} \, ,
\end{equation}
where $M_A$ denotes the CP-odd Higgs-boson mass.

The discussion so far has been completely general.
Particularizing to the MSSM, a perturbative matching calculation
relates the two theories. At tree level this trivially results in
\begin{equation}
\begin{aligned}
M^{(0)}_d &= \frac{v}{\sqrt{2}} \cos\beta\, Y_d, & 
M^{(0)}_u &= \frac{v}{\sqrt{2}} \sin\beta\, Y_u, \\
\kappa^{(0)} &= - \sin\beta\, Y_d, &
\tilde \kappa^{(0)} &= \cos\beta\, Y_u \\
m_{11}^{2^{(0)}} &= \left\vert \mu\right\vert ^{2}+m_{H_{d}}^{2} \equiv m_{1}^{2}, &
\lambda_{1}^{(0)} &= \lambda_{2}^{(0)}=-\lambda_{3}^{(0)}=
(g^{2}+g^{\prime2})/4 \equiv \tilde {g}^{2}/4, \\
m_{22}^{2^{(0)}} &= \left\vert \mu\right\vert ^{2}+m_{H_{u}}^{2} \equiv m_{2}^{2}, & 
\lambda_{4}^{(0)} &= g^{2}/2,\\
m_{12}^{2^{(0)}} &= B\mu, & 
\lambda_{5}^{(0)} &=\lambda_{6}^{(0)}=\lambda_{7}^{(0)}=0.
\end{aligned}
\label{2.4} 
\end{equation}
At this order $\kappa^{(0)}$ and $\tilde \kappa^{(0)}$ are aligned
with $M_d^{(0)}$ and $M_u^{(0)}$, respectively, so that no FCNC are
induced, as it must be in a model II.
At one loop, all couplings in \eq{2.1} are generated. Moreover,
the corrections to the Yukawa couplings have the more general form
\begin{equation}
\label{YukCorr}
M^{(1)}_d = \frac{v}{\sqrt{2}} \cos\beta \big[\Delta Y_d + \tan\beta\,
\Delta K \big] , \qquad
\kappa^{(1)} = -\sin\beta\, \big[\Delta Y_d - \cot\beta \Delta K \big] ,
\end{equation}
where $\Delta Y_{d_{ij}}$ and $\Delta K_{ij}$ parametrize the one-loop
vertices $\bar d_{Ri} H_d \cdot Q_{Lj}$ and $\bar d_{Ri} H_u^\dagger
Q_{Lj}$, respectively.
Diagonalizing $M_{d}$ rotates $\kappa^{(0)}$, giving rise to a
flavour-violating coupling $ \propto Y_d \tan\beta/(16\, \pi^2)$,
which can be of ${\cal O}(1)$ for $\tan\beta\sim 50$.

The origin of this explicit $\tan\beta$ enhancement (in addition to the mere
presence of large down-type Yukawa couplings), which can compensate the loop
factor $1/(16\, \pi^2)$, is the replacement of $v_d$ by $v_u \gg v_d$ in
the contribution of $\Delta K$ to $M_d$
\cite{hrs}.\footnote{We tacitly assume that the fermion kinetic terms in the
effective 2HDM have been made canonical. Such a field renormalization does
not contribute factors of $\tan\beta$ because it is determined by
dimensionless couplings. Cf.\ Sect.~\ref{sect:higgs} for a
  discussion of field renormalization. Our $\Delta K$ and $\Delta Y_d$
correspond to $\Delta_u Y_d $ and $-\Delta_d Y_d$, respectively, in the
first paper of Ref.~\cite{bcrs}.}  This removal of a $v_d$ suppression can happen
only in dimensionful quantities.  In the fermion mass terms, only one power of
$\tan\beta$ can appear because there was only one power of $v_d$ to begin
with. This is in agreement with the findings in~\cite{cgnw}. Our approach using
un-shifted Higgs fields (``unbroken-theory'') makes particularly evident that
this result holds to all orders, as the Yukawa Lagrangian only involves
dimensionless couplings and there are no hidden factors of $\tan\beta$.
Although we have integrated out only the sparticles -- as we assume a
hierarchy $v, M_A \ll M_{\rm SUSY}$ -- the argument continues to hold if
we also integrate over the Higgs fields, keeping only constant
background values of $\Phi, \Phi'$ (spurions). The reason is that for
determining the mass matrices, the relevant external four-momenta are of ${\cal
O}(m_q)$, providing an expansion parameter $m_q/v$ or $m_q/M_A$. Hence the
Higgs contributions to the effective potential (which on general grounds
respects the electroweak symmetry) can be organized into a (local) effective
Lagrangian, with $m_q$-suppressed corrections to the form \eq{eq:LY}
encoded in higher-dimensional operators with additional derivatives acting on
$d_{Ri}$ or $Q_{Lj}$.  The contribution from both Higgs and sparticle loops to
$M_d$ is then simply obtained upon substituting for $\Phi, \Phi'$ their vacuum
expectation values. This mass matrix is to be identified with a short-distance
(such as $\overline{\rm MS}$) mass in the effective QCD $\times$ QED at low
energies, where the dependence on the chosen scheme cancels against the
explicit form of the matching (of the 2HDM onto QCD $\times$ QED).

There is only one other place
where a similar $\tan\beta$ enhancement
can occur, namely in the dimensionful self-couplings of
the (shifted) Higgs fields, that is, their masses and trilinear couplings.
Indeed, at dimension two it is  
exhibited in the neutral Higgs mass
matrix \eq{eq:higgsmassneut}.
Explicitly, one has
(with $s_\beta \equiv \sin\beta$, $c_\beta \equiv \cos\beta$, and
$\lambda_k^r \equiv {\rm Re}\, \lambda_k$)
\begin{equation}
M_R^2 =v^{2}\left(
\begin{array}[c]{cc}
\lambda_{5}^{r} s_{\beta}^{2} + 2\lambda_{6}^{r} s_{\beta}c_{\beta} + \lambda_{1} c_{\beta}^{2}& 
\lambda_{7}^{r} s_{\beta}^{2} + \lambda_{3} s_{\beta}c_{\beta} + \lambda_{6}^{r} c_{\beta}^{2} \\
\lambda_{7}^{r} s_{\beta}^{2} + \lambda_{3} s_{\beta}c_{\beta} + \lambda_{6}^{r} c_{\beta}^{2} & 
\lambda_{2} s_{\beta}^{2} + 2\lambda_{7}^{r} s_{\beta}c_{\beta} + \lambda_{5}^{r} c_{\beta}^{2}\\
\end{array}
\right) + M^2_I \, ,
\label{Delta}%
\end{equation}
\begin{equation}
\label{CPoddMassMat}
M_{I}^{2}=M_{A}^{2}\left(
\begin{array}[c]{cc}
s_{\beta}^{2} & -s_{\beta}c_{\beta} \\
-s_{\beta}c_{\beta} & c_{\beta}^{2}
\end{array}
\right)\, ,
\end{equation}
where
$(m_{12}^{2})^r$ has been traded for $M_A^2$, with
\begin{equation}            \label{eq:madef}
s_{\beta} \, c_{\beta} \, M_{A}^{2}= (m_{12}^{2})^r
  -\frac{v^{2}}{2}(\lambda_{7}^{r}\,s_{\beta}^{2}
       + 2\lambda_{5}^{r}\,s_{\beta}c_{\beta}
       +\lambda_{6}^{r}\,c_{\beta}^{2}) .
\end{equation}
If CP is conserved, in the limit of infinite $\tan\beta$ ($c_\beta\to 0$)
the leading mass splitting $M_H^2 - M_A^2 = \lambda_5 v^2$, and the
leading correction to the tree-level result $\alpha=0$ is determined
by $\lambda_7$. In the former case, an enhancement by
two powers of $\tan\beta$ occurs
($M_H^2 -M_A^2 = {\cal O}(\cos^2\beta)$ at tree level), while
the loop correction to $\alpha$ is enhanced by a single power of $\tan\beta$
with respect to its tree-level value. Either effect is sufficient to remove the
cancellation in ${\cal F}^-$ in \eq{3.10}.
Moreover, a $1/\tan\beta$-unsuppressed CP-violating contribution
proportional to $\lambda_5^i$ and $\lambda_7^i$ appears to occur:
\begin{equation}
M_{RI}^{2}=\frac{v^{2}}{2}\left(
\begin{array}
[c]{cc}
\lambda_{5}^{i}\,s_{\beta}^{2}\,+\,2\lambda_{6}^{i}\,s_{\beta}c_{\beta}&
-\lambda_{5}^{i}\,s_{\beta}c_{\beta}\,-\,2\lambda_{6}^{i}\,c_{\beta}^{2}\\
2\lambda_{7}^{i}\,s_{\beta}^{2}\,+\,\lambda_{5}^{i}\,s_{\beta}c_{\beta} &
-2\lambda_{7}^{i}\,s_{\beta}c_{\beta}\,-\,\lambda_{5}^{i}\,c_{\beta}^{2} 
\end{array}
\right)  ,\label{CPVMassMat}%
\end{equation}
where $\lambda_k^i \equiv {\rm Im}\, \lambda_k$. However, as we
show in Sect.~\ref{sec:ltbeft} below, the individual
phases of $\lambda_5$ and $\lambda_7$ become
unphysical in the limit $\tan\beta \to \infty$, and
mixing between the CP-even and CP-odd sectors is described
by a single angle $\alpha'$, determined by the relative phase of $\lambda_5$
and $\lambda_7^2$.
Finally, the charged Higgs mass matrix is given by
\begin{equation}
M_+^{2}=\left(  1+\frac{v^{2}\left(  \lambda_{4}
+\lambda_{5}^{r}\right)  }{2M_{A}^{2}}\right)  \text{ }M_{I}^{2}
.\label{ChargedMassMat}
\end{equation}
Here no $\tan\beta$ enhancement due to loop-induced couplings occurs.

Unlike the case of the fermion mass matrix, the typical momentum
flowing through the effective Lagrangian \eq{2.1} for an on-shell
Higgs is itself of ${\cal O}(v)$ or ${\cal O}(M_A)$. Hence Higgs-loop
contributions to the Higgs masses cannot be included  in~\eq{2.1}, but
rather the full effective action would be needed. Higgs-loop
effects in $\kappa_{bq}$ and $\kappa_{qb}$ multiplying ${\cal F}^\pm$
could, however, be included via \eq{eq:LY}, since again the momenta
flowing through the vertices are much smaller than $v$, $M_A$. This is not
possible in Higgs boxes, where large momenta flow through
the FCNC vertices. We will present a systematic method to include all
Higgs-loop contributions in Sect.~\ref{sec:ltbeft}.

It is instructive to consider the explicit form of the numerator
in~\eq{3.14}, which is
\begin{align}  \label{eq:Fmnum}
\det A \; = & \; v^4 \Big[
  ( \lambda_2 \lambda_5^* - \lambda_7^{*2} )\, s^4_\beta
+ 2 \, (\lambda_2 \lambda_6^* - \lambda_3 \lambda_7^* + \lambda_5^*
          \lambda_7)\, s^3_\beta c_\beta \nonumber \\
& + ( \lambda_1 \lambda_2 - \lambda_3^2 + |\lambda_5|^2
          - 2 \lambda_6 \lambda_7^* + 4 \lambda_6^* \lambda_7 )\,
          s^2_\beta c^2_\beta \nonumber \\
& + 2 (\lambda_5 \lambda_6^* - \lambda_3 \lambda_6
          + \lambda_1 \lambda_7) s_\beta c^3_\beta
 + (\lambda_1 \lambda_5 - \lambda_6^2) c^4_\beta \Big] .
\end{align}
With \eq{2.4}, $\det A = v^4 \,  (\lambda_1 \lambda_2
- \lambda_3^2) \, s^2_\beta c^2_\beta = 0$,
reproducing the known vanishing of ${\cal F}^-$
employing the tree-level MSSM Higgs sector.
The cancellation  is removed already at the
leading-logarithmic level. For instance, $\lambda_2$ alone receives a large
additive correction $\propto y_t^4$ due to top-quark loops, which is
also responsible for the most important correction to the tree-level mass of
$h$. The corresponding corrections could be computed by RG-evolving
the tree-level couplings in the effective 2HDM. However, as we are
considering large $\tan\beta$,
we expect (and find below) the most important effect to be due to
$\lambda_5$ and $\lambda_7$, which remove the ${\cal O}(c^2_\beta)$
suppression of the leading-log result, as anticipated above.

\subsection{The case of minimal flavour violation}
\label{sec:higgsmed-mfv}
From the discussion so far it follows that
$|{\cal F}^+| = {\cal O}(1/M_A^2) \gg |{\cal F}^-| = {\cal O}(1/(16
\pi^2 M_A^2))$, implying $|C_2^{\rm LR}| \gg |C_1^{\rm SLL}|$
for generic $\kappa_{ij}$,\footnote{In
Ref.~\cite{D'Ambrosio:2002ex}, an argument based on $SU(2) \times
U(1)$ gauge invariance was used to infer that (in the present
notation) ${\cal F}^- = {\cal O}(v^2/M_A^4)$. This
statement, which clearly is respected by our \eq{3.14} in conjunction with
\eq{eq:Fmnum} (recall $M_h^2 = {\cal O}(v^2)$),
is about the asymptotic behaviour as $v/M_A \to 0$. The latter is
not necessarily a small number in practice. Indeed, many of the
analyses in the literature have dealt with the case $M_A \sim 200\,
{\rm GeV}$.
}
such that the motivation to consider ${\cal F}^-$ at all
is not very strong. The situation is fundamentally different
for MFV because then the contribution proportional
to ${\cal F}^+$ turns out to be
suppressed by a light quark mass, introducing a further small
parameter $m_q/m_b$
comparable to $1/(16 \pi^2)$ or
$1/\tan\beta$ for $q=s$ (and negligible for $q=d$).
For simplicity, in
this paper we consider the simplest version of
MFV, assuming flavour-universal soft
breaking terms $\tilde{m}^2_Q$, $\tilde{m}^2_u$ and
$\tilde{m}^2_d$ and trilinear SUSY-breaking terms $T_{u_{ij}}$,
$T_{d_{ij}}$ which are proportional to the Yukawa matrices and
therefore diagonal in the super-CKM basis 
(denoted with a hat): $\hat{T}_{u_{ij}} = a_{t}
y_{u_i} \delta_{ij}$ and $\hat{T}_{d_{ij}} = 
a_{b} y_{d_i} \delta_{ij}$,
see Appendix~\ref{sect:not} for details of our notation.
The structure of our results, however, does not depend on these
additional assumptions.
The $\tan\beta$-enhanced loop-induced FCNC couplings of the neutral Higgs
bosons in \eq{eq:LY} can be expressed as: 
\begin{eqnarray}
\kappa_{bq}&=&\epsilon_{Y}\text{ }y_{t}^{2}\lambda_{qb}\frac{\sqrt{2}\,m_{b}%
}{v\cos^{2}\beta}\frac{1}{1+\widetilde{\epsilon}_{3}\tan\beta}\frac
{1}{1+\epsilon_{0}\tan\beta}\, ,\label{3.16} \\
\kappa_{qb}&=&\epsilon_{Y}\text{ }y_{t}^{2}\lambda_{qb}^*\frac{\sqrt{2}\,m_{q}%
}{v\cos^{2}\beta}\frac{1}{1+\widetilde{\epsilon}_{3}\tan\beta}\frac
{1}{1+\epsilon_{0}\tan\beta}\, ,
\label{3.16bis}%
\end{eqnarray}
with $y_{t}=\sqrt{2}m_{t}/(v\sin\beta)$ and $\lambda_{qb} = V_{tq}V_{tb}^*$.
The effective couplings $\epsilon_{Y}$, $\epsilon_{0}$ and $\widetilde
\epsilon_{3}$, which depend on the MSSM parameters, have been analysed in the
decoupling limit $M_{\rm SUSY}\gg v$ in the limit $g=g^\prime=0$ in
Refs.~\cite{hpt,bk,ir} for the case that $\epsilon_{Y}$,
$\epsilon_{0}$ and $\widetilde \epsilon_{3}$ are real. We consider the
general case allowing $\mu$, the universal trilinear term $a_t$ and the
gaugino mass parameters to be complex. Effects from non-zero
$g,g^\prime$ have been taken into account in Ref.~\cite{bcrs}, where also
effects beyond the decoupling limit were considered. The corresponding
expressions for $M_{\rm SUSY}\gg v$, suited for our analysis, were derived in
Ref.~\cite{fgh}. We have recalculated the FCNC couplings of neutral Higgs
bosons including all CP-violating phases and found agreement with the results
for the FCNC self-energies given in Ref.~\cite{bcrs}, but encountered a
significant discrepancy with Ref.~\cite{fgh}.  In our results,
the phase conventions of the first five parameters can be inferred from
\eqsand{defchm}{deftril} of Sect.~\ref{sect:not}. The phase convention for
$M_3$ complies with that of $M_{1,2}$ and the gluino mass equals
$M_{\widetilde{g}}=|M_3|$.  Of course one can choose one of these parameters
(e.g.\ $M_3$) real.  Now the effective couplings of \eqsand{3.16}{3.16bis}
read:
\begin{align}
\epsilon_{0} &  =\frac{-2\alpha_{s}}{3\pi}\frac{\mu^*}{M_3}%
H_{2}\left(  \frac{M_{\widetilde{b}L}^{2}}{|M_3|^{2}}%
,\frac{M_{\widetilde{b}R}^{2}}{|M_3|^2}\right) \nonumber\\
&  +\frac{g^{\prime2}}{96\pi^{2}}\frac{\mu^*}{M_{1}}
\left[  H_{2}\left( \frac{M_{\widetilde{b}L}^{2}}{|M_1|^{2}},
                  \frac{|\mu|^{2}}{|M_1|^{2}}\right)
+2H_{2}\left(  \frac{M_{\widetilde{b}R}^{2}}{|M_{1}|^{2}},
             \frac{|\mu|^{2}}{|M_{1}|^{2}}\right)  \right] \nonumber\\
&  +\frac{g^{\prime2}}{144\pi^{2}}\frac{\mu^*}{M_{1}}H_{2}
\left(  \frac{M_{\widetilde{b}L}^{2}}{|M_{1}|^{2}},
        \frac{M_{\widetilde{b}R}^{2}}{|M_{1}|^{2}}\right)  
+\frac{3\, g^{2}}{32\pi^{2}} \frac{\mu^*}{M_{2}} H_{2}\left(  
        \frac{M_{\widetilde{b}L}^{2}}{|M_{2}|^{2}},
        \frac{|\mu|^{2}}{|M_{2}|^{2}} \right) ,\label{3.17}\\
\epsilon_{Y} &  =\frac{-1}{16\pi^{2}}\frac{a_{t}^*}{\mu}H_{2}\left(
        \frac{M_{\widetilde{t}L}^{2}}{|\mu|^{2}},
        \frac{M_{\widetilde{t}R}^{2}}{|\mu|^{2}}\right) \; +\; 
     \epsilon_{Y,v/M}\, ,\label{3.18}\\
\widetilde{\epsilon}_{3} &  =\epsilon_{0}+y_{t}^{2}\epsilon_{Y}\, .\label{3.19}%
\end{align}
Here 
\begin{equation}
H_{2}(x,y)=\frac{x\log x}{(1-x)(x-y)}+\frac{y\log y}{(1-y)(y-x)}.\label{3.20}%
\end{equation}
Numerically, the electroweak contributions in $\epsilon_0$ can
be of ${\cal O}(10\%)$. They improve the comparison with the results computed
with full chargino and squark mass matrices (see 
Eq.~(5.1) in the second paper in \cite{bcrs}). 

Ref.~\cite{bcrs} also discusses threshold corrections to the fermion
kinetic operators (wave function renormalizations). While these terms are
not $\tan\beta$-enhanced, the flavour-diagonal quark wave function
renormalization constants receive sizable contributions from squark-gluino
loops. One can parametrize these loops in terms of a new quantity
$\epsilon_0|_{\rm kin}$ which will add to $\epsilon_0$ in the relation
between the MSSM Yukawa coupling $y_{d_i}$ and the physical quark mass
$m_{d_i}$ (see \eq{defy} for the case of the bottom Yukawa coupling).
$\epsilon_0|_{\rm kin}$ will likewise appear in the relation between
$\kappa_{ij}$ and $y_{d_i}$, but it drops out once $\kappa_{ij}$ is
expressed in terms of $m_{d_i}$, so that it does not appear in
\eqsand{3.16}{3.16bis}.  This cancellation of the flavour-diagonal 
quark wave function renormalization can be verified by inserting
Eq.~(2.29) into Eq.~(2.26) of the second paper in \cite{bcrs}. This feature
can be traced back to the fact that the wave function renormalization
affects both the tree-level and the loop-induced Yukawa couplings with the
same multiplicative factor.

Comparing our result with Ref.~\cite{fgh}, we find different results for
$\epsilon_{0}$ and $\epsilon_{Y}$: In Ref.~\cite{fgh} the chargino-stop
contribution proportional to $g^2$ is erroneously assigned to
$\epsilon_{Y}$ rather than $\epsilon_0$. Since this piece does not
contain any Yukawa couplings (the chargino is a pure wino here), all
three generations contribute in the same way and the resulting overall
CKM structure combines to $V_{ub}^* V_{uq}+V_{cb}^* V_{cq}+ V_{tb}^*
V_{tq}$, which is zero for $q\neq b$ and equal to one for $q=b$. This
GIM cancellation eliminates the wino-stop loop from $\epsilon_{Y}$,
while this loop contributes to $\epsilon_0$ twice as much as the
corresponding loop with a neutral wino-like neutralino and a sbottom.
The two terms are combined into the last term in \eq{3.17}. Omitting the
chargino loop here would violate SU(2) gauge symmetry, which also
enforces $\widetilde t_L=\widetilde b_L$ in the decoupling limit. Since
$\epsilon_{Y}$ normalises all Higgs-induced FCNC couplings, one should
verify the accuracy of the $M_{\rm SUSY}\gg v$ limit: It is easy to
include the $\tan\beta$-enhanced contributions to $\epsilon_Y$ to all
orders in $v/M_{\rm SUSY}$. To this end one merely has to calculate the
FCNC $\ov b_R q_L$ self-energy using the exact chargino and up-squark
mass eigenstates. This self-energy renormalises the off-diagonal pieces
of the quark mass matrix and cause the mismatch between the flavour
structures of the latter with the Yukawa couplings leading to
$\epsilon_Y\neq 0$. In higher loop-orders $\tan\beta$-enhanced
contributions are suppressed by products of small CKM elements (and are
negligible) or are flavour-conserving and therefore contribute to
$\epsilon_0$ rather than to $\epsilon_Y$.  Using the $\ov b_R q_L$
self-energy $\lt( \Sigma^d_{mL} \rt)^{3i}$ (with $q=d_i$) from
Ref.~\cite{bcrs} one finds
\begin{equation}
\begin{split}
\epsilon_{Y,v/M} =& \frac{1}{16\pi^{2}}\frac{a_{t}^*}{\mu}H_{2}\left(
     \frac{M_{\widetilde{t}L}^{2}}{|\mu|^{2}},
     \frac{M_{\widetilde{t}R}^{2}}{|\mu|^{2}}\right) \; + \;
\frac{\sqrt{2}}{v y_t^2 \lambda_{bd_i} }  
\frac{\lt(\Sigma^d_{mL}  \rt)^{3i}}{y_b}  
\end{split} 
\label{epsyvm}
\end{equation}
(Note that $\lt( \Sigma^d_{mL} \rt)^{3i} \propto y_b $ and be aware of the
different sign conventions for $y_b$ in \eq{defy} and Ref.~\cite{bcrs}.) We
stress that \eq{epsyvm} must be evaluated for $i \neq 3$, so that the GIM
cancellation of the above-mentioned wino-stop loop takes place.
Numerically one finds a marginal impact of $\epsilon_{Y,v/M}$: Setting all
supersymmetric massive parameters equal to a common value $M_{\rm SUSY}$,
one finds that $\epsilon_{Y,v/M}$ amounts to a mere 1.4\% correction to
$\epsilon_Y$ for $M_{\rm SUSY}=400\gev$.  Even for $M_{\rm SUSY}=150\gev$,
for which the expansion in $v/M_{\rm SUSY}$ formally breaks down,
$\epsilon_{Y,v/M}$ depletes $\epsilon_Y$ by as little as 8\%.
$\epsilon_{Y,v/M}$ also enters $\epsilon_0$ through \eq{3.18}. It can be
inferred from Ref.~\cite{cgnw} that this procedure indeed leads to the correct
all-order resummation of the $\tan\beta$-enhanced corrections involving $y_t$.
Corrections to $\epsilon_0$ beyond the $M_{\rm SUSY}\gg v$ limit from $g,g^\prime$
and $y_b$ are considered in Refs.~\cite{cgnw} and \cite{bcrs}. We remark that
no terms proportional to $y_b^2$ occur in \eqsto{3.17}{3.19}, because the
corresponding loops violate hypercharge and involve a suppression factor of
$v^2/M_{\rm SUSY}^2$.

We verify from \eqsand{3.16}{3.16bis} that $\kappa_{qb}^* \kappa_{bq}$
multiplying ${\cal F}^+$ in~\eq{3.9} is suppressed by a factor $m_q/m_b$
relative to $\kappa_{bq}^2$, which multiplies ${\cal F}^-$. Hence
$C_1^{\rm SLL}$ is naively leading (over $C_2^{\rm LR}$)
from the point of view of MFV alone, and a
meaningful analysis of $B_{q}-\overline{B}_{q}$ mixing requires a
systematic investigation of all leading 
corrections to its vanishing ``tree'' value.  (The coefficient
$C_1^{\rm SRR}$ both undergoes a strong $m_{q}^{2}/m_{b}^{2}$ suppression {\em
and} involves ${\cal F}^{-*}$, and can thus be disregarded.)  It is
then useful to think of the $\Delta F=2$ amplitude as being a function
of the four small parameters identified so far:
\begin{equation}    \label{eq:smallpar}
l \equiv \frac{1}{(4 \pi)^2}, \quad \omega \equiv \frac{m_q}{m_b},
\quad \frac{1}{\tan\beta}, \quad \nu = \frac{v}{M_{\rm SUSY}} .
\end{equation}
The vanishing 2HDM tree diagram for ${\cal F}^-$ is (superficially)
${\cal O}( (\cot\beta)^{-2}\, l^2\,  \nu^0\, \omega^0) $,
i.e.\ ${\cal O}(1)$ when treating all expansion parameters on the same
footing. Conversely, ${\cal F}^+$ is nonzero at the tree level but is
suppressed by one power of $\omega$, which is non-negligible only for
$q=s$. We have already seen that ${\cal F}^-$
vanishes exactly for tree-level matching (or up to ${\cal O}(1/\tan^2
\beta)$ when including leading logs), so there are no ${\cal O}(1/\tan
\beta)$ corrections at first subleading order.
This leaves loop corrections (via sparticle corrections to the $\lambda_i$ as
well as loops in the effective 2HDM) and possible corrections due to
higher-dimensional operators, not written in~\eqsand{eq:LY}{2.1}.
We now discuss these contributions in turn.

\paragraph{Sparticle loops}
One-loop contributions from higgsinos, gauginos, and sfermions correct
the values of $\lambda_{1,2,3.4}$ in \eq{2.1} and induce
non-zero couplings $\lambda_{5,6,7}$.  As a technical result of 
our paper, we have computed
the $\lambda_i$ for general sparticle masses and flavour structure.
These results are reported in Appendix \ref{sec:matching_results}.
At tree-level in the effective theory and in the leading
order of $1/\tan\beta$ the quantitiy ${\cal F^-}$ receives only
contributions from $\lambda_2$, $\lambda_5$, and $\lambda_7$, cf.~\eq{eq:Fmnum}. 
The general results of
Eqs.~(\ref{eq:2-1}),(\ref{eq:decla}),(\ref{eq:12}),(\ref{eq:9}),(\ref{eq:20}),
(\ref{eq:lambda14sl}-\ref{eq:lambda14sq}),
and (\ref{eq:19-2}) for the MFV case read
\begin{equation}
\label{eq:lambda7MFV}
\begin{split}
 \lambda_7  =\bar\lambda_{7}= & \frac{1}{16 \pi^2} \Bigg \{
 \frac{1}{4} \mu  a_{\tau} |y_{\tau }|^2
 \left(2 g^{\prime ^2} C_0 \left(\tilde{m}_l,\tilde{m}_e,\tilde{m}_e \right)
   + \left(g^2-g^{\prime ^2}\right)
   C_0 \left(\tilde{m}_e,\tilde{m}_l,\tilde{m}_l\right) \right) \\
 & + \mu a_{\tau} |\mu|^2 | y_{\tau} |^4
 D_0\left(\tilde{m}_e,\tilde{m}_e,\tilde{m}_l,\tilde{m}_l\right) \\
 & -\frac{1}{4} \tilde{g}^2 \mu
 \bigg(3 a_b |y_b|^2 B'_0 \left(\tilde{m}_d,\tilde{m}_Q\right)
   + 3 a_t |y_t|^2 B'_0 \left(\tilde{m}_u,\tilde{m}_Q\right)
   + a_{\tau} |y_{\tau }|^2 B'_0 \left(\tilde{m}_e,\tilde{m}_l\right)
 \bigg) \\
 & + g^4 \left(3 \mu  M_2
   \tilde{D}_2\left(|M_2|,|M_2|,|\mu |,|\mu
     |\right)-\frac{3}{4} \mu  |M_2|
   B'_0 \left(|M_2|,|\mu |\right)\right) \\
 & -\frac{1}{4} g^2 g^{\prime ^2} \mu \bigg(
   |M_1| B'_0 \left(|M_1|,|\mu |\right)
   + 3 |M_2| B'_0 \left(|M_2|,|\mu |\right)
   \\ & \qquad \qquad
   -4 \left(M_1+M_2\right)
   \tilde{D}_2\left(|M_1|,|M_2|,|\mu |,|\mu|\right)
 \bigg) \\
 & + g^{\prime ^4} \left(\mu  M_1
   \tilde{D}_2\left(|M_1|,|M_1|,|\mu |,|\mu
     |\right)-\frac{1}{4} \mu  |M_1|
   B'_0 \left(|M_1|,|\mu |\right) \right) \Bigg \}\, ,
\end{split}
\end{equation}
\begin{equation}
\label{eq:lambda5MFV}
\begin{split}
  \lambda_{5}  =\bar\lambda_{5}=& -\frac{1}{16 \pi ^2} \mu ^2 \bigg \{
    3 a_b^2 \left|y_b\right|{}^4
    D_0 \left(\tilde{m}_d,\tilde{m}_d,\tilde{m}_Q,\tilde{m}_Q\right)
    + 3 a_t^2 \left|y_t\right|{}^4
    D_0 \left(\tilde{m}_Q,\tilde{m}_Q,\tilde{m}_u,\tilde{m}_u\right)
    \\ &
    + a_{\tau }^2 \left|y_{\tau }\right|{}^4
    D_0 \left(\tilde{m}_e,\tilde{m}_e,\tilde{m}_l,\tilde{m}_l\right)
    - 3 g^4 M_2^2 D_0\left(\left|M_2\right|,\left|M_2\right|,|\mu  |,|\mu|\right)
    \\ &
    - 2 g^2 g^{\prime ^2} M_1 M_2
    D_0\left(\left|M_1\right|,\left|M_2\right|,|\mu |,|\mu|\right)
    - g^{\prime ^4} M_1^2 D_0
    \left(\left|M_1\right|,\left|M_1\right|,|\mu |,|\mu|\right)
    \bigg \}\, ,
\end{split}
\end{equation}
and
\begin{equation}
\label{eq:lambda2MFV}
\begin{split}
 \lambda_{2} =\bar\lambda_{2} =& \frac{\tilde{g}^2}{4} + \frac{1}{16 \pi^2}
 \Bigg \{
 -\frac{3}{4} g^{\prime ^4} B_0 \left(\tilde{m}_e,\tilde{m}_e\right)
 -\frac{3}{8} \left(g^4+g^{\prime ^4}\right)
 B_0 \left(\tilde{m}_l,\tilde{m}_l\right)
 \\ &
 + \frac{1}{2} \left(g^{\prime ^2}-g^2\right) |\mu y_{\tau }|^2
 C_0 \left(\tilde{m}_e,\tilde{m}_l,\tilde{m}_l\right)
 - g^{\prime ^2} |\mu y_{\tau}|^2
 C_0 \left(\tilde{m}_l,\tilde{m}_e,\tilde{m}_e\right)
 \\ &
 -|\mu  y_{\tau }|^4
 D_0 \left(\tilde{m}_e,\tilde{m}_e,\tilde{m}_l,\tilde{m}_l\right)
 \\&
 -\frac{1}{4} g^{\prime ^4} B_0 \left(\tilde{m}_d,\tilde{m}_d\right)+
 \left(-3 |y_t|^4+2 g^{\prime ^2} |y_t|^2-g^{\prime ^4}\right)
 B_0 \left(\tilde{m}_u,\tilde{m}_u\right)
 \\ &
 + \frac{1}{8} \left(
   -9 g^4-24 |y_t|^4-g^{\prime ^4}
   -4 |y_t|^2 \left(g^{\prime ^2}-3 g^2\right)\right)
 B_0 \left(\tilde{m}_Q,\tilde{m}_Q\right)
 \\ &
 + \frac{1}{2} \left(3 g^2-12 |y_t|^2-g^{\prime ^2}\right) |a_t y_t| ^2
 C_0 \left(\tilde{m}_Q,\tilde{m}_Q,\tilde{m}_u\right)
 \\ &
 + 2 \left(g^{\prime ^2}-3 |y_t|^2\right) |a_t y_t|^2
 C_0 \left(\tilde{m}_Q,\tilde{m}_u,\tilde{m}_u\right)
 - g^{\prime ^2} |\mu y_b|^2
 C_0 \left(\tilde{m}_Q,\tilde{m}_d,\tilde{m}_d\right)
 \\ &
 - \frac{1}{2} \left(3 g^2+g^{\prime ^2}\right) |\mu y_b|^2
 C_0 \left(\tilde{m}_d,\tilde{m}_Q,\tilde{m}_Q\right) \\
 &
 -3 |a_t y_t|^4
 D_0 \left(\tilde{m}_Q,\tilde{m}_Q,\tilde{m}_u,\tilde{m}_u\right)
 -3 |\mu  y_b|^4
 D_0 \left(\tilde{m}_d,\tilde{m}_d,\tilde{m}_Q,\tilde{m}_Q\right) \\
 & + \frac{1}{2} \tilde{g}^2 \left(
   3 |\mu y_b|^2 B'_0 \left(\tilde{m}_d,\tilde{m}_Q\right)
   + |\mu y_{\tau }|^2 B'_0 \left(\tilde{m}_e,\tilde{m}_l\right)
   + 3 |a_t y_t|^2 B'_0 \left(\tilde{m}_u,\tilde{m}_Q\right)
 \right)
 \\ &
 + \frac{1}{24} \left[ -2 \log \frac{\tilde{m}_d^2}{\mu_0^2}  g^{\prime ^4}
   -6 \log \frac{\tilde{m}_e^2}{\mu _0^2} g^{\prime ^4}
   -8 \log \frac{\tilde{m}_u^2}{\mu _0^2} g^{\prime ^4} \right.
 \\ &
 \qquad \qquad \left.
   -3 \log \frac{\tilde{m}_l^2}{\mu_0^2} \left(g^4+g^{\prime  ^4}\right)
   -\log \frac{\tilde{m}_Q^2}{\mu _0^2} \left(9 g^4+g^{\prime  ^4}\right) \right]
 \\ &
 -\frac{1}{24} g^4 \left[
   -12 \tilde{D}_2 \left(|M_2|,|M_2|,|\mu |,|\mu|\right) |M_2|^2
   -60 \tilde{D}_4\left(|M_2|,|M_2|,|\mu |,|\mu |\right)
 \right.
 \\ &
 \qquad \qquad \left.
   + 9 W\left(|M_2|,|\mu |\right)
   + 4 \log \frac{|\mu |^2}{\mu _0^2}
   + 8 \log \frac{M_2^2}{\mu _0^2}
   + 14
 \right] \\
 & -\frac{1}{8} g^2 g^{\prime ^2}
 \left[
   -8 {\rm Re} \left(M_1 M_2^* \right)
   \tilde{D}_2\left(|M_1|,|M_2|,|\mu |,|\mu
   |\right)
 -8 \tilde{D}_4\left(|M_1|,|M_2|,|\mu|,|\mu |\right)
 \right. \\ & \qquad \qquad \left.
   + W\left(|M_1|,|\mu |\right)
   + 3 W\left(|M_2|,|\mu |\right)
   + 4
\right] \\
& -\frac{1}{24} g^{\prime ^4} \left[
 -12 \tilde{D}_2\left(|M_1|,|M_1|,|\mu |,|\mu |\right)
 |M_1|^2
 -12 \tilde{D}_4 \left(|\mu|,|\mu |,|M_1|,|M_1|\right)
  \right. \\ & \qquad \qquad \left.
+3 W\left(|M_1|,|\mu |\right)
 +4 \log \left(\frac{|\mu |^2}{\mu _0^2}\right)
 +6
\right] \Bigg \} \, ,
\end{split}
\end{equation}
where the loop functions $B_0$, $C_0$, $D_0$, $B'_0$, $\tilde D_2$, $\tilde D_2$,
and $W$ are defined in Appendix \ref{sec:loop-functions},
and the notation $\bar\lambda_i$ refers to the matching scheme as explained in Sect.~\ref{sect:higgs}.
Inspecting \eq{eq:Fmnum}, $\lambda_7$ enters quadratically, which
formally is of higher loop order. Nevertheless, it can be seen that
$\lambda_7^2  \propto y_t^8$ as opposed to $\lambda_2 \lambda_5^*
\propto \tilde g^2 y_t^4$, which can partly offset the additional loop
suppression. Indeed we find that, numerically, neglecting $\lambda_7$
is not always a good approximation (Sect.~\ref{sect:phen}).

The form of the matching result depends on the
renormalization schemes of both the full theory, i.e.,\ the MSSM,
and the effective theory, i.e.,\ the 2HDM. The latter cancels in
physical quantities, while explicit MSSM scheme
dependence cancels against the one implicit in the MSSM parameters, to
ensure that the couplings in the effective theory are independent of
the renormalization of the MSSM at any given order of perturbation
theory. The residual scheme dependence in both cases may, however,
be important as we are considering a leading effect.
We will discuss scheme issues in Sect.~\ref{sec:renorm},
paying special attention to the definitions of $\tan\beta$.

\paragraph{Higgs loops}
There is a considerable number of one-loop diagrams in the effective
2HDM that can contribute to
$\bb$ mixing amplitudes
(Fig.~\ref{fig:weakloop}, upper row).
\begin{nfigure}{tb}
\centerline{\includegraphics[width=16cm]{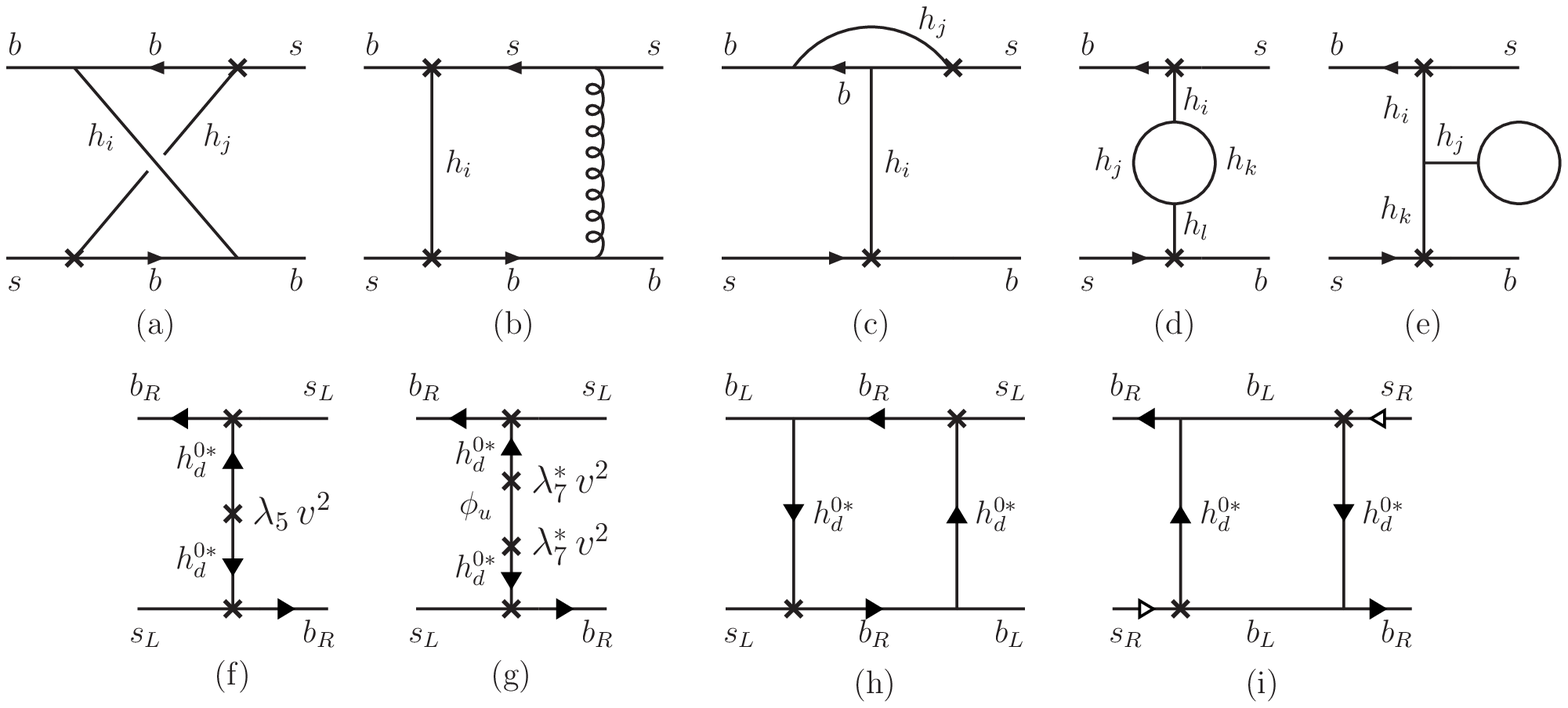}}
\caption{Upper row: A subset of one-loop diagrams for $B_q - \bar B_q$
mixing in the effective two-Higgs-doublet model. Lower row: 
Tree and one-loop diagrams contributing at large $\tan\beta$ when
employing the Lagrangian ${\cal L}_{\rm ltb}$ and tree-level couplings. The
crosses denote the flavor-changing neutral Higgs couplings and
(in diagrams (f) and (g)) loop-suppressed Higgs mass terms.
On the lower row, arrows designate
the flow of the conserved $U(1)$ charge discussed in Sect. \ref{sec:ltbeft}
\label{fig:weakloop}}
\end{nfigure}
These give the following contributions to the Wilson coefficients
multiplying $Q^{\rm VLL}_1$ and $Q^{\rm VRR}_1$:
\begin{eqnarray}
 C^{\rm VLL}_{1}|_{\rm{Higgs\, loops}} &=& -
\frac{1}{4} \frac{m_b^2}{v^2 \cos^2 \beta (1+\widetilde \epsilon_3^*\tan\beta)^2 }
\frac{\kappa_{bq}^2}{G_{F}^{2}M_{W}^{2}\lambda_{qb}^{2}} 
    C_0(M_A^2, M_A^2, 0) \, , \label{eq:higgsloopL}
\end{eqnarray}
\begin{eqnarray}
 C^{\rm VRR}_1|_{\rm Higgs\, loops} &=& - 
\frac{1}{4}\frac{m_b^2}{v^2 \cos^2 \beta (1+\widetilde \epsilon_3\tan\beta)^2 }
\frac{\kappa_{qb}^{*2}}{G_{F}^{2}M_{W}^{2}\lambda_{qb}^{2}}
    C_0(M_A^2, M_A^2, 0) \, .  \label{eq:higgsloopR}
\end{eqnarray}
In
these expressions, we have
neglected the small Yukawa coupling $y_q$ and employed tree-level MSSM
mass relations, in agreement with our approximation of working to leading
order in small parameters (in the present case, the loop factor $1/(16 \pi^2)$).
$C^{\rm VRR}_1$ is suppressed by two powers of $m_q/m_b$ inside
$\kappa_{qb}^{*2}$ in the MFV case, hence beyond our accuracy.
The results \eqsand{eq:higgsloopL}{eq:higgsloopR} involve a
great deal of cancellations, which can be understood in terms of symmetry
arguments, as explained in Sect.~\ref{sec:ltbeft} below.
We note the absence of charged-Higgs contributions in the
approximation considered here.

\paragraph{{\boldmath $v/M$}-suppressed effects}
All of the couplings given in \eq{eq:LY} correspond
to the zeroth order in the $v/M_{\rm SUSY}$ expansion, or
equivalently to the level of dimension-four operators. Gauge
invariance forbids dimension-five operators built from quark and
Higgs fields, so the leading higher-dimensional operators have
dimension six. This can lead to more general Higgs-fermion couplings than
those deriving from \eq{eq:LY}
and, in consequence, the cancellation leading to
$C^{\rm SLL}_1=0$ might be broken.
To see that this is indeed the case, consider the operator
\begin{equation}
Q^{(6)} = \frac{1}{M_{\rm SUSY}^2} (H_u^\dagger H_u) (\bar b_R H_u^\dagger Q_{L}),
\end{equation}
which gives rise, inter alia, to effective dimension-three and -four couplings
\begin{equation} \label{eq:vMcoup}
\frac{2 \sqrt{2}\, v_u^3}{M_{\rm SUSY}^2}\, \bar b_R s_L
+ \frac{2\,v_u^2}{M_{\rm SUSY}^2} (\bar b_R s_L h_u^{0} + 2\, \bar b_R  s_L h_u^{0*}).
\end{equation}
The first term is removed by a rediagonalization of the quark
mass matrices, but the two remaining terms, in general, are not.
The appearance of $h_u^0$ in addition to $h_u^{0*}$ leads
to a contribution to $C_1^{\rm SLL}$ proportional to $\kappa_{bq}\, C^{(6)}$.
However, because of $R$-parity, SUSY particles do not
contribute to tree graphs with external standard particles only,
such that $Q^{(6)}$ (or any other higher-dimension operator)
is only induced at the loop level,
and this loop-suppression factor is not compensated 
by factors of $\tan\beta$. (Recall that the ${\cal O}(1)$ FCNC couplings
at dimension four are nothing but rotated tree-level Yukawa couplings.)
Hence any $v/M_{\rm SUSY}$ corrections that break the cancellation in ${\cal F}^-$
involve an additional loop suppression, and can be neglected for
the present analysis.
On the other hand, as Eq.~(\ref{eq:vMcoup}) shows, the higher-dimensional
operators do have an impact on the rediagonalization of the quark mass
matrices and, consequently, on the size of the FCNC couplings
$\kappa_{bq}$. These effects preserve the cancellations in ${\cal
F}^-$ 
discussed above but have a mild impact on the 
FCNC couplings multiplying ${\cal F}^+$ in $C_2^{\textrm{LR}}$
(cf \eq{epsyvm} and the discussion around it).

\subsection
{\boldmath $U(1)_{\rm PQ}$ and effective Lagrangian for
  large $\tan \beta$  \unboldmath}
\label{sec:ltbeft}
To better understand the various types of cancellations in ${\cal
F}^-$ and in the Higgs-loop contributions to $C_1^{\rm VLL}$, as well
as the suppression of ${\cal F}^+$, we now introduce an effective
2HDM Lagrangian at large $\tan\beta$.
This will allow us, on the basis of simple symmetry arguments, to
clarify the role of the parameters $\lambda_5$ and
$\lambda_7$, the structure of Eqs.~(\ref{3.13bis}), (\ref{3.14}), and
(\ref{eq:Fmnum}), as well as the vanishing of ${\cal F}^-$ for
tree-level Higgs couplings at leading order  in $1/\tan\beta$.
It also provides a tool for computing loop diagrams involving
Higgs bosons efficiently and consistently,
which may be useful in other contexts such as collider processes
with Higgses in the initial or final state.

As before, we eliminate $m^2_{11}$, $m^2_{22}$,  and $(m^2_{12})^i$
by the minimization conditions
and trade $(m^2_{12})^r$ for $M_A^2$ via \eq{eq:madef}.
We then take the limit
\begin{equation}
v_d \to 0, \quad v_u \to v, \quad M_A^2 \mbox{ fixed},
\quad \lambda_i \mbox{ fixed},
\end{equation}
of the Lagrangian~(\ref{2.1}) in the broken phase.\footnote{This procedure will be justified
in Sect. \ref{sec:ltb_quantum}}
We also keep the Yukawa couplings fixed
when considering the couplings to fermions.
In this limit we have $\Phi = H_u$, $\Phi' = \epsilon H_d^*$, and
\begin{equation}
  h_u^0 = \frac{1}{\sqrt{2}} (v +  \phi_u + i\, G^0) , \quad
  h_d^0 = \frac{1}{\sqrt{2}} (\phi_d - i\, A^0), \quad
  h_u^+ = G^+, \quad
  h_d^+ = H^+ .
\end{equation}
If there were no mixing among neutral Higgses,
we would have $\phi_u = h^0$ and $\phi_d = H^0$, and $A^0$ would be
a mass eigenstate.
The mass matrices are compactly expressed by the quadratic potential
\begin{eqnarray}
V^{(2)}_{\rm ltb} &=& \Big[ m_A^2 + \frac{\lambda_5^r}{2} v^2 \Big] H_d^\dagger H_d
    + \frac{\lambda_4}{2} v^2 |h_d^+|^2 + \frac{\lambda_2}{2} v^2 \phi_u^2
\nonumber \\
&& + \Big[\frac{\lambda_5}{4} (h_d^{0*})^2
                      + \frac{\lambda_7}{\sqrt{2}} \phi_u h_d^{0*}
                      + \mbox{h.c.} \Big] v^2 \label{eq:V2ltb},
\end{eqnarray}
valid up to corrections of order $\cos \beta \sim 1/\tan\beta \ll 1$.
The trilinear terms are given in Appendix \ref{app:Vltb3}; the
quartic terms follow trivially from those in the symmetric Lagrangian
\eq{2.1}. Note that the
first line of~\eq{eq:V2ltb} is symmetric under the
$U(1)$ Peccei-Quinn (PQ) transformation
\begin{equation}
h_d^0 \to e^{-i \delta} h_d^0, \; h_d^+ \to e^{-i \delta} h_d^+,
\quad \mbox{or equivalently}, \quad  H_d \to e^{i \delta} H_d,
\label{U(1)Higgs}
\end{equation}
while the second line is not.
In the MSSM, the non-invariant terms appear only at the loop level.
We note that the $U(1)$ symmetry is not spontaneously broken in the
large-$\tan\beta$ limit, so there is no massless boson, in agreement
with our keeping $M_A^2$ fixed.\footnote{
Also at finite (but large) $\tan\beta$, there is no (pseudo-) Goldstone boson,
as $m_{11}^2 \sim M_A^2>0$ contributes to the
mass terms of both $\phi_d$ and $A^0$ (see also Sect.\ \ref{sec:ltb_quantum}).
}
Next, a PQ transformation makes $\lambda_5$ real, such that the
first term on the second line of \eq{eq:V2ltb} contributes
with opposite sign to the mass
terms for $\phi_d$  and $\chi_d = - A_0 + {\cal O}(\cos\beta)$,
splitting the two. There are
only two independent mixing angles that do not vanish: they can be
identified with the CP-conserving angle $\alpha = {\cal
 O}(\lambda_7^r)$ and a CP-violating 
$\alpha' = {\cal O}(\lambda_7^i)$; a third angle present in the general
2HDM is suppressed by ${\cal O}(\cot\beta; v/M)$.
All of these are symmetry-breaking effects.
To lowest order in the PQ-breaking couplings, the mass matrices
are diagonalized by
\begin{equation}
 \left( \begin{array}{c} H_1 \\ H_2  \\ H_3 \end{array} \right) =
 \left( \begin{array}{ccc}
   1 & -\frac{\lambda_7^r v^2}{M_A^2-\lambda_2 v^2} &
       \frac{\lambda_7^i v^2}{M_A^2 - \lambda_2 v^2} \\
   \frac{\lambda_7^r v^2}{M_A^2-\lambda_2 v^2} & 1 & 0 \\
   - \frac{\lambda_7^i v^2}{M_A^2 - \lambda_2 v^2} & 0 & 1
\end{array} \right)
 \left( \begin{array}{c} \phi_u \\ \phi_d  \\ A^0 \end{array} \right) ,
\end{equation}
\begin{equation}
 m_{1}^2 = \lambda_2 v^2, \qquad m_{2}^2 = M_A^2 + |\lambda_5| v^2,
\qquad    m^2_{3} = M_A^2 .
\end{equation}
In a general basis, CP-violating Higgs mixing is present
if and only if $\lambda_7^2/\lambda_5$ is complex. Note that there
is no mixing for the charged scalars according to \eq{eq:V2ltb},
i.e.\ no mixing between charged-Higgs and Goldstone bosons due
to sparticles in the large-$\tan\beta$ limit.

These considerations can be extended to the Higgs-fermion interactions.
The operators up to dimension four follow from~(\ref{eq:LY}), which, 
in the limit of infinite $\tan\beta$, becomes
\begin{equation}   \label{eq:LYltb}
\begin{split}
  {\cal L}^Y_{\rm ltb} =&
   - \frac{\sqrt{2}}{v} \bar d_{Ri}\, M_{d_{ij}}\, H_u^\dagger Q_{Lj}
   - \bar d_{Ri}\, \kappa_{ij}\, Q_{Lj} \cdot H_d \\
   &- \frac{\sqrt{2}}{v} \bar u_{Ri}\, M_{u_{ij}}\, Q_{Lj} \cdot \! H_u
   + \bar u_{Ri}\, \tilde \kappa_{ij} H_d^\dagger Q_{Lj}
  + \textrm{h.c.} \, .
\end{split}
\end{equation}
This can be made approximately invariant by extending the symmetry
transformation (\ref{U(1)Higgs}) to fermions. One judicious PQ charge
assignment is
\begin{equation}   \label{eq:U1ferm}
 d_{Ri} \to e^{i \delta} d_{Ri}, \quad Q_{Lj} \to Q_{Lj}, \quad
 u_{Rk} \to u_{Rk},
\end{equation}
which commutes with the SM
gauge group, implying that neutral and charged gauge boson couplings
respect the symmetry. It has been previously used
in~\cite{D'Ambrosio:2002ex}  to classify the Higgs-fermion couplings
in MFV.
However, since for MFV one has one more small parameter
$\kappa_{qb}/\kappa_{bq} \propto m_q/m_b$ for $q=s$ or $d$, 
it is useful to consider the following variant of \eq{eq:U1ferm}:
\begin{equation}   \label{eq:U1mfv}
 b_R \to e^{i \delta} b_R, \quad
 q_{R} \to q_{R}, \quad Q_{Lj} \to Q_{Lj}, \quad
 u_{Rk} \to u_{Rk}.
\end{equation}
Now $\kappa_{ij} \bar d_{Ri} Q_{Lj} \cdot H_d$  in \eq{eq:LYltb}
breaks the symmetry unless $d_{Ri}=d_{R3}=b_R$. However, all
$U(1)_{\rm PQ}$ breaking is still proportional to one of the small
parameters of \eq{eq:smallpar}:
$\kappa_{qj} = {\cal O}(\omega)$ and $\tilde \kappa_{ij} = {\cal O}(l)$.
The modified symmetry \eq{eq:U1mfv} forbids all
operators in the weak Hamiltonian \eq{eq:heff}
(Table \ref{tab:operatorcharges}),
including the would-be leading one, $Q_1^{\rm SLL}$, except for the
standard-model operator $Q_1^{\rm VLL}$ and for
$Q_{1,2}^{\rm SRR}$. The last two are, however, forbidden by the  original charge
assignment in \eq{eq:U1ferm}.
Hence the Wilson coefficients of these operators are suppressed
by $\omega=m_q/m_b$ or by factors of loop-induced effective couplings,
respectively.
\begin{ntable}
 \centering
\begin{tabular}{llll}
 Operator [field content] & $U(1)$ charge & \parbox{6cm}{Suppression of leading \\
   Higgs-mediated contribution} & Remark \\[4mm]
\hline \\
 $Q_{1,2}^{\rm SLL}\ [\bar b_R q_L \bar b_R q_L]$ & 2 &
       $\lambda_5$ (sparticle loop) & new \\[2mm]
 $Q_{1,2}^{\rm LR}\ [\bar b_R q_L \bar b_L q_R]$ & 1 [0] & $\omega$ & known \\[2mm]
 $Q_1^{\rm VLL}\ [\bar b_L q_L \bar b_L q_L]$ & 0 & 2HDM loop & SM operator \\[2mm]
 $Q_1^{\rm VRR}\ [\bar b_R q_R \bar b_R q_R]$ & 2 [0] & $\omega^2 \times$ 2HDM loop & tiny \\[2mm]
 $Q_{1,2}^{\rm SRR}\ [\bar b_L q_R \bar b_L q_R]$ & 0 [-2] & $\omega^2 \times $
 sparticle loop & tiny \\
\end{tabular}
\caption{Charges of the operators in the weak Hamiltonian under the
  approximate $U(1)$ symmetry discussed in the text, see
    Eq.~(\ref{eq:U1mfv}).
The number in brackets
  denotes the charge under the ``unmodified'' charge assignment
  of Eq.~(\ref{eq:U1ferm}).
\label{tab:operatorcharges}}
\end{ntable}

At the tree level (in the 2HDM), ${\cal F}^+$, which induces $Q_2^{\rm LR}$,
is multiplied by a factor $\kappa_{qb}^*$, which is a PQ-breaking
coupling. On the other hand, ${\cal F}^-$, which induces $Q_1^{\rm SLL}$,
is multiplied by the unsuppressed factor $\kappa_{qb}^2$. Hence
${\cal F}^-$ must be proportional to PQ-breaking couplings in the
Higgs potential (up to $1/\tan\beta$-suppressed terms). This also
seen from the fact that in the infinite $\tan\beta$ limit,
it is given by
$$
\int {\rm d}^4 x\, \Big\langle T\Big( h_d^0(x) h_d^0(0) \Big) \Big\rangle ,
$$
which vanishes if the PQ symmetry is unbroken. Explicitly, in
the large $\tan\beta$ limit one has:
\begin{align}
\mathcal{F}^{+} &  =\frac{2\lambda_{2}M_{A}^{2}+(\lambda_{2}\lambda_{5}%
^{r}-\left\vert \lambda_{7}\right\vert ^{2})v^{2}}{\lambda_{2}M_{A}%
^{4}+(\lambda_{2}\lambda_{5}^{r}-\left\vert \lambda_{7}\right\vert ^{2}%
)v^{2}M_{A}^{2}-(\lambda_{7}^{i}\operatorname{Im}(\lambda_{5}^{\ast}%
\lambda_{7})+%
\frac14
\lambda_{2}\lambda_{5}^{i2})v^{4}}\simeq\frac{2}{M_{A}^{2}},\label{3.21}\\
\mathcal{F}^{-} &  =\frac{-(\lambda_{2}\lambda_{5}^{\ast}-\lambda_{7}^{\ast
2})v^{2}}{\lambda_{2}M_{A}^{4}+(\lambda_{2}\lambda_{5}^{r}-\left\vert
\lambda_{7}\right\vert ^{2})v^{2}M_{A}^{2}-(\lambda_{7}^{i}\operatorname{Im}%
(\lambda_{5}^{\ast}\lambda_{7})+%
\frac14
\lambda_{2}\lambda_{5}^{i2})v^{4}}\simeq\frac{-(\lambda_{2}\lambda_{5}^{\ast
}-\lambda_{7}^{\ast2})v^{2}}{\lambda_{2}M_{A}^{4}} ,         \label{3.22}
\end{align}
where the rightmost expressions hold up to higher orders of
small couplings. For ${\cal F}^-$, this is identical
to the sum of the two leading diagrams in a ``mass-insertion
approximation'', where the PQ-breaking contributions to the
Higgs mass terms are treated as interactions (Fig. \ref{fig:weakloop}
(f) and (g)).

At the loop level (in the 2HDM), up to doubly suppressed
contributions one can employ the PQ-conserving parts of Eqs.~(\ref{eq:LYltb})
and (\ref{eq:V2ltb}), i.e.\ set $\lambda_5=\lambda_6=\lambda_7=0$,
as well as ignore $\kappa_{qb}$ and $\tilde \kappa_{ij}$.
The matching onto the weak Hamiltonian can be organized
according to one-light-particle-irreducible chirality amplitudes.
There are three amplitudes
\begin{eqnarray}
{\cal A}_{RR} &=& \Big\langle T \Big( b_R(x_1) b_R(x_2) \bar s_L(x_3) \bar
s_L(x_4) \Big) \Big\rangle, \\
{\cal A}_{RL} &=& \Big\langle T \Big(  b_R(x_1) b_L(x_2) \bar s_L(x_3) \bar
s_R(x_4) \Big) \Big\rangle, \\
{\cal A}_{VLL} &=& \Big\langle T \Big(  b_L(x_1) b_L(x_2) \bar s_L(x_3) \bar
s_L(x_4) \Big) \Big\rangle,
\end{eqnarray}
plus the parity conjugates of ${\cal A}_{RR}$ and ${\cal A}_{VLL}$.
(We have omitted amplitudes that
cannot match onto Lorentz-invariant local dimension-six operators.)
Only ${\cal A}_{VLL}$ is invariant under $U(1)_{PQ}$ (both versions) and can be
generated from a symmetric Lagrangian. It matches onto the standard-model
operator $Q_1^{\rm VLL}$.
There is a single diagram contributing, see Fig.~\ref{fig:weakloop}
(h). (Diagram (i) matches onto
$Q_1^{\rm VRR}$ and would be allowed for the unmodified PQ assignment
of \eq{eq:U1ferm}.)

The present discussion could  be extended to other
processes and to higher loop orders, by systematically treating the
PQ-breaking couplings as interactions and working to a fixed total
order in the small parameters; in practice, at such higher precision, one
might want to extend the effective 2HDM by higher-dimensional operators
to account for $v/M{\rm SUSY}$ corrections.

Finally, let us remark that because our choice of shift parameters $v_u$ and
$v_d$ minimize the potential $V$ in the Lagrangian of our effective theory
and not necessarily the full effective potential,
the one-point functions for the (shifted) Higgs fields
$\langle 0 | h_i | 0 \rangle$ ($h_i=\phi_u, \phi_d, A^0$) will,
in general, not vanish.
Hence also ``tadpole'' diagrams involving quark or Higgs loops
would have to be considered at the outset [Fig.~\ref{fig:weakloop}
(e)]. That they cancel in $\bb$ mixing in our
approximation follows from the fact
that no such diagrams are present when working with a complex $h_d^0$
field and the Lagrangian $V_{\rm ltb}$. Tadpoles may, however,
be relevant in other contexts.
We discuss our renormalization
of $v_u$, $v_d$, and $\tan\beta$ in detail in the following section.

\section{\boldmath Systematics of the large-$\tan\beta$ MSSM \unboldmath}
\label{sect:higgs}
The present section is devoted to certain technical aspects of
the large $\tan\beta$ limit. The first concerns the definition (i.e.\ 
renormalization) of $\tan\beta$ in the MSSM and in
the effective two-Higgs-doublet-model description of low-energy
(i.e., Higgs, electroweak, and flavour) phenomenology,
and the matching between the two. This is of phenomenological
importance, as $\tan\beta$ definitions used in the literature on the MSSM
are known to differ by parametrically large expressions
${\cal O}(\tan\beta \times {\rm loop\, factor})$.
This can lead to ambiguities in the value of $\tan\beta$ of 10-15
in certain regions of the MSSM parameter space between schemes that have
been extensively used in the study of radiative corrections to the
MSSM Higgs sector \cite{Freitas:2002um}.
Having clarified the connection between our ``full'' and ``effective''
$\tan\beta$, we justify the systematic expansion in
$1/\tan\beta$ at the Lagrangian level employed in Sect.\ \ref{sec:ltbeft}.

\subsection{\boldmath Renormalization of $\tan\beta$ \unboldmath}
\label{sec:renorm}
In the MSSM, $\tan\beta=v_u/v_d$ is defined as a ratio of vacuum
expectation values. This is an unambiguous notion at tree level,
because a preferred basis is provided by the chiral Higgs
supermultiplets of definite hypercharges $\pm 1/2$. Beyond tree level,
a scheme dependence arises as the bare parameters $p_i^0$
($p_i = m^2_1, m^2_2, B \mu, g, g'$, etc.) are renormalized,
$p_i^0 = p_i + \delta p_i$, as well as in the normalization of the fields
and in defining renormalized shift parameters $v_d$, $v_u$.
To formalize the renormalization program, we first define bare shifts that
minimize the bare effective potential including radiative corrections,
which is equivalent to requiring vanishing one-point functions for the
shifted fields, i.e.,
\begin{equation}      \label{eq:tadpole}
 \langle h_i^0 - \frac{1}{\sqrt{2}} v_i^0 \rangle \stackrel{!}{=}0 ,
\end{equation}
such that the $v_i^0$ are indeed vacuum expectation values.
Identifying (for any definition of renormalized shift parameters)
\begin{equation}      \label{eq:vevren}
v_i^0 = Z_i^{1/2} (v_i - \delta v_i) , \quad i=d,u ,
\end{equation}
scheme dependence arises through, and only through,
field renormalization and the counter\-terms $\delta v_i$.
Ref. \cite{Gamberini:1989jw} argued that for a stable perturbation
expansion it is desirable to define the
renormalized $v_i$ such as to minimize the renormalized effective
potential, i.e. $\delta v_i = 0$, and implemented this proposal
for $\overline{\rm DR}$ field renormalization and Landau gauge.
The same condition and gauge fixing was imposed in the computation of one-loop
corrections to the MSSM Higgs masses in
\cite{Okada:1990vk,Haber:1990aw,Ellis:1990nz,Brignole:1992uf}.
Refs. \cite{Chankowski:1992er,Dabelstein:1994hb} chose
to work with on-shell fields and in $R_\xi$ gauge instead, and their
shifts do not strictly minimize the one-loop effective potential. In fact,
in general gauges, for $\delta v_i=0$ the effective action
is not finite and the $v_i$ are both divergent
\cite{Weinberg:1973ua,Chankowski:1992er} and gauge
dependent \cite{Weinberg:1973ua,Nielsen:1975fs}
(as are the bare vevs $v_i^0$).\footnote{
This is in particular true in $R_\xi$ gauges if $\xi \not = 0$.
The apparent contradiction to the results
in \cite{Lee:1974zg}, whose authors are able to renormalize
the effective action with
purely ``symmetric'' counterterms, is resolved by noticing that
in the Lorenz gauge employed in \cite{Lee:1974zg} the gauge-fixed
Lagrangian still respects an invariance under constant (``global'')
gauge transformations. This is sufficient to forbid divergences that
cannot be removed by symmetric counterterms. Conversely, the
$R_\xi$ gauges break also this global invariance, for instance
through Goldstone and ghost mass terms, which are indeed responsible
for the ``non-symmetric'' divergences at one loop \cite{Weinberg:1973ua}.
The exception is the Landau gauge $\xi=0$, which has the invariance.
}
Hence to have finite renormalized $v_i$ and $\tan\beta$,
$\delta v_i \not = 0$, containing a gauge-dependent divergence, is
required.
For $\tan\beta$, we have
\begin{equation}   \label{eq:tanbetaren}
\tan \beta^0 \equiv \frac{v_u^0}{v_d^0} \stackrel{\rm 1-loop}{=}
\tan\beta \Big(1 + \frac{1}{2} \delta Z_u - \frac{1}{2} \delta Z_d
        - \frac{\delta v_u}{v_u} + \frac{\delta v_d}{v_d} \Big)
\equiv \tan\beta + \delta \tan\beta .
\end{equation}
Minimal subtraction for $Z_u, Z_d$, $\delta v_u/v_u$,
$\delta v_d/v_d$ defines $\tan\beta^{\overline {\rm DR}}$ \cite{Brignole:1992uf}.
It also  follows from \eq{eq:tanbetaren}
that a change between two schemes $R$ and $R'$ can be calculated from
$$
 \tan\beta^R - \tan\beta^{R'} = \delta \tan\beta^{R'} - \delta
 \tan\beta^R ,
$$
hence any scheme where $\delta \tan\beta $ is a pure divergence
has $\tan\beta=\tan\beta^{\overline{\rm DR}}$ regardless of
any nonminimal field
renormalizations as those employed in \cite{Freitas:2002um}.  In the
latter case,
however, $\delta v_u$, $\delta v_d$ are nonminimal
and the counterterm
for $\tan\beta$ has no simple relation to the field renormalization constants.

$\tan\beta^{\overline{\rm DR}}$ is gauge dependent \cite{Yamada:2001ck}, but
to one-loop order, the gauge-dependence drops out for the $R_\xi$ gauges.
In spite of its gauge dependence, the ${\overline{\rm DR}}$
scheme for $\tan\beta$ has been shown
to lead to a well-behaved perturbation expansion \cite{Freitas:2002um}
and is also used in the most recent version of the publicly available
computer programs FeynHiggs \cite{Frank:2006yh} and CPsuperH \cite{Lee:2007gn}.

A second issue is that a fully minimal subtraction scheme, where in
particular $\delta v_i^{\rm finite}=0$, generally
entails $v_i$ that do not minimize the (renormalized) tree
potential, such that the renormalized Lagrangian contains linear terms
\begin{equation}
{\cal L} \supset t_d \phi_d + t_u \phi_u  
\end{equation}
for the shifted (real parts of the) Higgs fields.
On the other hand, from \eq{eq:tadpole} and \eq{eq:vevren} it follows that
\begin{equation} \label{eq:tadpoleren}
\Gamma_i^{\rm ren} =  t_i + \Gamma^{(1)}_i + \delta t_i = 0
\end{equation}
always holds, if only $\delta v_u$ and $\delta v_d$ are included in
$\delta t_i$.
The presence of $t_u$, $t_d$ is perfectly fine, but
tadpole diagrams then have to be retained in the calculation.
(In particular, they appear in the expressions relating Higgs and
gauge boson mass parameters to the Lagrangian parameters. 
If all renormalization constants are minimal, \eq{eq:tadpoleren}
determines $t_i$ in terms of the bare proper one-point functions
$\Gamma^{(1)}$ \cite{Brignole:1992uf}.) Yet it may be more convenient to
perform additional finite renormalizations to work in a
scheme where $t_i=0$. This can be achieved either by suitable finite
terms in $\delta v_i$ or by finite renormalizations of the mass and
coupling parameters. The former shifts
$\tan\beta$ from its $\overline{\rm DR}$ value according to
\begin{equation}
\label{eq:tanbetatad}
\tan\beta^{\textrm{tad}} = \tan \beta^{\overline{\rm DR}} \left( 
  1 -
  \frac{\delta v_d^{\textrm{tad}}}{v_d} +
  \frac{\delta v_u^{\textrm{tad}}}{v_u} \right) \, .
\end{equation}
The latter option does not modify $\tan\beta$.

Going from the MSSM to a general 2HDM, $\tan\beta$
becomes -- strictly speaking --
an ill-defined notion as there is no preferred basis.
Identifying $H_1 = -\epsilon H_d^* $ and $H_2 = H_u$,
an $SU(2)$ rotation $H_i \to U_{ij} H_j$ removes the vacuum expectation
value of one
doublet; this corresponds to the $(\Phi, \Phi'$) basis introduced in
Sect.~\ref{sect:bbm}. Only $\Phi$ receives a vev, provides for the Higgs
mechanism, and has flavour-conserving couplings, while $\Phi'$ is an ordinary
scalar with FCNC couplings. To make contact with MSSM phenomenology, however,
it is useful to keep the notion of $\tan\beta$ in the effective theory. In
principle, we could fix a basis to enforce $\tan\beta^{\rm EFT} \equiv
\tan\beta^{\overline{\rm DR}}$, but find it technically simpler to
allow for a parametrically small (i.e. not $\tan\beta$-enhanced)
shift, as we discuss in the following.

In complete analogy with the MSSM case dicussed above, if
we employ a general gauge and $\overline{\rm MS}$ everywhere in the
effective theory, $v_u$ and $v_d$ will not minimize the tree-level
(nor the effective) potential. This would require a modification
of the formalism in Sect. \ref{sect:bbm}. In particular, in writing
the  mass matrices \eqsto{Delta}{eq:madef} and the
flavour structure of the scalar-fermion couplings in \eq{eq:LY} we assumed the
minimization conditions $t_1=t_2=0$. To
avoid such modifications, as well as changed
expressions for neutral meson mixing, we can either perform
renormalizations on the parameters $m^2_{11}$ and $m^2_{22}$ such
that $v_1$ and $v_2$ minimize the 2HDM potential, or achieve this through
nonminimal $\delta v_{1,2}$. We pursue the latter option,
keeping the symmetric parameters of the 2HDM minimally
subtracted.
This has the added virtue that the $\tan\beta$ such
defined is gauge independent at the order considered, as it is fully
determined by $\overline{\rm MS}$ mass and coupling parameters. These
are gauge invariant at one loop, which is clear from our explicit
matching calculation.
We presume this to hold also at higher orders, at least if the
appropriate wave-function renormalization is employed.
The $\delta v_i$ are determined entirely in terms of
``light''-particle loops and, at
least at one loop, do not lead to parametrically large
shifts $\propto \tan^2\beta \frac{1}{16 \pi^2}$, as can be verified
from the explicit expressions for the tadpoles in
\cite{Dabelstein:1994hb} or by considering tadpole diagrams in the
large-$\tan\beta$ effective Lagrangian.

To find the precise connection between $\tan\beta^{\overline{\rm DR}}$
and our effective $\tan\beta$, consider the total tree plus one-loop
contribution of the superpartners to the (MSSM) effective action for the
gauge and Higgs fields,
\begin{eqnarray}     \label{eq:effact}
S_{gh} &=& \int {\rm d}^4 x \Bigg[
          (1 + \Delta Z_W) (-\frac{1}{4}) W_{\mu\nu}^A W^{\mu\nu A}
          + (1 + \Delta Z_B) (-\frac{1}{4}) B_{\mu\nu} B^{\mu\nu}
\nonumber \\ & & \qquad \quad
          + (\delta_{ij} + \Delta Z_{ij}) (D_\mu H_i)^\dagger (D^\mu H_j)
 - \hat m^2_{ij} H_i^\dagger H_j
 - \sum_{k=1}^7 \hat \lambda_k O_k
          + \dots \Bigg] .
\end{eqnarray}
Here $O_i$ are the quartic terms
constructed from the Higgs fields appearing in \eq{2.1}, and the dots denote
higher-dimensional local terms. The precise values for the
coefficients depend on
the MSSM renormalization scheme. We assume the MSSM has been regularized by
dimensional reduction while the Higgs fields and $\tan \beta$ are minimally
subtracted ($\overline{\rm DR}$). The corresponding expressions 
$\hat \lambda_k$ are reported in \eq{2.4} (tree level) and in
Appendices \ref{sec:higgs-gaug-contr} and \ref{sec:sferm-contr-lambd} (one
loop).

\eq{eq:effact} can be identified with the classical action (ignoring
2HDM loops) for an effective
two-Higgs-doublet model with noncanonically normalized fields. To obtain from
this the $\overline{\rm MS}$-renormalized Lagrangian in the presence of
light-particle loops one simply has to add the contributions (which are local)
due to loops of $2\epsilon$ scalars present in DRED \footnote{Integrating over
the $2\epsilon$ scalars leaves a path integral over light fields that is
identical to that in the DREG-regularized effective theory, including the
$1/\epsilon$ divergence structure. We recall that the $2\epsilon$
scalars should be thought of as having a nonzero mass of ${\cal O}(M_{\rm
  SUSY})$ \cite{Jack:1994rk}.
}  and subsequently rescale the fields,
\begin{equation}  
\label{eq:5}
\left(
  \begin{array}{c}
    - \epsilon H_d^{\overline{\rm DR}} \\ H_u^{\overline{\rm DR}}
  \end{array}
\right)
=
\left(
  \begin{array}{cc}
    Z_{dd} & Z_{du} \\ Z_{ud} & Z_{uu} 
  \end{array}
\right)
\left(
  \begin{array}{c}
    H_1^{\overline{\rm eff}} \\ H_2^{\overline{\rm eff}}
  \end{array}
\right)
\; ,
\end{equation} 
subject to the condition $Z^\dagger (1 + \Delta Z) Z= {\bf 1}$.
This provides the relation between the $\overline{\rm DR}$ fields of the MSSM
and one out of an infinite choice of $\overline{\rm MS}$ fields in the
effective theory, labeled ``${\overline{\rm eff}}$''.
We fix the freedom to choose the Higgs basis in the effective theory by setting
$Z_{du}=0$ and $Z_{uu}^i=Z_{dd}^i=0$.
The relation between the shifts and $\tan\beta$ of the MSSM and of the 2HDM 
are now determined according to
\begin{equation}
\label{eq:vevmatch}
\begin{split}
\bar v_2(\mu) \equiv v_2(\mu)^{\overline{\rm eff}} &= 
v_u^{\overline{\rm DR}} - \delta Z_{ud} v_d
- \delta Z_{uu} v_u +
\delta v_2^{\rm tad} \\
\bar v_1(\mu) \equiv v_1(\mu)^{\overline{\rm eff}} &= 
v_d^{\overline{\rm DR}} - \delta Z_{dd} v_d + \delta v_1^{\rm tad}\\    
\tan\beta(\mu)^{\overline{\rm eff}} &= \tan\beta^{\overline{\rm DR}}
     \left(1 - \frac{\delta v_1^{\rm tad}}{v_1}
             + \frac{\delta v_2^{\rm tad}}{v_2}
             + \delta Z_{dd} - \delta Z_{uu}
             - \delta Z_{ud} \cot\beta \right) . \\
\end{split}
\end{equation}
Here we have expanded $Z_{uu/dd} = {\bf 1} + \delta Z_{uu/dd}$
and $Z_{ud} = \delta
Z_{ud}$, and the  $\delta Z_{ij}$ are related to the $\Delta Z_{ij}$ via
$\Delta Z_{11/22} = - 2 \delta Z_{uu/dd}$ and 
$\Delta Z_{12} = - \delta Z_{ud}^*$, with the explicit expressions
given in Appendix  \ref{sec:renorm-const}.
The shifts $\delta v_{1,2}^{\rm tad}$ are defined implicitly
as discussed above. In summary, we have constructed
a $\tan\beta$ which is
appropriate for effective weak interactions, gauge-independent
and, up to an ordinary (i.e.,\ not
$\tan\beta$-enhanced) loop correction, coincides with the widely used
$\tan\beta^{\overline{\rm DR}}$.  It means that the $\tan\beta$
measured in flavour physics, for instance through
$BR(B_s^0 \to \mu^+ \mu^-)$, and employed in our analysis,
can be identified with the corresponding DCPR parameter at large
$\tan\beta$, up to small corrections.

We note that our framework leads to a transparent expression for the
relation between the ${\overline{\rm DR}}$ scheme and the
so-called DCPR scheme employed in
\cite{Chankowski:1992er,Dabelstein:1994hb} in the limit $v \ll M_{\rm SUSY}$.
In the latter scheme, finite but, unlike in our effective 2HDM,
``diagonal'' wave function renormalizations of $H_u$, $H_d$
are performed, i.e., in our notation, $\delta Z_{ud}=\delta Z_{du}=0$.
Moreover, the renormalization conditions include
\begin{equation} \label{eq:DCPR}
  \frac{\delta v_u}{v_u} = \frac{\delta v_d}{v_d}, \qquad
  {\rm Re}\, \Sigma_{A^0 Z^0} ( M_A^2) = 0 ,
\end{equation}
where $\Sigma_{A^0 Z^0} (k^2)$ parameterizes the $A^0$-$Z^0$
mixing according to $\Sigma^\mu_{A^0 Z^0}(k) = k^\mu \,
\Sigma_{A^0 Z^0}(k^2)$.
Now, the sparticle contribution to $\Sigma_{A^0 Z^0}(k^2)$ reads
\begin{equation}
 \Sigma_{A^0 Z^0}(k^2) =
    \sin^2 \beta \, \Delta Z_{du}
    + \sin\beta \cos\beta (\delta Z_{uu} - \delta Z_{dd}) + \dots ,
\end{equation}
where the dots denote terms proportional to $\cos \beta$ but not involving
the wave-function renormalization constants. This follows
either by considering the mixed gauge boson-Higgs boson bilinear terms
resulting from the covariant kinetic operator for the Higgs fields in
\eq{eq:effact}, or via the Ward identity
\begin{equation}
  k_\mu \Sigma^\mu_{A^0 Z^0}(k^2) + M_Z \Sigma_{G^0 A^0}(k^2)
 = {\cal O}(k^2 - M_A^2)
\end{equation}
(which is trivially satisifed in our $SU(2)$-invariant formalism)
from the terms bilinear in the gauge fields in the same term.
The two conditions in \eq{eq:DCPR} then determine
\begin{equation}
  \delta\tan\beta^{\rm DCPR} =
   \tan\beta ( \delta Z_{uu} - \delta Z_{dd})
    = - \tan^2\beta \,{\rm Re}\,\Delta Z_{du} + \dots ,
\end{equation}
where the omitted terms are not $\tan\beta$-enhanced. This explains
the large numerical differences between $\tan\beta^{\overline{\rm DR}}$
and $\tan\beta^{\rm DCPR}$ found in \cite{Freitas:2002um} as a
parametrically large effect. Hence, $\tan\beta$
measured in flavour physics should not be identified with the
corresponding DCPR parameter at large $\tan\beta$.

As with the Higgs fields, we explicitly decouple the contributions of heavy
particles to the gauge field wave functions (hence to $g^{(\prime)}$) by a
finite renormalization ${g^{(\prime)}}^b = g^{(\prime)} + g^{(\prime)} \delta
g^{(\prime)}$,
cancelling the terms $\Delta Z_B$ and $\Delta Z_W$ in \eq{eq:effact} of the
gauge fields, $B_\mu^e=Z_B^{\frac{1}{2}} B_\mu$ and $W_\mu^e=Z_W^{\frac{1}{2}}
W_\mu$. For ${\overline{\rm DR}}$-subtracted MSSM couplings, this
gives $\overline{\rm MS}$-renormalized 2HDM gauge couplings.

We denote the quartic couplings in our 2HDM scheme by $\bar \lambda_i$.
The finite renormalizations leave $\lambda_5$  invariant,
$\bar \lambda_5 = \hat \lambda_5$, while the
other quartic coupling constants transform like
\begin{align}
\label{eq:2-1}
\bar \lambda_1 & = \hat \lambda_1 + \tilde g^2 \delta Z_{dd}^r + 
\frac{1}{2} \left(g^2 \delta g + g^{\prime ^2} \delta g' \right), &
\bar \lambda_2 & = \hat \lambda_2 + \tilde g^2 \delta Z_{uu}^r + 
\frac{1}{2} \left(g^2 \delta g + g^{\prime ^2} \delta g' \right),
\nonumber \\
\bar \lambda_3 & = \hat \lambda_3 - \frac{\tilde g^2}{2} \left(\delta Z_{dd}^r
  + \delta Z_{uu}^r \right) - \frac{1}{2} \left(g^2 \delta g + g^{\prime ^2}
  \delta g' \right), & 
\bar \lambda_4 & = \hat \lambda_4 + g^2 \left(\delta
  Z_{dd}^r + \delta Z_{uu}^r \right) + g^2 \delta g,
\nonumber \\
\bar \lambda_6 & = \hat \lambda_6 - \frac{\tilde g^2}{4} \delta Z_{ud}^*, &
\bar \lambda_7 & = \hat \lambda_7 + \frac{\tilde g^2}{4} \delta Z_{ud} 
\; ,
\end{align}
where $x^r$ and $x^i$ denote the real and imaginary part of $x$
respectively. The couplings $\bar \lambda_i$ are $\overline{\rm MS}$
couplings from the viewpoint of the effective theory.

The modification of the dimensionless couplings by the finite wave function
renormalizations affects the $B-\bar B$ mixing amplitudes as a 
formally higher-order effect, as does the scheme dependence of
$\tan\beta$. Unlike the latter, however, the former is never
$\tan\beta$ enhanced unless the wave function renormalization
constants themselves are.

\subsubsection*{\boldmath Invariance of $\bb$ mixing under field renormalization \unboldmath}
\label{sec:invariance-bb-mixing}

The effects of \eq{eq:5} on the Higgs-mediated FCNC
\eq{eq:LY} are twofold: (i) the values for $\cos\beta$ and $\sin\beta$ in 
\eq{eq:vevbasis} are
modified. This cancels the contributions to $\mathcal{F}^{\pm}$ from the
redefinition of the mass matrices up to a global factor:%
\begin{align}
\mathcal{F}^{+}(\lambda_{i},v_{u,d},M_{A}) &  \rightarrow\mathcal{F}%
^{+}(\lambda_{i}^{\prime},v_{u,d}^{\prime},M_{A}^{\prime})=\frac{v^{2}%
}{v^{\prime2}}\left\vert \det Z\right\vert ^{2}\mathcal{F}^{+}(\lambda
_{i},v_{u,d},M_{A}),\label{3.28}\\
\mathcal{F}^{-}(\lambda_{i},v_{u,d},M_{A}) &  \rightarrow\mathcal{F}%
^{-}(\lambda_{i}^{\prime},v_{u,d}^{\prime},M_{A}^{\prime})=\frac{v^{2}%
}{v^{\prime2}}\left(  \det Z^{\ast}\right)  ^{2}\mathcal{F}^{-}(\lambda
_{i},v_{u,d},M_{A}).\label{3.29}%
\end{align}
(ii) the $v_{u,d}$ renormalization in (i) comes with a modification of
$\kappa_{ij}$:%
\begin{equation}
\kappa_{ij}\rightarrow\kappa_{ij}^{\prime}=\frac{v^{\prime}}{v}\left(  \det
Z^{\ast}\right)  ^{-1}\kappa_{ij}\, .\label{3.31}%
\end{equation}
The above factors cancel each other out in the products $\kappa_{bq}^{2}%
$\thinspace$\mathcal{F}^{-}$
and $\kappa_{qb}^{\ast}\,\kappa_{bq}$ \thinspace$\mathcal{F}^{+}$, as they
should. In particular, our choice of wave-function renormalization
acting on the leading FCNC coupling \eq{3.16} produces an
extra term:
\begin{equation}
\delta\kappa_{bq}=-\kappa_{bq}\left(  s_{\beta}^{2}\,\delta Z_{dd}%
^{r}-s_{\beta}c_{\beta}\,\delta Z_{ud}^{r}+c_{\beta}^{2}\,\delta Z_{uu}%
^{r}\right)  .\label{3.32}%
\end{equation}
Considering Eqs.\ (\ref{3.16}), (\ref{3.16bis}), (\ref{eq:2-1}), and (\ref{3.32})
gives the same Wilson coefficients $C_2^{\rm LR}$ and $C_1^{\rm SLL}$
as does considering Eqs.\ (\ref{3.16}), (\ref{3.16bis}), and (\ref{eq:2-1})
with the finite parts of $\delta Z_{ij}$ set to
zero. While in practice, wave-function renormalization has to be
performed to relate the parameter $M_{A}$ to the physical Higgs
boson masses and to take $v =\left( \sqrt{2}G_{F}\right)
^{-1/2}\simeq246$\thinspace GeV beyond leading-order precision,
such renormalizations are not the source of a
non-vanishing of the $Q_1^{\rm SLL}$ amplitude, to be found instead in the
corrections to Higgs masses and mixings (via the self-couplings
$\lambda_{i}$, in particular $\lambda_{5}$);
wave-function-renormalization 
effects enter that amplitude only at higher orders
(as might have been expected). In this regard our findings disagree
with the conclusions of \cite{fgh}.

\subsection{\boldmath Health of the large-$\tan \beta$ \unboldmath
limit and fine-tuning}
\label{sec:ltb_quantum}
In Sect.\ \ref{sec:ltbeft} we took the limit $\tan\beta \to \infty$
($v_d \to 0$, $M_A^2={\rm const}$, $v_u^2 + v_d^2={\rm const}$,
$\lambda_i={\rm const}$, $v_u$ and $v_d$ defined as minima of the
tree potential) at the Lagrangian level. One might wonder
whether this procedure is valid at the quantum level. To justify it, we
show that
the $v_d=0$ case and the $v_d \not = 0$ case are analytically
connected, i.e.\ one can be
reached from the other without a phase transition.
It then follows that amplitudes are (in some neighbourhood of a
parameter point with $v_d=0$) analytic functions of the parameters (either
``symmetric'' or ``broken''). The renormalizability of the effective
Lagrangian $V_{\rm ltb}$ then follows by standard arguments
from the fact that it is equivalent
to the symmetric Lagrangian \eq{2.1} (for a certain choice of parameters),
which is renormalizable.

We first note that the number of independent minimization conditions
is unchanged in the $v_d=0$ limit. First, for general values of the
parameters, out of the four real (two complex) minimization
conditions, at most three are independent. This follows from the
$U(1)_Y$ invariance but is easy to verify explicitly. Fixing $v_u$ to
be real and positive, three polynomials of degree three determining
three unknowns $v_u$, $v_d^r$, $v_d^i$ remain. The system has a
solution $v_d=0$ if  
\begin{equation}
\lambda_2\, m_{12}^{2} + \lambda_7\, m_{22}^2=0 , \qquad \qquad
v_u^2 = - \frac{2 m^2_{22}}{\lambda_2} .
\end{equation}
Here the second equation determines $v_u=v$ as a function of $m^2_{22}$
and $\lambda_2$ similarly to the case of a single doublet, while the
first equation can be viewed as a fine-tuning condition between
$m^{2}_{12}$ and $\lambda_7$.
The dimensionless, complex parameter
\begin{equation}
 \label{eq:15}
 \epsilon = \frac{m_{12}^{2}}{m_{11}^2} +
 \frac{\lambda_7}{\lambda_2} \frac{m_{22}^2}{m_{11}^2} 
\end{equation}
parameterizes the deviation from the fine-tuning limit; we may trade
$m_{12}^2$ in favour of $\epsilon$. Clearly, at the limiting
point $\epsilon=0$ we indeed have three independent equations.
Now, it is easy to verify that, writing the
four real minimization conditions in the form
\begin{equation}
  f_i(\lambda_i, m^2_{ij}, v_u, v_d) = 0 ,
\end{equation}
the Jacobian matrix
\begin{equation}
   \frac{\partial (f_1, f_2, f_3, f_4)}{\partial(v_u, v_d^r, v_d^i)}
\end{equation}
has maximal rank (3) at any point with $v_d=0$. (Physically, this just
means that the neutral Higgs mass matrix has three nonzero eigenvalues there.)
Hence, by the implicit function theorem, we may solve for $(v_u, v_d)$
in a neighbourhood of it, where the solutions will be (real-)analytic
functions of $\epsilon$. In particular, $v_u$ behaves analytically
(and is strictly
positive) around $\epsilon=0$, i.e.\ no phase boundary is encountered.
Explicitly and to linear order, the real and imaginary parts of $v_d$ are
determined by
\begin{equation}    \label{eq:vdofepsilon}
  \Bigg( \begin{array}{cc}
  1 + \frac{\lambda_3+\lambda_5^r}{2}\frac{v_u^2}{m^2_{11}} &
        \frac{\lambda_5^i}{2}\frac{v_u^2}{m^2_{11}} \\
  \frac{\lambda_5^i}{2}\frac{v_u^2}{m^2_{11}} &
       1 + \frac{\lambda_3-\lambda_5^r}{2}\frac{v_u^2}{m^2_{11}}
\end{array} \Bigg)
  \Bigg( \begin{array}{cc}
    v_d^r \\ v_d^i
\end{array} \Bigg) = v_u 
  \Bigg( \begin{array}{cc}
    \epsilon^r \\ \epsilon^i
\end{array} \Bigg)  + {\cal O}(\epsilon^2) ,
\end{equation}
such that $\tan\beta = {\cal O}(1/|\epsilon|)$.
The nonsingular linear term allows us to change variables
from $m^2_{12}$ to a complex $v_d$. Of course, we may
always perform a field redefinition of $H_d$ such that $v_d$ is real.
Then, the mass parameters besides $m^2_{11}$ are power
series in $1/\tan\beta$, which read
\begin{equation}
\label{eq:16}
\begin{split}
  m_{12}^{2,i} &= 
  \frac{1}{2} \lambda_6^i v_d^2+
  \frac{1}{2} v_u \lambda_5^i v_d+
  \frac{1}{2} v_u^2 \lambda_7^i , \\
  m_{22}^2 &= 
  \frac{v_d \, m_{11}^2}{v_u}+
  \frac{1}{2} v_u v_d \left(\lambda_3+\lambda_5^r\right)+
  \frac{1}{2} v_u^2 \lambda_7^r + \mathcal{O}(v_d^2/v_u^2)  , \\
  M_A^2 &=
  m_{11}^2 + \frac{\lambda_3  - \lambda_5^r}{2} v_u^2 + {\cal O}(v_d/v_u) , \\
  M_h^2 &= \lambda_2 v^2 + {\cal O}(\lambda_7^2; \epsilon), \\
  M_H^2 &= m_{11}^2 + \frac{\lambda_3 + \lambda_5^r}{2} v^2
           + {\cal O}(\lambda_7^2; \epsilon) , \\
  M_{H^+}^2 &= m_{11}^2 + \frac{\lambda_3 + \lambda_4}{2} v^2
           + {\cal O}(v_d/v_u) .
\end{split}
\end{equation}
We see explicitly that we can continuously change the
dimensionful parameters in the Higgs potential 
from a situation where $v_d \neq 0$ to one where
$v_d=0$, keeping $M_A^2$ (and the dimensionless couplings) fixed,
as was assumed in Sect.~\ref{sec:ltbeft}.
The last three equations illustrate that
the large-$\tan\beta$ case is characterized by a ``primary'' doublet
$H_u$ which receives a large vev $v_u$ and a ``secondary'' doublet
$H_d$ with a positive gauge-invariant mass $m_{11}^2$ that
receives corrections of ${\cal O}(v^2)$ and ${\cal O}(\epsilon)$,
respectively, due to its dimensionless and dimensionful couplings to $H_u$.
Those corrections differ among the physical components of the
doublet, approximately to be identified with $H^0$, $A^0$, $H^\pm$,
due to electroweak symmetry breaking.
In principle, $m^2_{11}$ could be negative, but in that case,
$v_d \approx 0$ will typically not be the global minimum of the potential.

We close this section by considering the fine-tuning which is
necessary to obtain a large $\tan \beta$ while keeping the mass $M_A$
fixed.
For $v_d$ real, \eq{eq:vdofepsilon} implies
\begin{equation}
   m^{2,r}_{12} 
     = - \frac{\lambda_7^r}{\lambda_2} m^2_{22}
       + \cot\beta (m^2_{11} + \frac{\lambda_3 + \lambda_5^r}{2} v_u^2) ,
\end{equation}
which illustrates the tuning that is known to be necessary to have 
large $\tan\beta$ in the MSSM. For the generic
situation $m_{11}^2 \sim M_{\rm SUSY}^2 \gg M_Z^2$, the right-hand
side is dominated by the $m^2_{11}$ term: $\lambda_7^r$ is down by a
loop factor relative to $\lambda_2$, and $m^2_{22} \sim v^2 \ll M_{\rm
  SUSY}^2$ (the
little hierarchy). Hence,
\begin{equation}
  m^{2,r}_{12}/M_{\rm SUSY}^2 \sim 1/\tan\beta ,
\end{equation}
which implies an extra tuning beyond the one to achieve the correct
weak scale.
For smaller $m_{11}^2 \sim M_A^2 \sim M_Z^2$, which is interesting
from the point of view of $B$-physics phenomenology, the required
tuning gets even worse -- unless, of course, the whole SUSY scale is
lowered to the weak scale, which is, however, problematic since then
$M_h^2 \approx \lambda_2 v^2$ is generally below the experimental
lower limit. On the other hand, as we have seen, $M_A^2 \sim
m^2_{11}$, such that no extra tuning is required to keep $M_A^2$
finite, while one might have expected otherwise from the well-known
tree-level formula
\begin{equation}
   M_A^2 = (\tan\beta + \cot\beta) m^2_{12} ,
\end{equation}
which is generalized by \eq{eq:madef}.
Also, while a small $m^2_{12}$ is indeed sensitive to radiative corrections,
those are automatically correlated with shifts of $v_d$ and in
consequence of $\tan\beta$ in such a way that $M_A^2$ receives only mild
corrections.

\section{Phenomenology}\label{sect:phen}

In Sect.~\ref{sect:bbm}, we performed a detailed study of the supersymmetric
contributions to $\Delta M_{d}$ and $\Delta M_{s}$ in the generic framework of an effective 2HDM.
The corresponding matching coefficients were computed at the one-loop level in Sect.~\ref{sect:higgs} and Appendix~\ref{sec:matching_results}.
In this section, we assess the maximal size of the various types of effects identified in the MFV case
taking into account the existing constraints on the supersymmetric parameter space,
in particular from the $\BsMM$, $\BTN$, and $\bsg$ branching fractions.
For convenience, we start with a compendium of the formulas derived in Sect.~\ref{sec:higgsmed-mfv}:
\begin{equation}
\label{eq:4.1}
\Delta M_{q}=\left\vert \Delta M_{q}^{\rm SM}+\Delta M_{q}^{\rm LR}+\Delta M_{q}^{\rm LL}
+\Delta M_{q}^{\rm HL}\right\vert \equiv\left\vert 1+h_{q}\right\vert \Delta M_{q}^{\rm SM},
\end{equation}
where the standard-model, the left-right and left-left Higgs-pole, and the neutral Higgs-loop contributions read
\begin{equation}
\label{eq:4.2}
\begin{split}
\Delta M_{q}^{\rm SM}  &  =\left\vert V_{tq}V_{tb}^{\ast}\right\vert ^{2}f_{B_{q}
}^{2}M_{B_{q}}P_{1}^{\rm VLL}\left[  \frac{G_{F}^{2}M_{W}^{2}}{6\pi^{2}}
S_{0}(m_{t}^{2}/M_{W}^{2})\right]  ,\\
\Delta M_{q}^{\rm LR}  &  =\left\vert V_{tq}V_{tb}^{\ast}\right\vert ^{2}f_{B_{q}
}^{2}M_{B_{q}}P_{2}^{\rm LR}\text{ }\,\left[  \frac{-1}{3}\frac{m_{b}m_{q}}{v^2}\left\vert
\overline{\kappa}\right\vert ^{2}\mathcal{F}^{+}\right]  ,\\
\Delta M_{q}^{\rm LL}  &  =\left\vert V_{tq}V_{tb}^{\ast}\right\vert ^{2}f_{B_{q}
}^{2}M_{B_{q}}P_{1}^{\rm SLL}\hspace{0.03cm}\left[  \frac{-1}{6}\frac{m_{b}^{2}}{v^2}
\overline{\kappa}^{2}\mathcal{F}^{-}\right]  ,\\
\Delta M_{q}^{\rm HL}  &  =\left\vert V_{tq}V_{tb}^{\ast}\right\vert ^{2}f_{B_{q}
}^{2}M_{B_{q}}P_{1}^{\rm VLL}\left[  \frac{1}{12}\frac{y_{b}^{*2}}{16\pi^{2}}\frac{m_{b}^{2}}{v^2}\overline{\kappa}
^{2}\frac{1}{M_{A}^{2}}\right],
\end{split}
\end{equation}
respectively. The Inami-Lim function $S_0$ is given by
$S_0(x)=(x-11x^2/4+x^3/4)(1-x)^{-2}-(3x^3\log (x)/2)(1-x)^{-3}$
and $v=(\sqrt{2}G_F)^{-1/2}=246\gev$.
The flavour-changing and flavour-conserving quark-Higgs couplings were
defined in Sects.~\ref{sec:fcnc-higgs-tree} and~\ref{sec:higgsmed-mfv}:
\begin{equation}
\label{eq:4.3}
\overline{\kappa}\equiv\frac{\kappa_{bq}v}{\lambda_{qb}m_{b}}=\frac{\kappa
_{qb}v}{\lambda_{qb}^{\ast}m_{q}}=\frac{y_{t}^{2}\sqrt{2}}{c_{\beta}^{2}
}\,\frac{\epsilon_{Y}}{\left(  1+\widetilde{\epsilon}_{3}t_{\beta}\right)
\left(  1+\epsilon_{0}t_{\beta}\right)  },\text{\quad\quad}y_{b}=\frac
{\sqrt{2}m_{b}}{vc_{\beta}}\,\frac{1}{1+\widetilde{\epsilon}_{3}t_{\beta}},
\end{equation}
with $\epsilon_{0,Y}$ and $\widetilde{\epsilon}_{3}$ given
in Eqs.~(\ref{3.17}-\ref{3.19}) and $y_t$ in Appendix~\ref{sect:not}.
The $\mathcal{F}^{\pm}$ factors describing the propagation of the neutral Higgses
were defined in Eqs.~(\ref{3.13bis}) and (\ref{3.14}),
with the effective couplings $\lambda_i$ entering the neutral Higgs mass matrix
computed in Sect.~\ref{sec:renorm} and Appendix \ref{sec:matching_results}.
For large $\tan\beta$, we have in very good approximation:
\begin{equation}
\label{eq:4.4}
\mathcal{F}^{+}\simeq\frac{2}{M_{A}^{2}},\text{\quad\quad}
\mathcal{F}^{-}\simeq\frac{(-\lambda_{5}^{\ast}+\lambda_{7}^{\ast2}/\lambda_{2})v^{2}}{M_{A}^{4}}
\end{equation}
(exact formulas were used in our numerical analysis though).
Explicit expressions for $\lambda_{5}$, $\lambda_{7}$, and $\lambda_{2}$
in the MFV case were given in Eqs.~(\ref{eq:lambda2MFV}-\ref{eq:lambda7MFV}). Altogether,
counting $\epsilon_{Y},\mathcal{F}^{-}\sim (16\pi^2)^{-1}$ and $M_A\sim 120\ \gev$
to get an idea of the naive size of the various effects in the absence of constraints, we obtain:
\begin{equation}
\label{eq:4.5}
\begin{split}
h_{s}  &  =\bigg\{  -2.40\left[  \frac{m_{s}/m_{b}}{0.053/2.75}\right]
\left[  \frac{P_{2}^{\rm LR}/P_{1}^{\rm VLL}}{3.2/0.71}\right] \\[1mm]
&+0.35\ \frac{16\pi^{2}(-\lambda_{5}^{\ast}+\lambda_{7}^{\ast2}/\lambda_{2})(120\ \gev)^{2}
e^{2i\phi_{\overline{\kappa}}}}{M_{A}^{2}}
\left[  \frac{P_{1}^{\rm SLL}/P_{1}^{\rm VLL}}{-1.36/0.71}\right]  \\
&+\,0.01\ \frac{e^{2i\phi_{\overline{\kappa}}}}{\left(
1+\widetilde{\epsilon}_{3}t_{\beta}\right)  ^{2}}\left[  \frac{t_{\beta}}
{40}\right]  ^{2}\left[  \frac{m_{b}}{2.75}\right]  ^{2}\bigg\}\
\frac{\left\vert 16\pi^{2}\epsilon_{Y}\right\vert ^{2}(120\ \gev)^{2}}{\left\vert
1+\widetilde{\epsilon}_{3}t_{\beta}\right\vert ^{2}\left\vert 1+\epsilon
_{0}t_{\beta}\right\vert ^{2}M_{A}^{2}}\left[  \frac{t_{\beta}}{40}\right]
^{4}\left[  \frac{m_{b}}{2.75}\right]^{2},
\end{split}
\end{equation}
where $m_b$ is in GeV and $\phi_{\overline{\kappa}}\equiv\arg(\overline{\kappa})$.
$h_{d}$ is given by the same expression with $m_s$ replaced by $m_d$,
so that the first term becomes subleading.

A first obvious remark is that $\Delta M_{q}^{\rm HL}$ cannot compete
with $\Delta M_{s}^{\rm LR}$ or $\Delta M_{q}^{\rm LL}$ unless $y_b$
becomes non-perturbative.  This is rather accidental (notice the small
loop factor in Eq.~(\ref{eq:4.2}) as well as the smallness of $P^{\rm
  VLL}_{1}$ with respect to $P^{\rm LR}_{2}$ and $P^{\rm SLL}_{1}$).
Further, the contribution of $\Delta M_{q}^{\rm LL}$ seems somewhat
limited.  However, the loop functions $\lambda_5$ and $\lambda_7$ could
be enhanced for large $\mu$ or $a_{t,b}$, see
Eqs.~(\ref{eq:lambda5MFV},\ref{eq:lambda7MFV}). A more quantitative
analysis is thus desirable.  In the next two sections, we perform a
random scan of the MFV-MSSM parameter space to find the maximal $\Delta
M_{q}^{\rm LL}$ and $\Delta M_{q}^{\rm LR}$ values allowed by current
experimental data. Eqs.~(\ref{eq:4.1}-\ref{eq:4.5}) do allow for new
CP-violating phases\footnote{Let us recall that in that case $M^2_A$,
  defined as the non-zero eigenvalue of the CP-odd mass matrix $M^2_I$
  in Eq.(\ref{eq:higgsmassneut}), is no more an eigenvalue of the full
  Higgs mass matrix.}, yet these will be set to zero in the scan.
CP-violating effects within the MFV scenario will be shortly discussed
in Sect.~\ref{sect:CPV}.

\subsection{Scan of the parameter space}\label{sect:scan}

The values of the various input parameters used in the scan are collected in Table \ref{tab:4-1}.
Note that only the products $P_{2}^{\rm LR}m_s$ and $P_{2}^{\rm LR}m_d$, or alternatively
$P_{2}^{\rm LR}m_s$ and $m_{d}/m_{s}$, are needed, see Eq.~(\ref{eq:4.2}).
We scan over $P_{2}^{\rm LR}m_s$ but keep $m_{d}/m_{s}$ fixed as $\Delta M_{d}^{\rm LR}$ is doomed to be small anyway.
The decay constants $f_{B_q}$ and CKM factors $\left\vert V_{tq}V_{tb}^{\ast}\right\vert$ are not specified.
Instead, outputs are formulated in terms of ratios free from these
rather poorly known parameters.
Finally, we take $\alpha=1/127.9$, $\sin^{2}\theta_{W}=0.231$, and $M_{Z}=91.1876\ \gev$.

For simplicity, the gaugino mass parameters are assumed to have the same sign
(which we can choose positive),
as well as the trilinear terms (positive or negative).
Note that the absolute scale of $M_{\mathrm{SUSY}}$ plays no role as supersymmetric parameters
enter $\epsilon_{0,Y}$ and $\lambda_i$ by means of ratios.
Only the spread of the interval chosen for $M_{\mathrm{SUSY}}$ matters.
Still, $M_{\mathrm{SUSY}}$ should not be taken too large to help satisfy the $b\rightarrow s\gamma$ constraint
in the case $\mu<0$.
We will come back to this point later.
We allow for rather large values of $M_A$, close to the lower end of the interval chosen for $M_{\mathrm{SUSY}}$.
Still, the matching performed in Sect.~\ref{sect:higgs} and Appendix \ref{sec:matching_results} remains valid as the corrections from
higher dimension operators at the electroweak scale are ruled by the ratio
$v/M_{\mathrm{SUSY}}$ and not $M_A/M_{\mathrm{SUSY}}$.
The formulas for the various observables at the $B$ mass scale are thus unaffected.

\begin{table}[!h]
\centering
{\setlength{\tabcolsep}{0.3cm}{\renewcommand{\arraystretch}{1.1}
\begin{tabular}{|l|l|l|}
\hline
Quark masses and $\alpha_s\ $ & Bag factors & SUSY parameters\\
\hline
$m_{t}=164$ GeV & $P_{1}^{\rm VLL}\in[0.66,0.76]$ & $\tan\beta\in[10,60]$\\
$m_{b}=2.75$ GeV & $P_{2}^{\rm LR}m_{s}\in[0.12,0.22]$ GeV& $M_{A}\in[120,600]$ GeV\\
$m_{d}/m_{s}=1/19$ & $P_{1}^{\rm SLL}\in[-1.48,-1.24]$ & $M_{\mathrm{SUSY}}\in[600,1800]$ GeV\\
$\alpha_{s}=0.108$ &&\\
\hline
\end{tabular}}}
\caption{Input values.
Here $M_{\mathrm{SUSY}}$ stands for any of the supersymmetric parameters
$|\mu|$, $M_{\widetilde t_L}$, $M_{\widetilde t_R}$, $M_{\widetilde b_R}$,
$M_{\widetilde \tau_L}$, $M_{\widetilde \tau_R}$,
$|a_{t}|$, $|a_{b}|$, $|a_{\tau}|$,
$M_{1}$, $M_{2}$, $M_{3}$.
As explained in Sect.~\ref{sec:renorm}, the renormalized parameters $M_A$ and $\tan\beta$
are identical in the MSSM and effective 2HDM for $M_{\mathrm{SUSY}}\gg v$.
The quark masses and $\alpha_{s}$ are defined in the $\overline{\rm MS}$ scheme at the scale $m_{t}$.
The bag factors, defined at the scale $m_{t}$ as well, are discussed in Appendix \ref{sect:rghp}.}
\label{tab:4-1}
\end{table}

The constraints imposed on the points generated inside the above ranges are summarized in Table~\ref{tab:4-2}.
We now discuss them in turn:

\vspace{1mm}
i) The bottom Yukawa coupling $y_b$ is maintained small enough, say, $y_b<2$, to guarantee the validity of perturbation theory.
This condition removes possible fine-tuned points in parameter space for which the denominators in Eq.~(\ref{eq:4.3}) are close to zero.

\vspace{1mm}
ii) The lightest Higgs boson mass $M_h$ has to come out large enough to comply
with the LEP II experimental lower bound.
$M_h$ is obtained from the CP-even Higgs mass matrix in Eq.~(\ref{Delta}),
with the effective couplings $\lambda_i$ computed at the one-loop level.
Higher order corrections to $\lambda_2$ are known to be important \cite{Okada:1990vk,Haber:1990aw,Ellis:1990nz}.
However, $h^0$ comes up in the FCNC vertices $\kappa_{ij}$ of Eq.~(\ref{eq:LY})
along with a $\cot\beta$ suppression factor.
The $\tan\beta$-enhanced effects considered here are thus largely uncorrelated with $M_h$.
For this reason we do not correct the one-loop formulas and simply impose $M_h>115\gev$.

\vspace{1mm}
iii) The following bounds are imposed on $a_t$ and $a_b$ to avoid the occurence of color
symmetry-breaking vacua at tree-level \cite{Casas:1996de}:
\begin{equation}
\label{eq:StabBounds}
\begin{split}
|a_t|^2&<3(M^2_{\widetilde t_L}+M^2_{\widetilde t_R}+m^2_{22}),\\
|a_b|^2&<3(M^2_{\widetilde t_L}+M^2_{\widetilde b_R}+m^2_{11}).
\end{split}
\end{equation}
The corresponding bound for $a_\tau$ is not imposed as sleptonic
parameters have very little impact on the quark FCNC considered here anyway.

\vspace{1mm}
iv) The most stringent constraint on the FCNC coupling $\overline{\kappa}$
comes from the $B_{s}\rightarrow\mu^{+}\mu^{-}$ branching fraction,
which we normalize to $\Delta M_s$ to avoid the occurence of the parameters $f_{B_s}$
and $V_{ts}V_{tb}^{\ast}$.
This time the Higgs-pole contribution overcomes the standard-model and Higgs-loop pieces.
In addition these last two interfere destructively, so we will neglect them.
The counterpart of Eq.~(\ref{eq:4.2}) then reads
(with $m_{\mu}^{2}/M_{B_{q}}^{2}=0$ for simplicity):
\begin{equation}
\label{eq:4.6}
\begin{split}
\mathcal{B}(B_{q}\rightarrow\mu^{+}\mu^{-})
&=\tau_{B_{q}}\left\vert V_{tq}V_{tb}^{\ast}\right\vert^{2}
f_{B_{q}}^{2}M_{B_{q}}^{5}\,
\frac{m_{\mu}^{2}}{64\pi v^{4}} 
\frac{\left\vert \overline{\kappa}\right\vert ^{2}\left[  \left\vert \mathcal{F}_{P}\right\vert
^{2}+\left\vert \mathcal{F}_{S}\right\vert ^{2}\right]}
{\cos^{2}\beta |1+\epsilon_\mu t_\beta|^2}\\
&\equiv R_{q}\,\mathcal{B}(B_{q}\rightarrow\mu^{+}\mu^{-})^{SM},
\end{split}
\end{equation}
where $\mathcal{F}_{P}$ and $\mathcal{F}_{S}$ refer to the Wilson coefficients
of the effective operators $Q_P=(\bar{b}_Rs_L)(\bar{\ell}\gamma_5\ell)$ and
$Q_S=(\bar{b}_Rs_L)(\bar{\ell}\ell)$ arising from neutral Higgs exchanges
and
\begin{eqnarray}
\epsilon_\mu &=& \frac{g^{\prime 2}}{16 \pi^2} \frac{\mu^*}{M_1} 
\lt[ -\frac{1}{2} 
    H_{2}\left(  \frac{M_{\widetilde{\mu}L}^{2}}{|M_{1}|^{2}},
                    \frac{|\mu|^{2}}{|M_{1}|^{2}}\right)  
   + H_{2}\left(  \frac{M_{\widetilde{\mu}R}^{2}}{|M_{1}|^{2}},
                    \frac{|\mu|^{2}}{|M_{1}|^{2}}\right) \rt] \nn  
&&  - \frac{g^{\prime 2}}{16 \pi^2} \frac{\mu^*}{M_1}      
H_{2}\left(  \frac{M_{\widetilde{\mu}L}^{2}}{|M_{1}|^{2}},
                   \frac{M_{\widetilde{\mu}R}^{2}}{|M_{1}|^{2}}
                    \right)   
 -\frac{3 g^2}{32 \pi^2} \frac{\mu^*}{M_2} 
  H_{2}\left(  \frac{M_{\widetilde{\mu}L}^{2}}{|M_{2}|^{2}},
                    \frac{|\mu|^{2}}{|M_{2}|^{2}}\right)  
.\label{defepsmu}
\end{eqnarray}  
This result agrees with \cite{mmns} but disagrees with
\cite{Itoh:2004ye}. The loop function $H_2$ was defined in
Eq.(\ref{3.20}) and $M_{\widetilde{\mu}L(R)}=M_{\widetilde{\tau}L(R)}$
in our MFV scenario. In the large $\tan\beta$ limit and at
tree-level in the Higgs potential, we have:
$\mathcal{F}_{P}=-\mathcal{F}_{S}=\mathcal{F}^+/2=1/M_{A}^{2}$, so that
$\mathcal{B}(B_{q}\rightarrow\mu^{+}\mu^{-})$ is tightly correlated with
$\Delta M_q^{\rm LR}$ \cite{bcrs}. 
Going beyond the
tree-level and large $\tan\beta$ approximations we obtain:
$\mathcal{F}_{P}=s_{\beta}(\mathcal{F}^{+}-\mathcal{F}^{-})/2$, with
$\mathcal{F}^{\pm}$ given in Eqs.~(\ref{3.13bis}) and (\ref{3.14}).
This formula is actually valid in any 2HDM, including arbitrary
CP-violating phases (in the CP-conserving case it reduces to the usual
identity $\mathcal{F}_{P}=s_{\beta}/M_A^2$).  We did not find such a
general and simple form for $\mathcal{F}_{S}$, yet it is straightforward
to write it in terms of $M_A$, $\tan\beta$, and the $\lambda_i$'s
(alternatively one can of course express it in terms of the neutral
Higgs masses and mixing angles).  Note that one still has
$\mathcal{F}_{S}=-(\mathcal{F}^{+}+\mathcal{F}^{-})/2$ up to
$\cot\beta$-suppressed terms.  Sparticle loop corrections to the Higgs
self-energies turn out to be relevant in the case of $\mathcal{F}_{S}$:
they can be as large as 15\% for small $M_A$ after all constraints are
taken into account, as we will see in
  Sect.~\ref{sect:correlation}.  Numerically, Higgs-mediated effects
can easily be very large:
\begin{equation}
\label{eq:4.7}
R_{s}=9930
\left[1+\frac{(-\lambda_5^r+|\lambda_7|^2/\lambda_2)v^2}{M_A^2}\right]
\frac{\left\vert 16\pi^{2}\epsilon_{Y}\right\vert ^{2}(120\ \gev)^{4}}
{\left\vert 1+\widetilde{\epsilon}_{3}t_{\beta}\right\vert ^{2}\left\vert
1+\epsilon_{0}t_{\beta}\right\vert^{2} {|1+\epsilon_\mu t_\beta|^2} M_{A}^{4}}\left[\frac{t_{\beta}}{40}\right]^{6}
\end{equation}
and $R_d\simeq R_s$. The first correction factor above captures the bulk
of the effects from the Higgs self-energies, yet the exact formula for
$\mathcal{F}_{S}$ should be used for better precision.  In practice, the
looser constraint $\mathcal{B}(B_{s}\rightarrow\mu^{+}\mu^{-})/\Delta
M_s<5.7\times10^{-9}\ ps$, obtained from
$\mathcal{B}(B_{s}\rightarrow\mu^{+}\mu^{-})^\textrm{exp}<10^{-7}$
\cite{BsMuMuExpOldBound} and $\Delta M_s^\textrm{exp}=17.77\pm 0.1\pm
0.07\,ps^{-1}$ \cite{Abulencia:2006ze}, is built-in in the scan
procedure, then the current 95\% C.L. bound
$\mathcal{B}(B_{s}\rightarrow\mu^{+}\mu^{-})/\Delta
M_s<3.3\times10^{-9}\ ps$ corresponding to
$\mathcal{B}(B_{s}\rightarrow\mu^{+}\mu^{-})^\textrm{exp}<5.8\times
10^{-8}$ \cite{:2007kv} is imposed.  We also checked the bound
$\mathcal{B}(B_{d}\rightarrow\mu^{+}\mu^{-})/\Delta
M_d<3.6\times10^{-8}\ ps$, corresponding to
$\mathcal{B}(B_{d}\rightarrow\mu^{+}\mu^{-})^\textrm{exp}<1.8\times
10^{-8}$ \cite{:2007kv} and $\Delta M_d^\textrm{exp}=0.507\pm
0.005\,ps^{-1}$ \cite{PDG08}.  This provides no additional constraint.
Neither do $\mathcal{B}(B_{s,d}\rightarrow\mu^{+}\mu^{-})$ and $\Delta
M_{s,d}$ taken separately due to the large parametric uncertainties from
$f_{B_q}$.

\vspace{1mm} v) The $b\to s\gamma$ branching fraction with the energy
cut $E_\gamma>1.6\gev$ is computed using the fortran code SusyBSG
\cite{Degrassi:2007kj}.  Higgs-mediated effects now appear at loop-level
with smaller powers of $\tan\beta$, so that purely supersymmetric loop
corrections (scaling as $1/M_{\mathrm{SUSY}}$) are comparatively more
important. For $a_t\mu<0$ and relatively light $M_{\mathrm{SUSY}}$,
chargino and charged-Higgs loops can interfere destructively and more
room is left for New Physics.  This interplay is welcome when $\mu<0$ as
the charged Higgs contribution then tends to overshoot the experimental
branching fraction.  On the other hand, in that case, the discrepancy
between the $(g-2)_\mu$ standard-model prediction and its present
measurement \cite{Bennett:2006fi} increases (for a recent discussion,
see e.g. \cite{mmns} and references therein).  The significance of this
discrepancy, however, is questioned by the new
$e^+e^-\to\pi^+\pi^-\gamma$ BABAR data \cite{Davier08}.  We therefore
still include the situation $\mu<0$ in our
considerations. The $\mathcal{B}(b\to s\gamma)$ experimental world
average reads: $\mathcal{B}(b\to
s\gamma)^\textrm{exp}=(3.52\pm0.23\pm0.09)\times10^{-4}$ \cite{HFAG08}.
The standard-model central value of the SusyBSG program
agrees well with the next-to-next-to-leading order prediction
$\mathcal{B}(b\to s\gamma)=(3.15 \pm 0.23)\times10^{-4}$
\cite{Misiak:2006zs}. We combine the experimental error with the
uncertainties discussed in Ref.~\cite{Degrassi:2007kj} and obtain the
following two-sigma range:
$2.71\times10^{-4}<\mathcal{B}(b\to s\gamma)<4.33\times10^{-4}$.

\vspace{1mm}
vi) The $B^{+}\to\tau^{+}\nu_{\tau}$ branching fraction is given by
\begin{equation}
\label{eq:4.8}
\mathcal{B}(B^{+}\rightarrow\tau^{+}\nu_{\tau})=\frac{G_{F}^{2}}{8\pi}
\tau_{B^{+}}\left\vert V_{ub}\right\vert ^{2}f_{B_{d}
}^{2}M_{B^{+}}m_{\tau}^{2}\left(  1-\frac{m_{\tau}^{2}}{M_{B^{+}}^{2}}\right)
^{2}\left\vert 1-g_{P}\right\vert ^{2},
\end{equation}
where
\begin{equation}
\label{eq:4.9}
g_{P}=\frac{M_{B^{+}}^{2}t_{\beta}^{2}}{\left(  1+\epsilon_{0}t_{\beta
}\right)  \left(  1+\epsilon_{\tau}t_{\beta}\right)  M_{H^{+}}^{2}}
\end{equation}
parametrizes Higgs-mediated effects.  $\epsilon_{\tau}$ is obtained
from $\epsilon_\mu $ in \eq{defepsmu} by the replacement
$M_{\widetilde{\mu}L(R)}\to M_{\widetilde{\tau}L(R)}$.  Corrections to the
Higgs potential merely change the value of $M_{H^+}$, which becomes a function
of $M_A$, $\tan\beta$, and the various supersymmetric parameters.  Again, we
include these corrections in our numerical analysis.
Given the large theoretical and experimental uncertainties,
we impose: $g_P<0.36\ \cup\ 1.64<g_P<2.73$.
The constraint from $\mathcal{B}(B\to D \tau\nu)$ allows to reduce the second interval, and we end up with
$g_P<0.36\ \cup\ 1.64<g_P<1.79$ \cite{Trine:2008qv}.

\begin{table}[!h]
\centering
{\setlength{\tabcolsep}{0.3cm}{\renewcommand{\arraystretch}{1.1}
\begin{tabular}{|l|l|}
\hline
Built-in constraints & Additional constraints\\
\hline
$\mathcal{B}(B_{s}\rightarrow\mu^{+}\mu^{-})/\Delta M_{s}<5.7\times10^{-9}\ ps$&
$\mathcal{B}(B_{s}\rightarrow\mu^{+}\mu^{-})/\Delta M_{s}<3.3\times10^{-9}\ ps$\\
$y_{b}<2$ &
$2.71\times10^{-4}<\mathcal{B}(b\rightarrow s\gamma)<4.33\times10^{-4}$\\
$M_{h}>115\ \gev$ &
$g_P<0.36\ \textrm{or}\ 1.64<g_P<1.79$ \\
Stability bounds, see Eq.~(\ref{eq:StabBounds})&\\
\hline
\end{tabular}}}
\caption{Constraints built-in in the scan procedure (left) and imposed afterwards (right).}
\label{tab:4-2}
\end{table}

\subsection{Size of the new contributions}
\label{sect:correlation}

\begin{figure}
\centering
\begin{tabular}{cc}
\includegraphics[width=0.46\linewidth,keepaspectratio=true,angle=0]{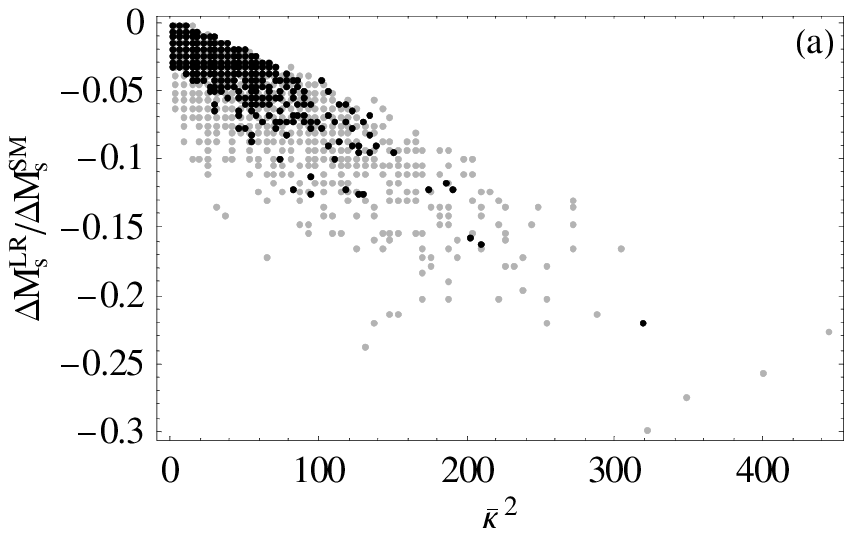} &
\includegraphics[width=0.46\linewidth,keepaspectratio=true,angle=0]{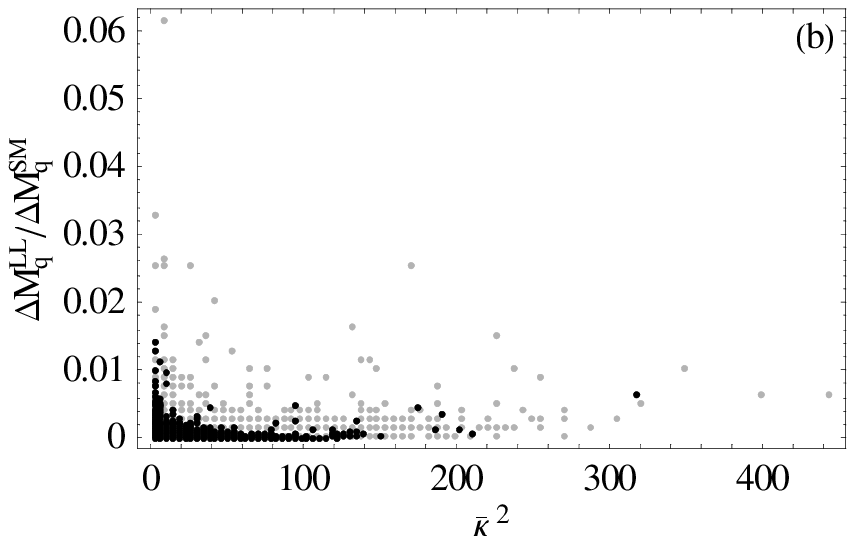} \\
\includegraphics[width=0.46\linewidth,keepaspectratio=true,angle=0]{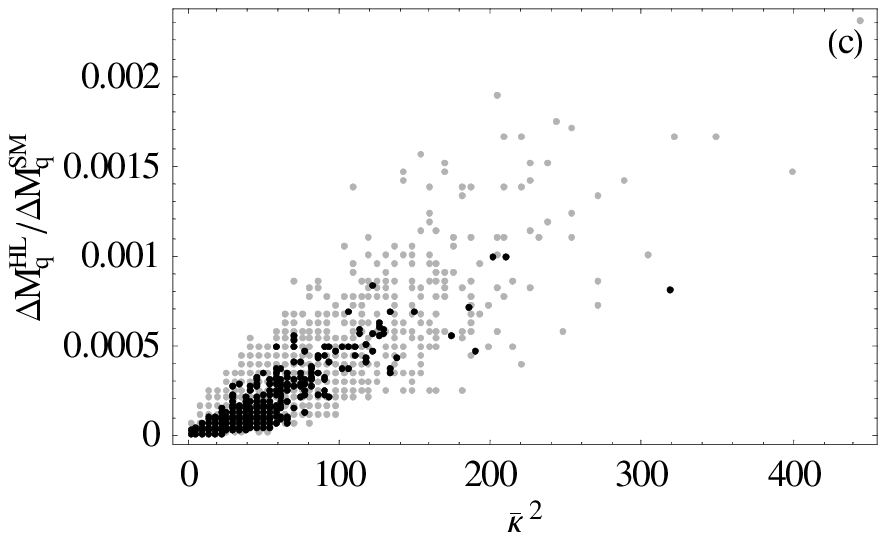} &
\includegraphics[width=0.46\linewidth,keepaspectratio=true,angle=0]{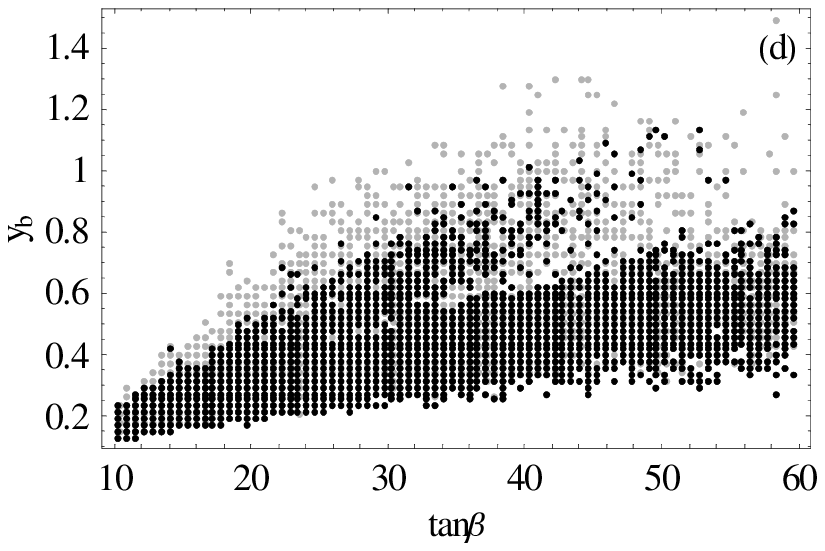} \\
\includegraphics[width=0.46\linewidth,keepaspectratio=true,angle=0]{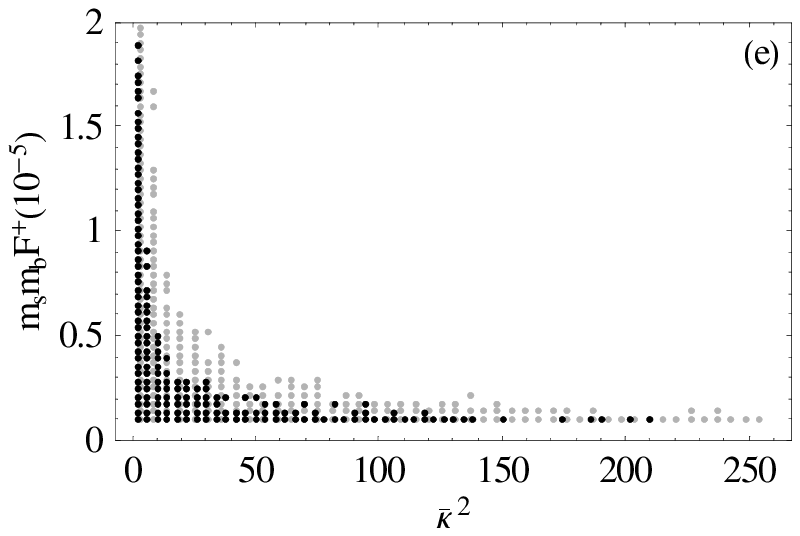} &
\includegraphics[width=0.46\linewidth,keepaspectratio=true,angle=0]{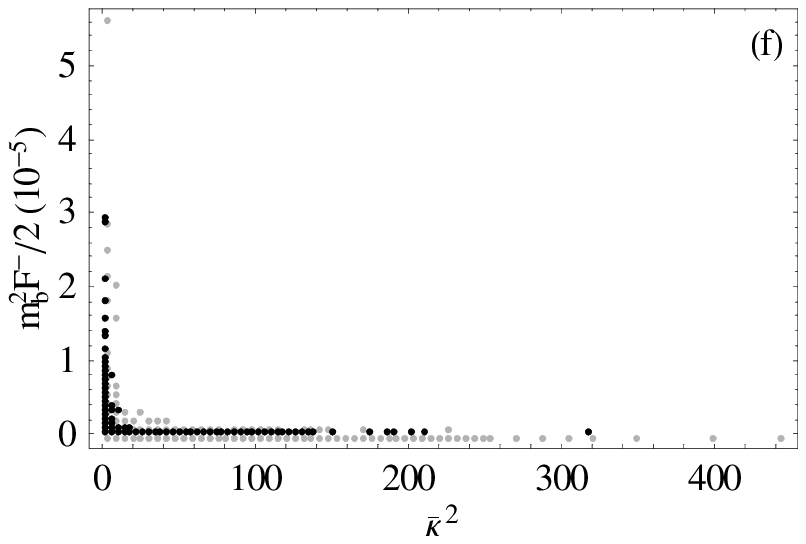} \\
\includegraphics[width=0.46\linewidth,keepaspectratio=true,angle=0]{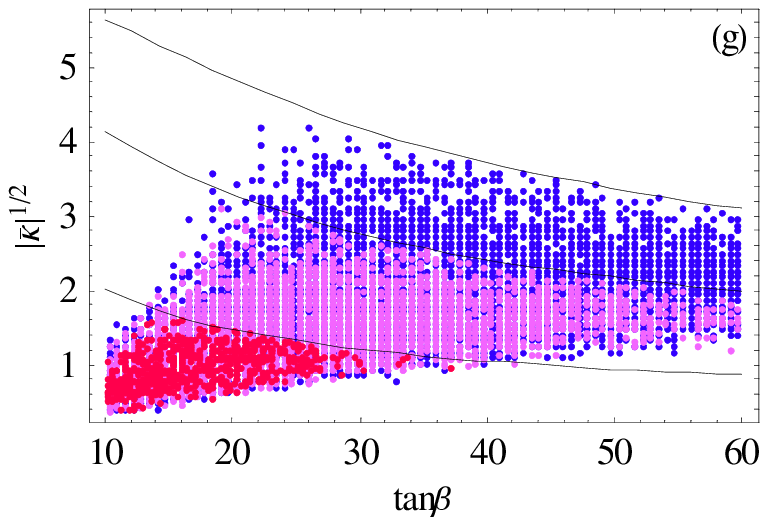} &
\includegraphics[width=0.46\linewidth,keepaspectratio=true,angle=0]{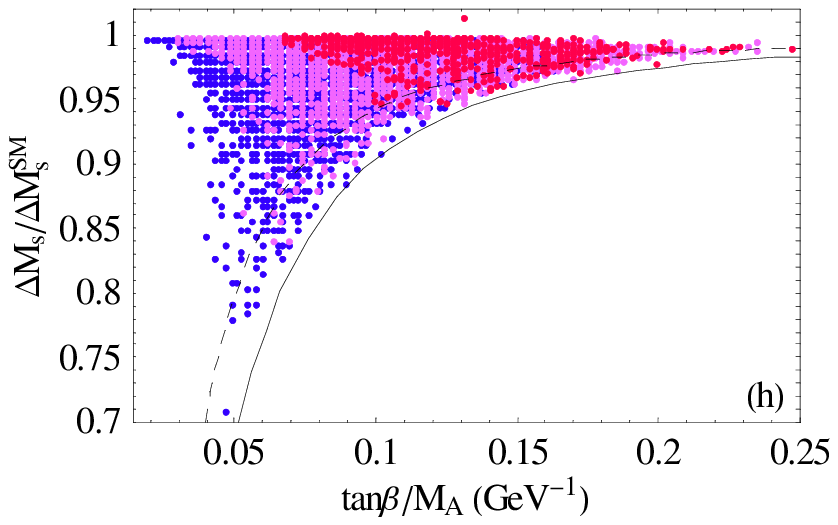}
\end{tabular}
\caption{Study of Higgs-mediated contributions to $\Delta M_q$ (see text).
Black dots denote the points in parameter space that satisfy all constraints,
while grey dots refer to those that only satisfy the initial constraints (see Table~\ref{tab:4-2}).
In plots (g) and (h), blue/magenta/red (dark grey / light grey / grey) points correspond
to $M_A=550/350/150\ \gev$, respectively.
Plain lines indicate the $\mathcal{B}(B_{s}\rightarrow\mu^{+}\mu^{-})$ constraint.
In plot (h), the dashed line corresponds to the more stringent $\mathcal{B}(B_{s}\rightarrow\mu^{+}\mu^{-})$
constraint in the right column of Table \ref{tab:4-2}.}
\label{fig:HiggsContr}
\end{figure}

\begin{figure}
\centering
\begin{tabular}{cc}
\includegraphics[width=0.46\linewidth,keepaspectratio=true,angle=0]{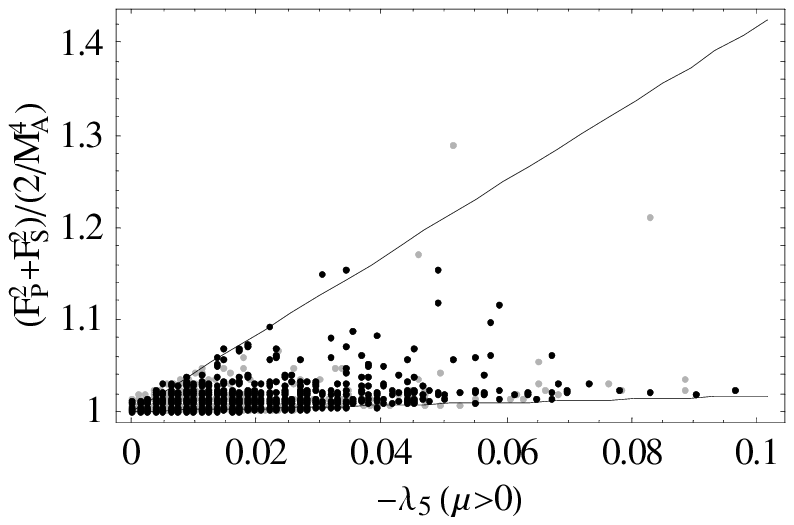} &
\includegraphics[width=0.46\linewidth,keepaspectratio=true,angle=0]{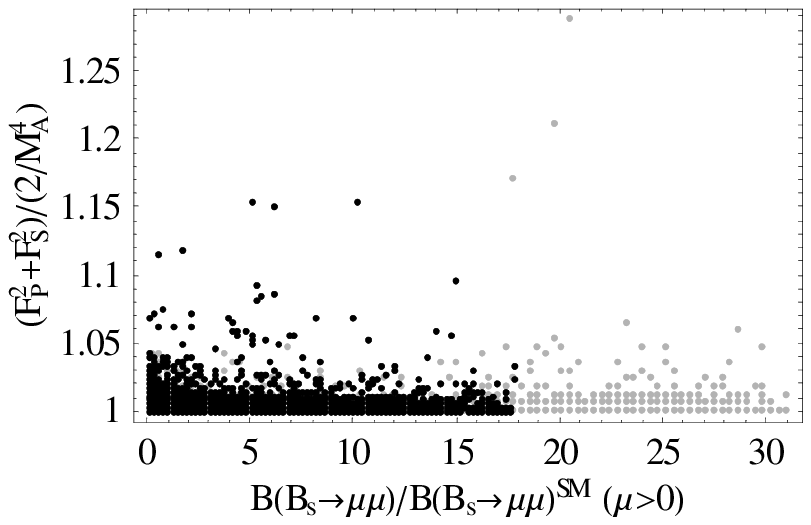}
\end{tabular}
\caption{Correction factor to $\mathcal{B}(B_{q}\to\mu^{+}\mu^{-})$ arising from supersymmetric
loop effects in the Higgs self-energies as a function of the effective coupling $\lambda_5$ (left)
and the total Higgs-mediated effects in $B_{s}\to\mu^{+}\mu^{-}$ (right), for $\mu>0$.
On the left-hand side, the upper (lower) line corresponds to $M_A=120\ (600)\ \gev$,
assuming the approximate formula of Eq.~(\ref{eq:4.7}) for the correction factor.}
\label{fig:BsMuMuCorrection}
\end{figure}

The various Higgs-mediated contributions $\Delta M_s^{\rm LR}$, $\Delta
M_q^{\rm LL}$ and $\Delta M_q^{\rm HL}$ normalized to the Standard-Model
prediction $\Delta M_q^{\rm SM}$ are displayed in
Fig.~\ref{fig:HiggsContr}(a-c) as a function of the FCNC coupling
$\overline{\kappa}$.  As expected from Eq.~(\ref{eq:4.5}), Higgs-loop
effects are very small (the bottom Yukawa coupling actually does not
reach its upper bound $y_b=2$ in the presence of the other constraints,
see Fig.~\ref{fig:HiggsContr}(d).  The upper and lower branches
correspond to $\mu<0$ and $\mu>0$, respectively).  Further, the
contribution of $\Delta M_q^{\rm LL}$ appears to be much smaller than
that of $\Delta M_s^{\rm LR}$ despite the fact that $m_b\mathcal{F}^{-}$
can compete with $m_s\mathcal{F}^{+}$, see
Fig.~\ref{fig:HiggsContr}(e,f).  This suppression is a consequence of
the $B_{s}\to\mu^{+}\mu^{-}$ constraint.  Indeed, large values of
$\mathcal{F}^{-}$ are obtained for small values of $M_A$, to which
$\mathcal{B}(B_{s}\to\mu^{+}\mu^{-})$ is particularly sensitive.  As a
result, the recent CDF bound \cite{:2007kv} only leaves room for very
small $\overline{\kappa}$ couplings, killing practically all effects in
$\Delta M_q^{\rm LL}$ (and actually also in $\Delta M_s^{\rm LR}$ for
such small $M_A$ values).  In Fig.~\ref{fig:HiggsContr}(g), we
illustrate this decrease of the maximal $\overline{\kappa}$ value
allowed by the $B_{s}\to\mu^{+}\mu^{-}$ constraint with $M_A$.
Blue/magenta/red (dark grey / light grey / grey) points correspond to
$M_A=550/350/150\ \gev$ (the constraints in the right column
of Table~\ref{tab:4-2} were not imposed to keep the focus on
$B_{s}\to\mu^{+}\mu^{-}$).  As one can see, for $M_A$ fixed, the largest
possible $\overline{\kappa}$ first increases with $\tan\beta^2$, as
expected from Eq.~(\ref{eq:4.3}), saturates the
$\mathcal{B}(B_{s}\to\mu^{+}\mu^{-})$ experimental upper bound for some
$\tan\beta$ value, and is then forced to decrease.  For smaller $M_A$,
the $B_{s}\to\mu^{+}\mu^{-}$ constraint is more stringent and only a
smaller $\overline{\kappa}_{max}$ can be achieved. This growth of
$\overline{\kappa}_{max}$ with $M_A$ is sufficient to overcome the
$1/M_A^2$ suppression factor in $\Delta M_s^{\rm LR}$ but not the
$1/M_A^4$ one in $\Delta M_q^{\rm LL}$.  Overall, Higgs-mediated effects
in $\Delta M_q$ are of the LR type and the room for such effects
increases with $M_A$.  The correlation between $\Delta M_s$ and
$\mathcal{B}(B_{s}\to\mu^{+}\mu^{-})$ pointed out in Refs.~\cite{bcrs}
is thus preserved, up to the relatively small Higgs self-energy
corrections to $B_{s}\to\mu^{+}\mu^{-}$ mentioned above
Eq.~(\ref{eq:4.7}).  These are only relevant for $\mu>0$,
$\tan\beta\lesssim 25$, small $M_A$, and large $\lambda_5$, though (see
Fig.~\ref{fig:BsMuMuCorrection}). The mass difference in the $B_d$
system, on the other hand, remains unaffected.  These results seem to
contradict those of Ref. \cite{Parry:2006vq}, where large LL-type
effects were claimed.  Without attempting a close numerical comparison
(the sign of $\Delta M^{\rm LL}_s/\Delta M^{\rm LR}_s$ in
\cite{Parry:2006vq} is actually reversed with respect to ours), let us
point out that, as shown by Figs.~3(e,f), a large $\Delta M^{\rm
  LL}_s/\Delta M^{\rm LR}_s$ ratio does not automatically lead to large
non-standard effects in $\Delta M_s$ due to the
$\mathcal{B}(B_{s}\to\mu^{+}\mu^{-})$ constraint.

\begin{figure}
\centering
\begin{tabular}{cc}
\includegraphics[width=0.46\linewidth,keepaspectratio=true,angle=0]{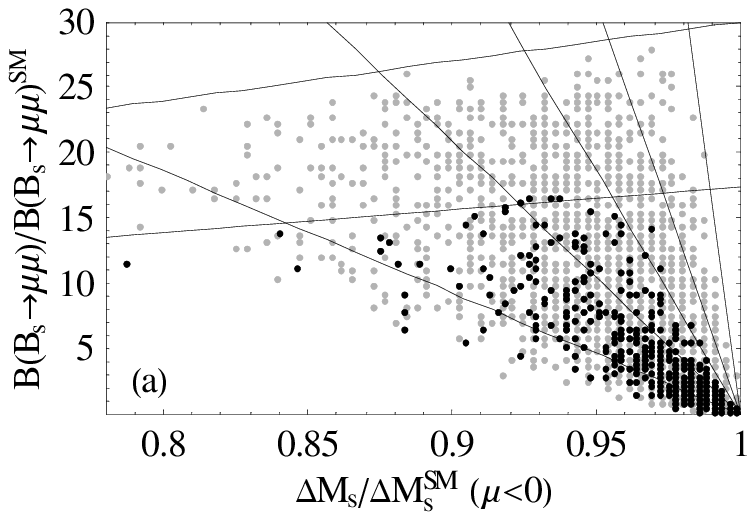} &
\includegraphics[width=0.46\linewidth,keepaspectratio=true,angle=0]{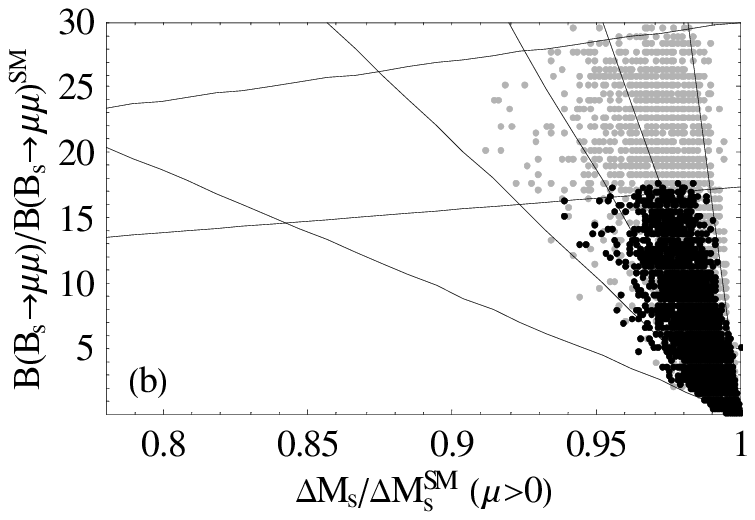} \\
\includegraphics[width=0.46\linewidth,keepaspectratio=true,angle=0]{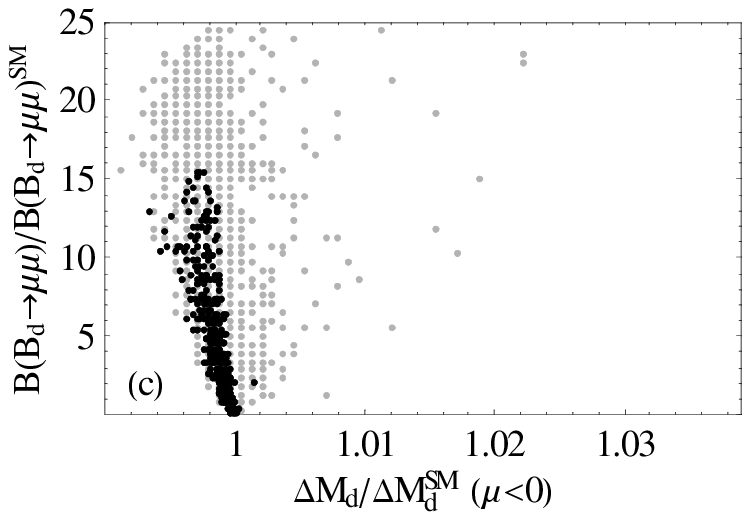} &
\includegraphics[width=0.46\linewidth,keepaspectratio=true,angle=0]{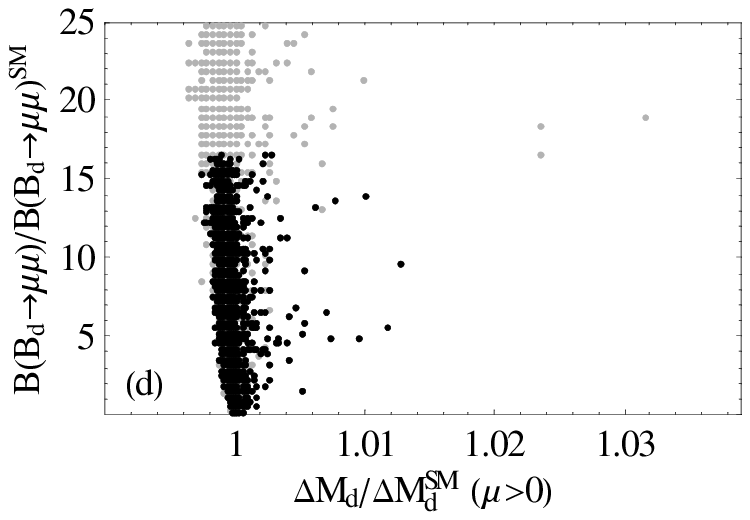}
\end{tabular}
\caption{Correlation between $\Delta M_q$ and
  $\mathcal{B}(B_{q}\to\mu^{+}\mu^{-})$: (a) $q=s$, $\mu<0$; (b) $q=s$,
  $\mu>0$; (c) $q=d$, $\mu<0$; (d) $q=d$, $\mu>0$.  The descending lines
  correspond to a fixed value of the ratio $\tan\beta/M_A$.  From left
  to right: $\tan\beta/M_A=0.05, 0.075, 0.10, 0.13, 0.21\ \gev^{-1}$.
  The ascending lines refer to the
  $\mathcal{B}(B_{s}\rightarrow\mu^{+}\mu^{-})/\Delta M_{s}$
  constraints, see Table~\ref{tab:4-2}.  These lines do not take into
  account the uncertainties on the quark masses and bag factors, nor the
  effects from sparticle loop corrections to the Higgs potential in
  $\Delta M_{s}$, $\mathcal{B}(B_{s}\rightarrow\mu^{+}\mu^{-})$, and to
  the lepton Yukawa couplings in
  $\mathcal{B}(B_{s}\rightarrow\mu^{+}\mu^{-})$, so that the actual
points do not follow them exactly but are somewhat scattered.}
\label{fig:Correlation}
\end{figure}

Being of the LR type, the maximal effect allowed in $\Delta M_{s}$ is
essentially determined by the current
$\mathcal{B}(B_{s}\to\mu^{+}\mu^{-})$ experimental upper bound for a
fixed (but large enough) value of the ratio $\tan\beta/M_A$.  This is
illustrated in Fig.~\ref{fig:HiggsContr}(h) for a slightly larger bound
(cf. left-hand side column in Table~\ref{tab:4-2}).  The correlation
itself is displayed in Fig.~\ref{fig:Correlation}, where each diagonal
corresponds to a fixed value of the ratio $\tan\beta/M_A$.  We
distinguish the cases $\mu>0$ and $\mu<0$ as the latter leads to larger
effects due to smaller denominators in Eq.~(\ref{eq:4.3}) but is
disfavoured by the measurement of $(g-2)_\mu$, as mentioned previously.
The sign of the various $a$-terms, on the other hand, has only little
impact.  Still, in the case $\mu<0$, $a_t>0$ helps satisfy the $b\to
s\gamma$ constraint.  Note that the effect of the $B^+\to\tau^+\nu_\tau$
constraint is particularly transparent on Fig.~\ref{fig:Correlation}: it
removes the points with large $\tan\beta/M_A$ ratios, i.e., the steepest
diagonals.  Altogether, for $M_A<600\ \gev$, Higgs-mediated effects in
$\Delta M_s$ can reach $\sim7\%$ ($\sim20\%$) for $\mu>0$ ($\mu<0$).
These findings agree with those of Ref.~\cite{Altmannshofer:2007rj}.
They merely follow from the $\mathcal{B}(B_{s}\to\mu^{+}\mu^{-})$
constraint, as one can see from Figs.~\ref{fig:HiggsContr}(g,h) and
\ref{fig:Correlation}.

\begin{figure}
\centering
\begin{tabular}{cc}
\includegraphics[width=0.46\linewidth,keepaspectratio=true,angle=0]{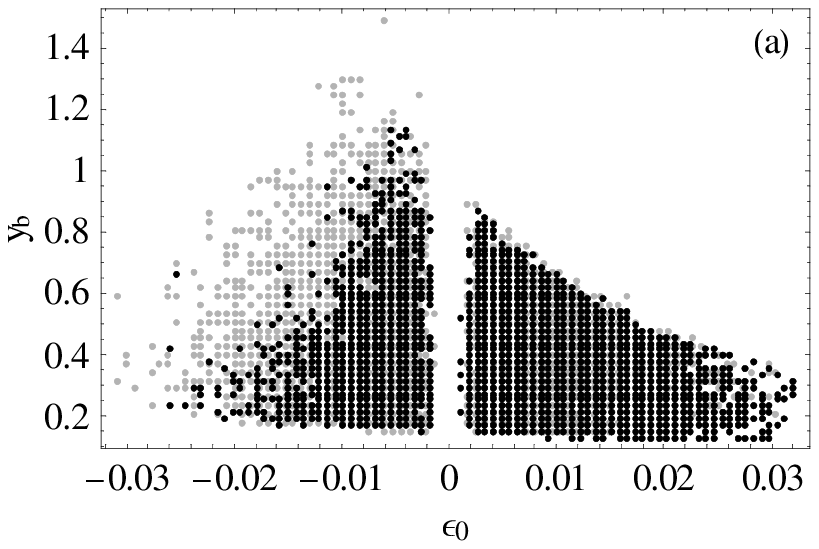} &
\includegraphics[width=0.46\linewidth,keepaspectratio=true,angle=0]{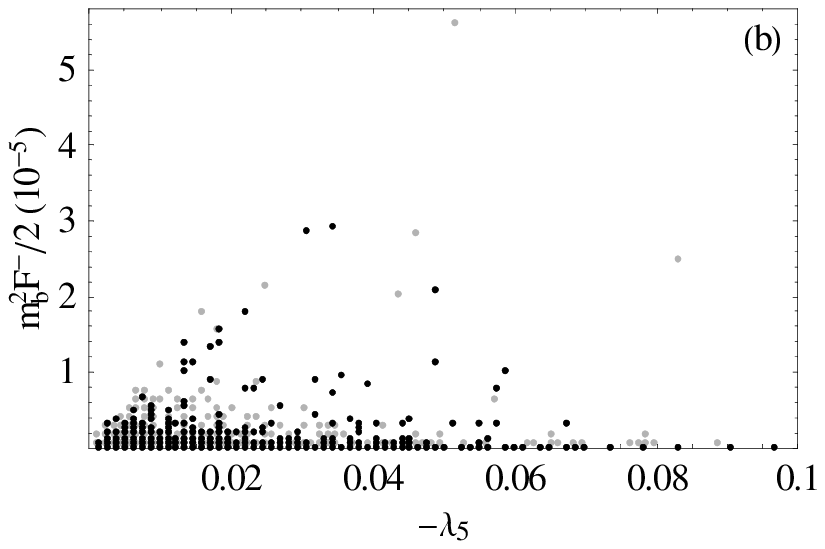} \\
\includegraphics[width=0.46\linewidth,keepaspectratio=true,angle=0]{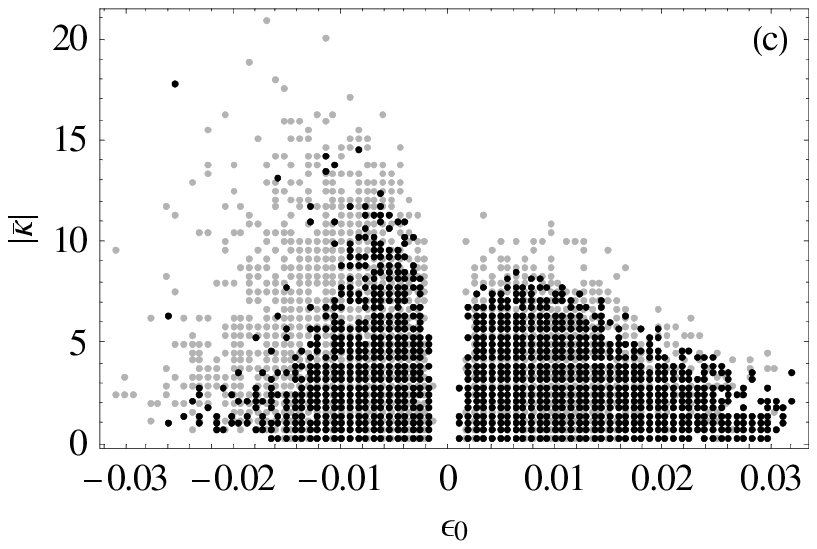} &
\includegraphics[width=0.46\linewidth,keepaspectratio=true,angle=0]{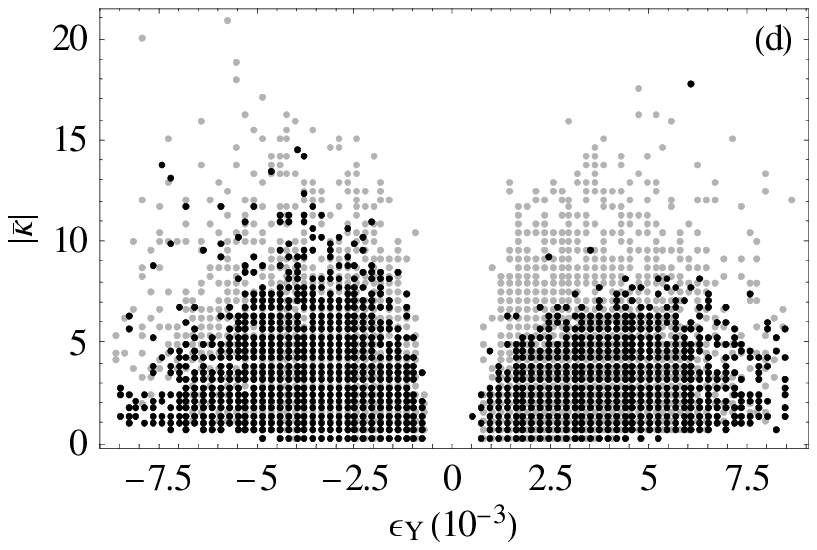} \\
\includegraphics[width=0.46\linewidth,keepaspectratio=true,angle=0]{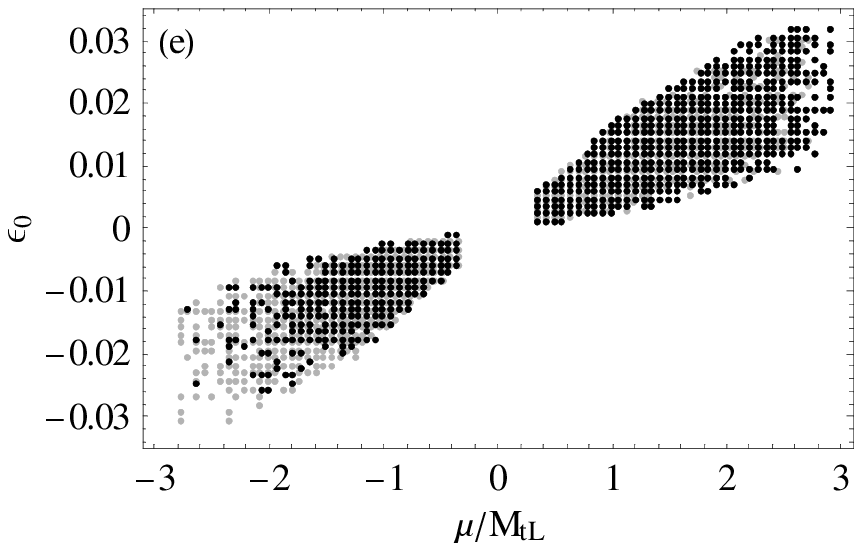} &
\includegraphics[width=0.46\linewidth,keepaspectratio=true,angle=0]{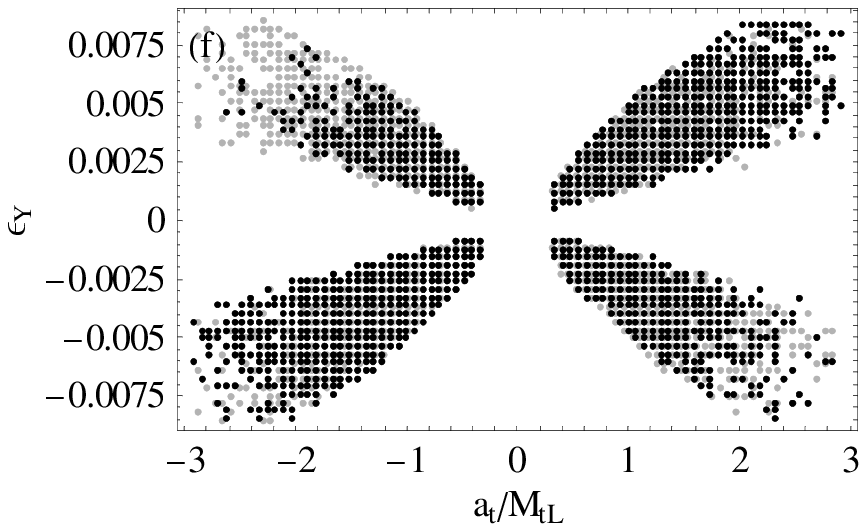}\\
\includegraphics[width=0.46\linewidth,keepaspectratio=true,angle=0]{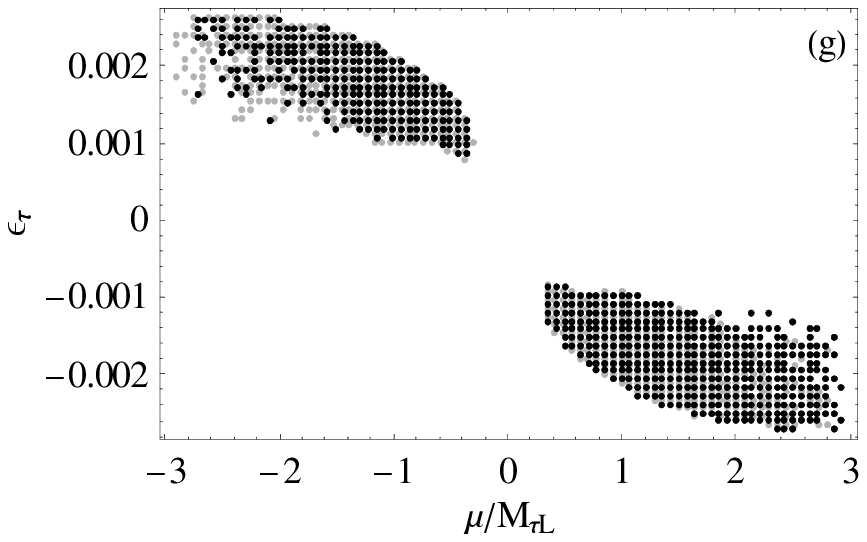} &
\includegraphics[width=0.46\linewidth,keepaspectratio=true,angle=0]{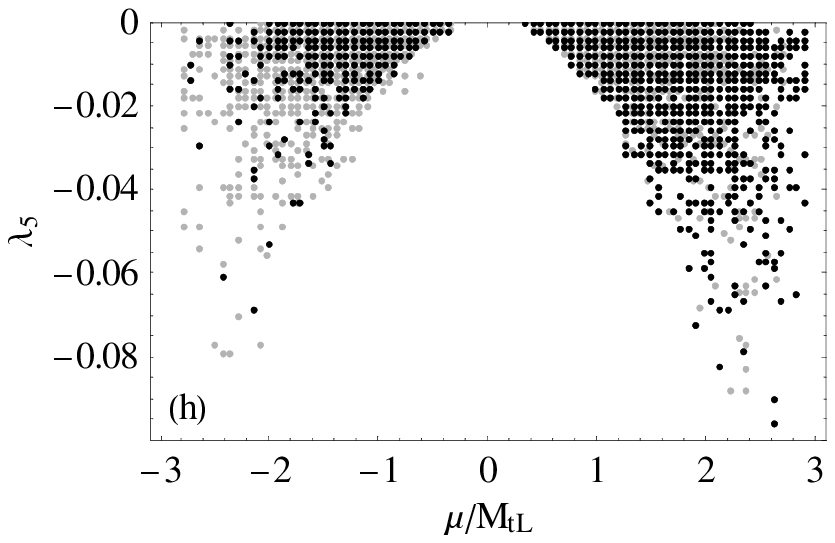}
\end{tabular}
\caption{Dependence of various quantities on effective couplings or supersymmetric parameters.}
\label{fig:VariousDependences}
\end{figure}

Finally, for completeness (or out of curiosity), we display in
Fig.~\ref{fig:VariousDependences} the dependence of various quantities
on effective couplings or supersymmetric parameters.  In particular, in
the last four plots, we illustrate how the loop functions
$\varepsilon_{0}$, $\varepsilon_{Y}$, $\varepsilon_{\tau}$, and
$\lambda_{5}$ increase with the range chosen for $M_{\mathrm{SUSY}}$
(more precisely, they increase with the trilinear and $\mu$ terms and
decrease with the squark and slepton mass parameters $M_{\widetilde
  f_{L}}$ and $M_{\widetilde f_{R}}$ with $f=t,b,\tau$).

\subsection{CP-violating effects}\label{sect:CPV}

The Higgs-mediated $\bb$ mixing amplitudes studied here can in principle
generate new contributions to the CP-violating phases measured in the
$B_d\to J/ \psi K_S$ time-dependent CP asymmetry and the $B_s\to
J/\psi\phi$ time-dependent angular distribution. 
The coefficients of the $\sin(\dm_q \, t)$ terms are
\begin{equation}
\label{eq:4.3.1}
\begin{split}
S_{J/ \psi K_S}=&\sin(2\beta+\phi_d^\Delta),\\
S_{J/ \psi\phi}=&-\sin(-2\beta_s+\phi_s^\Delta),
\end{split}
\end{equation}
where
$\beta\equiv\arg [-(V_{td}^{*}V_{tb})/(V_{cd}^{*}V_{cb})]$,
$\beta_s\equiv-\arg [-(V_{ts}^{*}V_{tb})/(V_{cs}^{*}V_{cb})] $, and
\begin{equation}
\label{eq:4.3.2}
\phi_q^\Delta=\arg(M_{12}^q/M_{12}^{q,\rm SM})\equiv\arg(1+h_q).
\end{equation}
In $B_s\to J/\psi \phi$ an angular analysis separates 
the different CP components, the sign quoted for $S_{J/\psi\phi}$ 
in \eq{eq:4.3.1} refers to the dominant CP-even component.
These phases have received a lot of attention recently.  In particular,
the new measurements of $-2\beta_s+\phi_s^\Delta$ by the CDF and D0
collaborations \cite{:2008fj}, both more than 1.5 sigma above its SM
prediction \cite{ln}, have triggered speculations about the validity of
the SM \cite{Bona:2008jn}.  A possible tension between the value of
$\sin2\beta$ obtained from $S_{J/ \psi K_S}$ and the amount of CP
violation in the kaon system was also pointed out \cite{Buras:2008nn}.

Looking back at Eqs.~(\ref{3.9}-\ref{3.14}), it is clear that the new phases $\phi_q^\Delta$,
when associated with the $Q^{\rm LR}_2$ effective operator,
have to be brought up by the quark-Higgs couplings $\kappa_{ij}$ as $\mathcal{F}^+$ cannot develop an imaginary part.
When associated with $Q^{\rm SLL}_1$ or $Q^{\rm SRR}_1$, on the other hand,
they can arise from both the Yukawa sector and the Higgs potential via $\mathcal{F}^-$.
Within MFV, $\kappa_{qb}^{\ast}\,\kappa_{bq}=|\bar\kappa|^2\lambda^2_{qb}m_qm_b/v^2$
and only $C^{\rm SLL}_1$ can produce a new phase (via $\epsilon_{0,Y}$ or $\lambda_{5,7}$).
However, the $B_{s}\rightarrow\mu^{+}\mu^{-}$ branching fraction is barely affected
by CP-violating effects, so that its constraints on $|\bar\kappa|$ are still very well approximated
by the plain lines in Fig.~\ref{fig:HiggsContr}(g) (for some representative $M_A$ values).
As a result, just like in the CP-conserving case, the net effect of the suppression of $|\bar\kappa|$
and enhancement of $\mathcal{F}^-$ for small $M_A$ is quite small.
The MSSM with large $\tan\beta$ and MFV is thus not able to account for a large non-standard phase
in $\bbs$ (or $\bbd$) mixing, if the evidence for such a phase were confirmed.
Let us emphasize, however, that the formulation of MFV adopted here does not coincide exactly
with the full symmetry-based definition of Ref.~\cite{D'Ambrosio:2002ex}.
In the formalism of Ref.~\cite{D'Ambrosio:2002ex}, it was shown recently that new phases could appear
in the $\delta_{LL}^{13,23}$ sector, in addition to those in the $(\delta_{LR}^d)^{13,23}$ sector \cite{Colangelo:2008qp}.
The possible impact of these MFV phases via $\kappa_{qb}^{\ast}\,\kappa_{bq}$ in $C^{\rm LR}_2$ is a priori rather limited
due to the $B_{s}\rightarrow\mu^{+}\mu^{-}$ constraint, yet a more quantitative analysis is desirable.

Beyond MFV, the $Q^{\rm LR}_2$ contribution is expected to dominate. As said before,
supersymmetric loop corrections to the Higgs propagator $\mathcal{F}^+$ do not bring in any new phases.
These can only enter via the quark-Higgs couplings $\kappa_{qb}^{\ast}$ and $\kappa_{bq}$.
The possible size of CP-violating effects
generated in this way without violating the existing constraints
deserves a study on its own. We will not discuss this further here.

\section{Conclusions}
\label{sect:ccl}
 We have studied supersymmetric loop corrections to the MSSM Higgs
 sector. While the tree-level Higgs sector of the MSSM is a 2HDM of
 type II, the soft supersymmetry-breaking terms lead to new
 loop-induced couplings which result in a generic 2HDM with FCNC
 couplings of neutral Higgs bosons to quarks, even if the
 supersymmetry-breaking sector is minimally flavour-violating. The
 strength of these couplings grows with $\tan\beta$ and precision
 observables of flavour physics are known to severely constrain
 large-$\tan\beta$ scenarios of the MSSM. The appropriate tool for such
 studies is an effective Lagrangian, which is derived by integrating
 out the heavy supersymmetric particles. The abundant literature on the
 subject has primarily focused on the flavour-changing Yukawa couplings
 \cite{hrs,hpt,bk,ir,bcrs,ltb,cgnw}. Among the FCNC quantities, 
 \bbm\ plays a special role, because the apparently dominant
 contribution of \fig{fig:bbtree} vanishes. Therefore \bbm\ is
 sensitive to subleading effects, whose systematic study was the main 
 motivation for this paper. Pursuing this goal we have derived several 
 conceptual and analytic results which can be applied well beyond   
 this topic.  They can be classified into three categories:
\bigskip\newline
\textit{1. MSSM Higgs sector}
\smallskip\newline
We have matched the complete MSSM Higgs sector, i.e.\ both the
Yukawa interactions and the Higgs potential, onto an effective 2HDM.  Our
results for the effective Yukawa couplings are valid for arbitrary CP
phases of $\mu$, $a_t$, and the gaugino masses; and \eqsand{3.17}{3.18}
correct the gaugino contributions to $\epsilon_0$ and $\epsilon_Y$
quoted in Ref.~\cite{fgh}.  The complete one-loop matching corrections
for the quartic Higgs couplings for the most general MSSM are explicitly 
listed in one place for the first time. This result goes beyond minimal flavour
violation and beyond the large-$\tan\beta$ limit. 
It is well-known
that improper choices of the MSSM renormalization scheme can lead to
radiative corrections which grow with $\tan\beta$ rendering perturbative
results unreliable \cite{Freitas:2002um}.  At the heart of this problem
is the feature that $\tan\beta$ is an ill-defined parameter in the
general 2HDM, which permits arbitrary rotations among the two Higgs
doublets. In the matching of the MSSM onto the 2HDM this feature enters
through the wave function renormalization, and we propose an
$\overline{\textrm{MS}}$ renormalization of $\tan \beta$ in the 2HDM
which 
is stable in the limit of large $\tan\beta$.
The relation to a
$\overline{\textrm{DR}}$-renormalized $\tan \beta$ in the MSSM is
discussed including electroweak corrections. We identify the places in
the effective Higgs potential where physical $\tan \beta$-enhanced
effects occur. The coefficients $\lambda_2$, $\lambda_5$ and
$\lambda_7$, which are important for \bbm, are explicitly specified for
the MFV case in \eqsto{eq:lambda2MFV}{eq:lambda7MFV}. 
Some loop corrections to the Higgs potential at large $\tan\beta$ and their
impact on $\tan\beta$ itself and the Higgs-fermion couplings have
also been considered in an effective-field-theory framework
in Ref.~\cite{brs}, which
appeared during completion of this paper. Part of the results therein
overlap with Section \ref{sect:higgs} of this paper. We disagree
with \cite{brs} in some points (cf. Section \ref{sect:higgs}),
notably in that we find a
$\tan\beta$-enhanced term in the relation of the $\overline{DR}$ and
DCPR $\tan\beta$ parameters. We stress that, in general, only the
former is numerically close to
the $\tan\beta$ parameter extracted from $B$-physics observables.
\bigskip\newline
\textit{2. Large $\mathit{\tan\beta}$ phenomenology}
\smallskip\newline
The prime application of our results is \bbm.
We have identified
a global $U(1)$ symmetry of the $\bar b_R q_L$ Higgs-mediated FCNC
transitions and the tree-level Higgs potential in the large-$\tan\beta$
limit which suppresses the superficially leading contribution of
\fig{fig:bbtree}. A systematic study of \bbm\ has required the analyses
of four subleading contributions, which are governed by the small
parameters $m_{d,s}/m_b$, $1/\tan\beta$, $v/M_{\rm SUSY}$ and the loop
factor $1/(4 \pi)^2$.  These parameters either provide a breaking of the
$U(1)$ symmetry or allow for a contribution proportional to the
$U(1)$-conserving standard-model effective operator.  Prior to this
work, only corrections involving $m_{d,s}/m_b$ had been studied
\cite{bcrs} (with the exception of Ref.\cite{fgh}).  The $v/M_{\rm
  SUSY}$ corrections are found numerically small. The new loop
contributions include all non-decoupling SUSY corrections to the quartic
Higgs interactions $\lambda_1$--$\lambda_7$ and the contribution of
neutral Higgs box diagrams in the effective theory. In the complex MSSM
the results for ${\cal F}^{\pm}$ comprising the neutral Higgs
propagators become cumbersome. We have expressed ${\cal F}^{\pm}$ in
terms of sub-determinants of the neutral Higgs mass matrix. These
expressions are easy to implement and clearly reveal
the invariance of the Higgs-mediated amplitudes
under rotations of the basis
$(H_{u},-\varepsilon H_{d})$. The results for the Higgs
sector are also used to refine the MSSM predictions for the $B^+ \to
\tau^+ \nu$ and the $B_{s,d} \to \mu^+ \mu^-$ branching ratios.
In this context we stress that loop corrections to the
Higgs potential do not give rise to additional $\tan\beta$-enhanced
contributions to the charged-Higgs-fermion couplings beyond those
known before Ref.~\cite{fgh} appeared. Hence no modification of the
charged-Higgs contributions to $B^+ \to \tau^+ \nu$ or $B \to X_s \gamma$
relative to Ref.~\cite{bcrs} occurs.

While the MSSM corrections to \bbm\ in the large $\tan \beta$ scenario
could be dominated by the contribution of $\lambda_5$ and $\lambda_7$,
the size of this piece is limited by the experimental upper bound on
${\cal B}(B_s\to \mu^+\mu^-)$. After performing an exhaustive analysis
of this quantity, ${\cal B} (\bar B \to X_s\gamma)$, ${\cal B} (B^+ \to
\tau^+ \nu)$ and the mass of the lightest neutral Higgs boson $M_{h}$,
we find that the impact of the corrections to the Higgs potential on
$\dm_s$ is always weaker than that of the $m_s/m_b$ correction
identified in Ref.~\cite{bcrs}. Assessing the total Higgs-mediated MSSM
corrections to $\dm_s$ we find an upper limit of 7\% of the SM
contribution for $\mu>0$ and $M_A<600\gev$. If $\mu$ is negative,
the upper bound is around 20\%. This is in
contrast with Refs.~\cite{Parry:2006vq} and
Ref.\cite{fgh}, which claim large effects of the Higgs potential
on \bbm. The corrections to ${\cal
  B}(B_q\to \mu^+\mu^-)$ from the Higgs potential are typically also
small, but can reach 15\% in some corners of the parameter space. In
summary, the correlation between an enhancement of ${\cal B}(B_q\to
\mu^+\mu^-)$ and a (moderate) depletion of $\dm_s$ found in
Ref.~\cite{bcrs} remains essentially intact.

We finally note that our new contributions can alter the CP phase of the
\bbm\ amplitude, while the previously known Higgs contribution
proportional to $m_s m_b{\cal F}^+$ has the same phase as the SM term (in MFV
scenarios).  While the maximal possible CP phase is clearly below the
sensitivity of the current Tevatron experiments, it is an open question
whether future \bbm\ experiments can help to unravel the CP structure of 
the MSSM Higgs potential.   
\bigskip\newline
\textit{3. Heavy-quark relations and bag parameters}
\smallskip\newline
We have transformed the NLO anomalous dimensions computed in Ref.\cite{bmu}
to an operator basis and a renormalization scheme typically used in
lattice calculations. The according anomalous dimensions are needed to
evaluate the `bag' parameters, which parametrise the hadronic matrix
elements, at the electroweak scale. We further employed a heavy-quark
relation to improve the numerical prediction of the bag parameter
$B_1^{\textrm{SLL}\, \prime}$ entering the SUSY contributions to \bbm.  The
heavy-quark relation essentially determines $B_1^{\textrm{SLL}\, \prime}$ in
terms of the bag parameter $B_1^{\rm VLL}$, which is needed for the SM
prediction \cite{ln}. We found $P^{\rm SLL}_1=-\frac{5}{8}B_1^{\rm SLL\, \prime}(m_t)=-1.36\pm0.12$.
\section*{Acknowledgements}
M.G.\ and S.J.\ appreciate helpful conversations with Dominik St\"ockinger,
Martin Beneke, John Ellis,
Gian Giudice, Uli Haisch, Janusz Rosiek, Pietro Slavich, Jure Zupan, and other
members of the CERN theory group on various aspects of this work.
U.N.\ appreciates  fruitful discussions
with Lars Hofer und Dominik Scherer. S.J.\ acknowledges the
hospitality of the University of Karlsruhe.
We are grateful to A.J. Buras for comments on the manuscript.

This work was supported by the DFG grant No.~NI 1105/1--1, by project C6
of the DFG
Research Unit SFB--TR 9 \emph{Computergest\"utzte Theoretische
Teilchenphysik}\, by the EU Contract No.~MRTN-CT-2006-035482, \lq\lq
FLAVIAnet'',
the DFG cluster of excellence ``Origin and Structure of the Universe'',
and the RTN European Program MRTN-CT-2004-503369.

\begin{appendix}
\section{Notations and conventions}\label{sect:not}

To state our phase conventions for $\mu$ and $M_2$ we quote the chargino mass
matrix:
\begin{equation}
{\cal M}_{\chi^+} =
\lt(
\begin{array}{cc}
\ds M_2                    & \ds \frac{g\ v \sin\beta}{\sqrt{2}} \\ 
\ds \frac{g\ v \cos\beta}{\sqrt{2}} & \ds \mu 
\end{array} \rt) 
\label{defchm}
\end{equation}
with $v=246\gev$ and the chargino mass term in the Lagrangian
\begin{equation}
{\cal L}_{\chi^+}^{\rm mass} = - (\lambda^-,
\widetilde h{}_d^2 ) {\cal M}_{\chi^+} (\lambda^+, \widetilde h{}_u^1)^T\, .
\end{equation}

For the case of a general flavour
structure of the squark mass matrices we define the trilinear couplings
$\hat{T}_{u_{ij}}$ and $\hat{T}_{d_{ij}}$ (with flavour indices $i,j$) such that the
squark mass matrices read
\begin{eqnarray}
\hat{\cal M}^2_{\widetilde u} &=&
\lt(
\begin{array}{cc}
\ds \hat{M}^2_{\widetilde u_L} & 
\ds \frac{v \sin\beta}{\sqrt 2} \lt[ \hat{T}^\dagger_u - 
                                   \mu\, \hat{Y}^\dagger_u
                           \cot\beta \rt] \\[2mm]
\ds \frac{v \sin\beta}{\sqrt 2} \lt[ \hat{T}_u - 
                                   \mu^* \hat{Y}_u \cot \beta \rt] & 
\ds \hat{M}^2_{\widetilde u_R}     
\end{array} \rt), \no \\[5mm]
\hat{\cal M}^2_{\widetilde d} & = &
\lt(
\begin{array}{cc}
\ds \hat{M}^2_{\widetilde d_L} & 
\ds \frac{v \cos\beta}{\sqrt 2} 
\lt[ \hat{T}^\dagger_d - \mu\, \hat{Y}^\dagger_d \tan \beta \rt] \\[2mm]
\ds \frac{v \cos\beta}{\sqrt 2} 
\lt[\hat{T}_d - \mu^* \hat{Y}_d \tan \beta \rt]   & 
\ds \hat{M}^2_{\widetilde d_R}     
\end{array} \rt),
\label{deftril}
\end{eqnarray}
in the super-CKM basis, where the 
($\overline{\rm DR}$-renormalized) Yukawa matrices 
are diagonal: $\hat{Y}_q=\mbox{diag}\, (y_{q_1},y_{q_2},y_{q_3})$, 
$q=u,d$. 
The mass matrices in \eq{deftril} correspond to the squark mass term
\begin{equation}
{\cal L}^{\rm mass}_{\widetilde q} = 
-\Phi^\dagger_{\widetilde u} \hat{\cal M}^2_{\widetilde u} 
 \Phi_{\widetilde u}
-\Phi^\dagger_{\widetilde d} \hat{\cal M}^2_{\widetilde d} 
 \Phi_{\widetilde d}
\end{equation}
with $\Phi_{\widetilde u} = (\widetilde u_L,\widetilde c_L, \widetilde
t_L, \widetilde u_R, \widetilde c_R, \widetilde t_R)^T$ and
$\Phi_{\widetilde d} = (\widetilde d_L,\widetilde s_L, \widetilde b_L,
\widetilde d_R, \widetilde s_R, \widetilde b_R)^T$.

Our sign convention for the MSSM Yukawa couplings implies the
following relations between 2HDM quark masses and MSSM Yukawa 
couplings in the up sector:
\begin{eqnarray}
y_{u_i} &=& \frac{\sqrt 2}{v\sin\beta} \,m_{u_i}.
                    \label{defyt}
\end{eqnarray} 
In general, the analogous relations in the down sector involve complex phases
associated with the $\tan\beta$-enhanced threshold corrections in Eq.~(\ref{YukCorr}).
In particular,
Ref.~\cite{cepw}, which  
discusses the complex MSSM for the case without flavor mixing, relates the
$b$ Yukawa coupling to the $b$ quark mass as 
\begin{eqnarray}
y_b & =& \frac{\sqrt 2}{v\cos\beta} \, 
   \frac{m_b}{ 1+\widetilde \epsilon_3\tan\beta  }, 
                    \label{defy}
\end{eqnarray} 
rendering $y_b$ complex for complex $\widetilde\epsilon_3$ in Eq.~(\ref{3.19}).
Our approach of matching the MSSM to an
effective 2HDM permits different phase conventions, because the quark fields
in the MSSM and the 2HDM can
be chosen to differ by a phase factor. We can
rephase the $b_R$ super-field of the MSSM in such a way that $y_b$ is real and
positive and $1+\widetilde \epsilon_3\tan\beta $ in \eq{defy} is replaced by
$|1+\widetilde \epsilon_3\tan\beta |$. The (physical) phase of $1+\widetilde
\epsilon_3\tan\beta $ will then, however, appear explicitly in the Higgs and
higgsino couplings to bottom (s)quarks.
Introducing $3\times 3$ flavour mixing,
the relation between $\hat{Y}_d$ and $m_d$, $m_s$, $m_b$ is found from
Eq.~(\ref{YukCorr}).
Now the quark fields in the 2HDM differ from those in
the MSSM by a complex rotation in flavour space and a particular choice for the
phases of MSSM fields appears less obvious. In particular, one could render
all $y_{d_i}$ real and positive by suitable rephasings of the right-handed
superfields.
Note that within MFV $y_b$ is still related
to $m_b$ via Eq.~(\ref{defy}) in good approximation without such rephasings.
The analogous relation for the first two generations reads
\begin{eqnarray}
y_{d,s} & =& \frac{\sqrt 2}{v\cos\beta} \, 
   \frac{m_{d,s}}{ 1+\epsilon_0\tan\beta  }\, ,
                    \label{defyq}
\end{eqnarray}
while in the lepton sector, we have
\begin{eqnarray}
y_\ell & =& \frac{\sqrt 2}{v\cos\beta} \, 
   \frac{m_\ell}{ 1+ \epsilon_\ell \tan\beta  } 
\qquad\qquad \mbox{with $\ell=e,\mu,\tau$}\, .
                    \label{defyell}
\end{eqnarray}
In (most of) the paper
we express our results in terms of fermion masses (i.e.\ avoiding
$y_{d_i,\ell}$) to achieve formulae which are independent of such phase conventions.
Note that the phases of $\epsilon_0$ and $\epsilon_Y$ are physical and no
phase convention other than that of the CKM matrix matters for $\kappa_{ij}$
in \eqsand{3.16}{3.16bis}. While the phase convention of $y_{q_i}$ enters the
phases in $\hat{T}_q$, it drops out from the MFV parameters $a_q$ in
\eq{defatri}.

Finally, the quadratic squark soft-breaking terms are defined as follows:
\begin{equation}
\begin{split}
(\hat{M}^2_{\widetilde u_L})_{ij}
=&\ (V_{CKM}^0\, \tilde{m}^2_Q V_{CKM}^{0\dagger})_{ij}
+ \frac{v^2 \sin^2\beta}{2}\delta_{ij} |y_{u_i}|^2 
+ \delta_{ij} M^2_Z \cos 2\beta (1/2-2\sin^2\theta_W/3),\\
(\hat{M}^2_{\widetilde d_L})_{ij}
=&\ (\tilde{m}^2_Q)_{ij}
+ \frac{v^2 \cos^2\beta}{2}\delta_{ij} |y_{d_i}|^2  
+ \delta_{ij} M^2_Z \cos 2\beta (-1/2+\sin^2\theta_W/3),\\
(\hat{M}^2_{\widetilde u_R})_{ij}
=&\ (\tilde{m}^2_u)_{ij}
+ \frac{v^2 \sin^2\beta}{2}\delta_{ij} |y_{u_i}|^2  
+ 2\delta_{ij} M^2_Z \cos 2\beta \sin^2\theta_W/3, \\
(\hat{M}^2_{\widetilde d_R})_{ij}
=&\ (\tilde{m}^2_d)_{ij}
+ \frac{v^2 \cos^2\beta}{2}\delta_{ij} |y_{d_i}|^2 
- \delta_{ij} M^2_Z \cos 2\beta \sin^2\theta_W/3 \, ,
\end{split}
\end{equation}
where $V_{CKM}^0$ corresponds to the relative rotation of left-handed $u$-type
and $d$-type quark fields performed when diagonalizing the Yukawa matrices.  It
differs from the actual CKM matrix, defined by the rotations that diagonalize
the 2HDM mass matrices rather than the MSSM Yukawa couplings, by
loop-suppressed (but $\tan\beta$-enhanced) corrections. In particular, within
MFV, we have:
\begin{equation}    \label{eq:ckmcorr}
V_{CKM_{ij}}=
\begin{cases}
\Big| \frac{\ds 1+\epsilon_0\tan\beta}{
                 \ds 1+\widetilde\epsilon_3\tan\beta}\ \Big|
V^0_{CKM_{ij}}  & {\rm for}\ (i,j)= (u,b),(c,b), (t,d), (t,s),\\[3mm]
V^0_{CKM_{ij}} & {\rm otherwise}.
\end{cases}
\end{equation}
The relations \eq{eq:ckmcorr} take a particularly compact form in the
(exact) Wolfenstein parametrization, where one has
\begin{equation}
A = \Big| \frac{\ds 1+\epsilon_0 \tan\beta}
                    {\ds 1+\widetilde\epsilon_3 \tan\beta}\ 
    \Big| A^0,
\quad \lambda = \lambda^0,
\quad \bar \rho = \bar \rho^0,
\quad \bar \eta = \bar \eta^0 .
\end{equation}

Whenever we consider
the case of MFV we write
\begin{equation}
\hat{T}_{u_{ij}} = a_{t} y_{u_i} \delta_{ij},\quad 
\hat{T}_{d_{ij}} = a_{b} y_{d_i} \delta_{ij},\quad
\tilde{m}^2_{Q_{ij}} = \tilde{m}^2_{Q} \delta_{ij},\quad 
\tilde{m}^2_{u_{ij}} = \tilde{m}^2_{u} \delta_{ij},\quad
\tilde{m}^2_{d_{ij}} = \tilde{m}^2_{d} \delta_{ij}.\label{defatri}
\end{equation}
The SU(2) relation between $\hat{M}^2_{\widetilde d_L}$ and $\hat{M}^2_{\widetilde u_L}$ then implies for the third generation:
\begin{equation}
M^2_{\widetilde t_L} \;=\; M^2_{\widetilde b_L} \, +\,  
  m_t^2 \,-\,  \frac{m_b^2}{|1+\widetilde \epsilon_3\tan\beta|^2}
                     \, +\, M^2_Z \cos 2\beta (1-\sin^2\theta_W).
\label{massrel}
\end{equation}
In the strict SU(2) limit (i.e., $v/M_{\rm SUSY}\to 0$) one has $M^2_{\widetilde t_L} \;=\;
M^2_{\widetilde b_L}$, but for small $M^2_{\widetilde b_L}$ the term
involving $m_t^2$ can be relevant. Also FCNC $\widetilde
W$-$\widetilde u_{Li}$ loops vanish (for 
universal $M^2_{\widetilde u_L}$) by the GIM mechanism up to the
$m_t^2$ term in \eq{massrel}.

Finally, it is convenient to define the so-called superflavour basis,
obtained from a generic electroweak interaction eigenstate basis
by rotating the supermultiplets $Q_L$, $u_R$ and $d_R$ such that
the quadratic squark soft-breaking terms are diagonal.
We denote the corresponding entries by $\tilde{m}^2_{Q_{i}}$,
$\tilde{m}^2_{u_{i}}$, and $\tilde{m}^2_{d_{i}}$.
For $M_{\rm SUSY}\gg v$, these are just the squark masses,
and the computation of the effective couplings $\lambda_i$
induced by heavy squark loops for arbitrary flavour and CP structure is greatly simplified.
The Yukawa matrices and trilinear terms in this basis are simply written $Y_{u,d}$ and $T_{u,d}$, respectively.
They are given in terms of $\hat{Y}_{u,d}$ and $\hat{T}_{u,d}$ as follows:
\begin{equation}
Y_{u}^T = U_u \hat{Y}_{u} V_{CKM}^0 V_d^\dagger,\quad 
Y_{d}^T = U_d \hat{Y}_{d} V_d^\dagger,\quad
T_{u}^T = U_u \hat{T}_{u} V_{CKM}^0 V_d^\dagger,\quad 
T_{d}^T = U_d \hat{T}_{d} V_d^\dagger,
\end{equation}
where the matrices $V_d$, $U_u$ and $U_d$ are defined such that
\begin{equation}
\textrm{diag}(\tilde{m}^2_{Q_i}) = V_d \tilde{m}^2_{Q} V_d^\dagger,\quad 
\textrm{diag}(\tilde{m}^2_{u_i}) = U_u \tilde{m}^2_{u} U_u^\dagger,\quad
\textrm{diag}(\tilde{m}^2_{d_i}) = U_d \tilde{m}^2_{d} U_d^\dagger.
\end{equation}
Assuming MFV, one is allowed to choose $V_d=U_u=U_d=1$.

Our conventions comply with the Les Houches accord \cite{Skands:2003cj}.
In particular, our $Y_{u,d}$ and $T_{u,d}$ matrices correspond to a
particular choice of the generic $Y_{u,d}$ and $T_{u,d}$ matrices of
Ref.\cite{Skands:2003cj}.  Our conventions also agree with those of
Ref.~\cite{rosiek}, except that the sign convention of our 
$\hat{Y}_d$ in \eq{defy} is
opposite.  Besides, our $\hat{T}_u$ equals $-A_u$ and our $\hat{T}_d$
equals $A_d$ of Ref.~\cite{rosiek}, respectively.

\section{Matching of the MSSM on a 2HDM}
\label{sec:matching_results}

The notation of Sect.~\ref{sect:higgs} distinguishes between the
coefficients $\hat \lambda_i$ and $\bar \lambda_i$. The former
quantities contain the results from the supersymmetric loop
corrections to the quadrilinear Higgs couplings, whose tree-level
values are given in \eq{2.4}. The latter coefficients also include the
effect of the wave function and gauge coupling renormalization
constants in Sect.~\ref{sec:renorm-const}. 

In the following we summarize the one-loop matching
corrections for the quartic Higgs-coupling constants in the general
MSSM. While the calculation of loop corrections to the Higgs sector of the MSSM
has a long history, see for example
\cite{Okada:1990vk,Haber:1990aw,Ellis:1990nz,Brignole:1992uf,Dabelstein:1994hb,Chankowski:1992er,cepw},
and a determined reader could extract part of the
matching coefficients below from these works, the results collected in this
appendix as a service to the reader are more complete than
those in the literature, capturing the effects of the full set of
mass, flavour-violation, and CP-violation parameters of the
most general MSSM.

The general results are quite lengthy hence we start with $\bar
\lambda_{2,5,7}$, where all renormalization constants are included,
in the approximation of third generation dominance and degenerate
soft-breaking parameters
$\tilde m^2_{u,d,Q,e,l}=\tilde m^2$
and $M_{1,2} = M_{1/2}$. The quartic coupling constant
$\lambda_2$ is already present at tree level and as such depends on
the renormalization scale $\mu_0$ at one-loop. It reads:
\begin{equation}
\label{eq:lambda2simp}
\begin{split}
  \bar \lambda_2 =& \frac{\tilde g^2}{4} + \frac{1}{16 \pi^2} \Bigg\{
  - \frac{
    \frac{1}{2} \left|a_t y_t\right|{}^4
    + \frac{1}{2} |\mu |^4 \left|y_b\right|{}^4
    + \frac{1}{6} |\mu |^4 \left|y_{\tau }\right|{}^4 }
  {\tilde m^4}
  \\ &
  \qquad + \frac{ 
    \left( 6 \left|y_t\right|{}^2 -\frac{1}{2} \tilde g^2 \right) 
    \left|a_t y_t\right|{}^2
    + \tilde g^2 |\mu |^2 \left|y_b\right|{}^2
    + \frac{1}{3} \tilde g^2 |\mu |^2 \left|y_{\tau }\right|{}^2 }
  {\tilde m^2} 
  \\ &
   \qquad +\left(
    g^4-\frac{3}{2} \left|y_t\right|{}^2 g^2
    +\frac{5 g^{\prime ^4}}{3}
    +6 \left|y_t\right|{}^4
    -\frac{3}{2} g^{\prime ^2} \left|y_t\right|{}^2
  \right) \log \left(\frac{\tilde m^2}{\mu _0^2} \right) 
  \\ &
  \qquad
  + g^4 \left(\frac{\log \left(M_{\mu }\right) \left(13-3 M_{\mu }\right)
      M_{\mu }^2}{4 \left(M_{\mu }-1\right){}^3}-\frac{11}{12} \log
    \left(\frac{|\mu |^2}{\mu _0^2}\right)-\frac{M_{\mu } \left(5
        M_{\mu }+14\right)+1}{8 \left(M_{\mu }-1\right){}^2}\right)
  \\ &
  \qquad
  + g^2 \, g^{\prime ^2} \left(-\frac{2 \log \left(M_{\mu }\right)
      \left(M_{\mu }-4\right) M_{\mu }^2}{\left(M_{\mu
        }-1\right){}^3}-\frac{\left(M_{\mu }+5\right) M_{\mu
      }}{\left(M_{\mu }-1\right){}^2}-2 \log \left(\frac{|\mu |^2}{\mu
        _0^2}\right)\right) 
  \\ &
  \qquad
  + g^{\prime ^4} \left(-\frac{15}{4} \log
    \left(\frac{|\mu |^2}{\mu _0^2}\right)-\frac{M_{\mu } \left(17
        M_{\mu }+158\right)+5}{24 \left(M_{\mu
        }-1\right){}^2}+
  \right. \\ &
  \qquad \qquad \left.
    \frac{\log \left(M_{\mu }\right) \left(M_{\mu }
        \left(\left(141-43 M_{\mu }\right) M_{\mu
          }-12\right)+4\right)}{12 \left(M_{\mu }-1\right){}^3}\right)
  \Bigg\} \, ,
\end{split}
\end{equation}
where we have defined the mass ratio $M_{\mu}=|M|^2_{1/2}/|\mu|^2$.
Its renormalization-scale dependence is cancelled by the
inclusion of electroweak corrections in the effective 2HDM. The other
coupling constants important in the large $\tan \beta$ limit
are
\begin{equation}
\label{eq:lambda5simp}
\begin{split}
  \bar \lambda_5 = \frac{1}{16 \pi^2} &\Bigg\{
  -\frac{\mu ^2 \left(3 a_b^2 \left|y_b\right|{}^4+3
      \left|y_t\right|{}^4 a_t^2+\left|y_{\tau }\right|{}^4
      a_{\tau }^2\right)}{6 \tilde{m}^4}
  \\ &
  \qquad
  +    \frac{\left(g^4+2 g^{\prime ^2} g^2+3 g^{\prime ^4}\right) \left(\log
 \left(M_{\mu }\right)+\left(\log \left(M_{\mu }\right)-2\right)
 M_{\mu }+2\right)}{\left(M_{\mu }-1\right){}^3}
  \Bigg\}
\end{split}
\end{equation}
and
\begin{equation}
\label{eq:lambda7simp}
\begin{split}
  \bar \lambda_7 = 
  \frac{1}{16 \pi^2} & \Bigg\{
  \frac{1}{\tilde{m}^2}
  \left( -3 \mu  a_t \left|y_t\right|^4-
    \frac{1}{12} \mu \tilde{g}^2 
    \left(6 a_b \left|y_b\right|^2-3 \left|y_t\right|^2 a_t +
      2 \left|y_{\tau }\right|^2a_{\tau }\right) \right) 
  \\ & + \frac{\mu}{6 \tilde{m}^4}  \left(3
    \left|a_t\right|^2 a_t \left|y_t\right|^4+|\mu |^2
    \left(3 a_b \left|y_b\right|^4+\left|y_{\tau
        }\right|^4 a_{\tau }\right)\right)
  \\ & +
  \left(\left(g^4+4 g^{\prime ^2} g^2+3 g^{\prime ^4}\right) \frac{\mu}{|\mu|}
      +8 \left(g^4+2 g^{\prime ^2} g^2+3 g^{\prime ^4}\right) \frac{M_{1/2}}{\mu ^*}\right) \times
  \\ & \qquad 
  \frac{
    \left(-M_{\mu }^2+2 \log \left(M_{\mu }\right) M_{\mu
      }+1\right)}{8 \left(M_{\mu }-1\right){}^3}
  \Bigg\} \, .
\end{split}
\end{equation}

In the following subsections we quote the results for $\hat \lambda_i$
in the general MSSM. The effective potential $V$ in
\eq{2.1} must be used with $\lambda_i=\bar\lambda_i$ and the relation
between $\bar\lambda_i$ and $\hat\lambda_i$ is given in
\eq{eq:2-1}; the renormalization constants needed in this relation are given in Sect.~\ref{sec:renorm-const}. We decompose $\hat \lambda_{1-7}$ as
\begin{equation}
\label{eq:decla}
\begin{split}
\hat \lambda_i = & \lambda_i^{\rm tree} \, +\,  
        \frac{\lambda_i^{\rm ino} + \lambda_i^{\rm sferm}}{16 \pi^2}. 
\end{split}
\end{equation}
The tree-level values $\lambda_i^{\rm tree}$ are given in \eq{2.4}.
$\lambda_i^{\rm ino}$ and $\lambda_i^{\rm sferm}$, given in
Sect.~\ref{sec:higgs-gaug-contr} and \ref{sec:sferm-contr-lambd},
contain the contributions from higgsino and gaugino loops and from
sfermion loops, respectively. Finally we also list the relevant loop
functions in Sect.~\ref{sec:loop-functions}. All these results are
given in the superflavour basis including the most general
soft-breaking terms.

\subsection{Renormalization constants}
\label{sec:renorm-const}

The renormalization of $g^{(\prime)}$ is related only to the field
renormalization of $W$ and $B$, $Z_{W,B} = 1 + \delta Z_{W,B}$, 
if we decouple the sfermionic, higgsino and gaugino contributions:
$\delta g' = - \delta Z_B/2$ and $\delta g = - \delta Z_W/2$.
The finite part of the
one-loop wavefunction renormalization constants of the gauge bosons
are
\begin{equation}
\label{eq:10}
\begin{split}
  \delta Z_W &= 
  \frac{g^2}{16 \pi^2} \frac{1}{6} 
  \left[
    4 \log \frac{|\mu |^2}{\mu_0^2} +
    8 \log \frac{M_2^2}{\mu_0^2} +
    \sum_{i=1}^3 \left( \log \frac{\tilde{m}_{l_i}^2}{\mu_0^2 } +
      N_C \log \frac{\tilde{m}_{Q_i}^2}{\mu_0^2 } \right) - 4
  \right]
  \\
  \delta Z_B &= 
  \frac{g^{\prime ^2}}{16 \pi^2} \frac{1}{3}
  \left[
    2 \log \frac{|\mu |^2}{\mu_0^2 } +
    \sum_{i=1}^3 \left( 
      \log \frac{\tilde{m}_{e_i}^2}{\mu_0^2 } +
      \frac{1}{2} \log \frac{\tilde{m}_{l_i}^2}{\mu_0^2 } + 
    \right. \right. \\ 
  & \qquad \qquad \qquad \qquad \qquad \left. \left. 
      \frac{4 N_C}{9} \log \frac{\tilde{m}_{u_i}^2}{\mu_0^2 } +
      \frac{N_C}{9} \log \frac{\tilde{m}_{d_i}^2}{\mu_0^2 } +
      \frac{N_C}{18} \log \frac{\tilde{m}_{Q_i}^2}{\mu_0^2 } \right)
  \right] \, ,
\end{split}
\end{equation}
where $\mu_0$ is the renormalization scale
and the soft-breaking terms are written in the superflavour basis (see Appendix A).

The sfermionic contributions to the wavefunction renormalization constants of the
Higgs bosons are
\begin{equation}
\label{eq:19}
\begin{split}
\delta Z_{dd} &= {  \frac{1}{32 \pi ^2}} \sum_{ij} \left[ 3
    B_0'\left(\tilde{m}_{d_i},\tilde{m}_{ Q_j}\right) {
      T_{d_{ji}}} { T_{d_{ji}}^*} + 3 |\mu|^2
    B_0'\left(\tilde{m}_{u_i},\tilde{m}_{ Q_j}\right)
    Y_{u_{ji}} Y_{u_{ji}}^* \right. \\
  & \qquad \qquad \qquad \qquad \qquad \qquad + \left.
    B_0'\left(\tilde{m}_{e_i},\tilde{m}_{l_j}\right) { T_{e_{ji}}}
    { T_{e_{ji}}^*} \right] \\
 \delta Z_{ud} &= {  -\frac{1}{16 \pi ^2}} 
   \sum_{ij} \left[ 3 \mu^{ *}
    B_0'\left(\tilde{m}_{d_i},\tilde{m}_{ Q_j}\right) {
      T_{d_{ji}}^*} Y_{d_{ji}} + 3 \mu^{ *}
    B_0'\left(\tilde{m}_{u_i},\tilde{m}_{ Q_j}\right) {
      T_{u_{ji}}^*}
    Y_{u_{ji}} \right. \\
  & \qquad \qquad \qquad \qquad \qquad \qquad + \left.  \mu^{ *}
    B_0'\left(\tilde{m}_{e_i},\tilde{m}_{l_j}\right) { T_{e_{ji}}^*}
    Y_{e_{ji}} \right] \\
 \delta Z_{uu} &= {  \frac{1}{32 \pi ^2}} 
    \sum_{ij} \left[ 3 B_0'
    \left(\tilde{m}_{u_i},\tilde{m}_{ Q_j}\right) {
      T_{u_{ji}}} { T_{u_{ji}}^*} + 3 |\mu|^2
    B_0'\left(\tilde{m}_{d_i},\tilde{m}_{ Q_j}\right)Y_{d_{ji}}
    Y_{d_{ji}}^* \right. \\
  & \qquad \qquad \qquad \qquad \qquad \qquad + \left.  |\mu|^2
    B_0'\left(\tilde{m}_{e_i},\tilde{m}_{l_j}\right) Y_{e_{ji}}^* Y_{e_{ji}}
  \right] \, ,
\end{split}
\end{equation}
while the respective contributions of the gaugino and higgsino loops
read:
\begin{equation}
\label{eq:11}
\begin{split}
\delta Z_{dd} &= - \frac{1}{16 \pi ^2} \frac{1}{8} \left(
    g^{\prime ^2} W(|M_1|, |\mu|) +
    3 g^2 W(|M_2|, |\mu|) 
  \right) \\
\delta Z_{ud} &= - \frac{1}{16 \pi ^2} \mu^* \left(
    g^{\prime ^2} M_1^* B_0'(|M_1|, |\mu|) +
    3 g^2 M_2^* B_0'(|M_2|, |\mu|) 
  \right) \\
\delta Z_{uu} &= - \frac{1}{16 \pi ^2} \frac{1}{8} \left(
    g^{\prime ^2} W(|M_1|, |\mu|) +
    3 g^2 W(|M_2|, |\mu|)
  \right) \, .\\
\end{split}
\end{equation}

\subsection{Higgsino-gaugino contributions  to $\lambda_1$--$\lambda_7$}
\label{sec:higgs-gaug-contr}

The situation of $\bar \lambda_5=\hat \lambda_5$ is particularly
simple:  The matching correction only involves 
the box function and 
$\lambda_5^{\rm ino}$ can be written in a compact form:
\begin{equation}
\label{eq:12}
\begin{split}
\lambda_5^{\rm ino} =  &
3 g^4 \mu^2 M_2^2 D_0 \left( M_2, M_2, |\mu|, |\mu| \right)
+ \\ &
2 g^2 g^{\prime ^2} \mu^2 M_1 M_2 
  D_0 \left( M_1, M_2, |\mu|, |\mu| \right) + \\ &
g^{\prime ^4} \mu^2 M_1^2
  D_0 \left( M_1, M_1, |\mu|, |\mu| \right) \, ,
\end{split}
\end{equation}
if we use the loop functions defined in Sect.~\ref{sec:loop-functions}.

We find for $\lambda^{\rm ino}=\lambda_1^{\rm ino},\ldots \lambda_4^{\rm ino},
\lambda_6^{\rm ino}, \lambda_7^{\rm ino}$:
\begin{equation}
\label{eq:9}
\begin{split}
\lambda^{\rm ino} = &
g^4 \left(
  a_s +
  a_2 \tilde D_2 \left( M_2, M_2, |\mu|, |\mu| \right) +
  a_4 \tilde D_4 \left( M_2, M_2, |\mu|, |\mu| \right)
\right) + \\ &
g^2 g^{\prime ^2} \left(
  a'_s +
  a'_2 \tilde D_2 \left( M_1, M_2, |\mu|, |\mu| \right) +
  a'_4 \tilde D_4 \left( M_1, M_2, |\mu|, |\mu| \right)
\right) + \\ &
g^{\prime ^4} \left(
  a''_s +
  a''_2 \tilde D_2 \left( M_1, M_1, |\mu|, |\mu| \right) +
  a''_4 \tilde D_4 \left( M_1, M_1, |\mu|, |\mu| \right)
\right) \, ,
\end{split}
\end{equation}
where the coefficients $a_s, \ldots a''_4$ depend on the 
index $i$ labeling $\lambda_i$ in \eq{eq:decla}, which we suppress 
throughout this appendix.
The coefficients $a_{2,4}^{(\prime (\prime))}$ are given in Table
\ref{tab:higgsgauge}, while 
\begin{equation}
\label{eq:29}
a_s = \left\{
  \begin{array}{c}
    - \frac{3}{4} \\ 0 \\ 0
  \end{array}
  \right. ,
  \quad a_s' = \left\{
    \begin{array}{c}
    - \frac{1}{2} \\ 1 \\ 0
  \end{array}
  \right. ,
  \quad a_s'' = \left\{
  \begin{array}{c}
    - \frac{1}{4} \\
    0 \\ 
    0 
  \end{array}
  \right.
  \quad \textrm{for} \quad
  \left\{
  \begin{array}{l}
    \lambda_1 \; \textrm{to} \; \lambda_3 \\ 
    \lambda_4 \\ 
    \lambda_5 \; \textrm{to} \; \lambda_7
  \end{array}
  \right.
\end{equation}
\begin{table}[h]
\centering
\renewcommand{\arraystretch}{1.2}
\begin{tabular}{|c|c|c|c|c|c|c|}
\hline
$\lambda$ & $a_2$ & $a_4$ & $a'_2$ & $a'_4$ & $a''_2$ & $a''_4$ \\ \hline 
$\lambda_1$ & 
$\frac{1}{2} |M_2|^2$ & $\frac{5}{2}$ & 
$\frac{1}{2} (M_1 M_2^* + M_1^* M_2)$ & $1$ & 
$\frac{1}{2} |M_1|^2$ & $\frac{1}{2}$ \\
$\lambda_2$ & 
$\frac{1}{2} |M_2|^2$ & $\frac{5}{2}$ & 
 $\frac{1}{2} (M_1 M_2^* + M_1^* M_2)$ & $1$ & 
$\frac{1}{2} |M_1|^2$ & $\frac{1}{2}$ \\
$\lambda_3$ & 
$3 |\mu |^2+\frac{5}{2} |M_2|^2$ & $\frac{1}{2}$ & 
$2 |\mu |^2+\frac{1}{2} (M_1 M_2^* + M_1^* M_2)$ & $1$ & 
$|\mu |^2+\frac{1}{2} |M_1|^2$ & $\frac{1}{2}$ \\
$\lambda_4$ & 
$-3 |\mu |^2-2 |M_2|^2$ & $2$ & 
$2 |\mu |^2- M_1 M_2^* - M_1^* M_2$ & $-2$ & 
$-|\mu |^2$ & $0$ \\
$\lambda_6$ & 
$3 \mu  M_2$ & $0$ &  
$\mu  \left(M_1+M_2\right)$ & $0$ &
$\mu  M_1$ & $0$ \\
$\lambda_7$ & 
$3 \mu  M_2$ & $0$ & 
$\mu  \left(M_1+M_2\right)$ & $0$ &
$\mu  M_1$ & $0$ 
\\ \hline
\end{tabular}
\caption{Coefficients entering 
      $\lambda_1^{\rm ino}$--$\lambda_4^{\rm ino}$, 
      $\lambda_6^{\rm ino}$, and $\lambda_7^{\rm ino}$ in 
      \eq{eq:9}.}
\label{tab:higgsgauge}
\end{table}

\subsection{Sfermion contributions to $\lambda_1$--$\lambda_7$}
\label{sec:sferm-contr-lambd}

The sfermion contribution to $\lambda_{1-7}$ are
products of loop functions and flavour dependent coefficients if we
sum over the generation index of the internal sfermions.
For $\hat \lambda_5$ our results then take the simple form
\begin{equation}
\label{eq:20}
\begin{split}
\lambda_5^{\rm sferm} =& 
 d_1^{ijkl}
 D_0\left(\tilde{m}_{e_i},\tilde{m}_{e_j},\tilde{m}_{l_k},
          \tilde{m}_{l_l}\right)
 + d_2^{ijkl}
 D_0\left(\tilde{m}_{d_i},\tilde{m}_{d_j},
          \tilde{m}_{Q_k},\tilde{m}_{Q_l}\right)
 + \\ & d_3^{ijkl}
 D_0\left(\tilde{m}_{Q_i},\tilde{m}_{Q_j},\tilde{m}_{u_k},
          \tilde{m}_{u_l}\right),
\end{split}
\end{equation}
where the slepton contribution is contained in $d_1^{ijkl}$ listed in Table
\ref{tab:d0lambdaslepton}
\begin{table}[h]
\centering
\begin{tabular}{|c|c|}
  \hline & $d_1^{ijkl}$ \\ \hline
 $\lambda_1$ & $-T_{e_{ki}} T_{e_{lj}} T^*_{e_{kj}}
 T^*_{e_{li}}$  \\
 $\lambda_2$ & $ -|\mu |^4 Y_{e_{kj}} Y_{e_{li}}
 Y^*_{e_{ki}} Y^*_{e_{lj}}$  \\
 $\lambda_3$ & $ -|\mu |^2 T_{e_{li}} Y_{e_{kj}}
 \left(T^*_{e_{lj}}
 Y^*_{e_{ki}}+T^*_{e_{ki}}
 Y^*_{e_{lj}}\right)$ \\
$\lambda_4$ & $ |\mu |^2 T_{e_{li}} T^*_{e_{ki}}
 Y_{e_{kj}} Y^*_{e_{lj}}$ \\
$\lambda_5$ & $ -\mu ^2 T_{e_{kj}} T_{e_{li}}
 Y^*_{e_{ki}} Y^*_{e_{lj}}$ \\
$\lambda_6$ & $ \mu  T_{e_{ki}} T_{e_{lj}} T^*_{e_{li}}
 Y^*_{e_{kj}}$ \\
$\lambda_7$ & $ \mu  |\mu |^2 T_{e_{lj}} Y_{e_{ki}}
 Y^*_{e_{kj}} Y^*_{e_{li}}$ \\ \hline
\end{tabular}
\caption{Slepton contribution to 
         $\lambda_1^{\rm sferm}$--$\lambda_7^{\rm sferm}$ 
         in Eqs.~(\ref{eq:20}), (\ref{eq:lambda14sl}), and (\ref{eq:19-2}).}
\label{tab:d0lambdaslepton}
\end{table}
and $d_{2-4}^{ijkl}$ comprises the squark contribution
(Table \ref{tab:d0lambdasquark}).
\begin{table}[t]
\centering
\begin{tabular}{|c|c|c|}
  \hline & $d_2^{ijkl}$ & $d_3^{ijkl}$ \\ \hline
 $\lambda_1$ & $-3 T_{d_{ki}}
 T_{d_{lj}} T^*_{d_{kj}}
 T^*_{d_{li}}$ & $-3 |\mu |^4 Y_{u_{il}}
 Y_{u_{jk}} Y^*_{u_{ik}}
 Y^*_{u_{jl}}$ \\
 $\lambda_2$ & $-3 |\mu
 |^4 Y_{d_{kj}} Y_{d_{li}}
 Y^*_{d_{ki}} Y^*_{d_{lj}}$ & $-3
 T_{u_{ik}} T_{u_{jl}} T^*_{u_{il}}
 T^*_{u_{jk}}$ \\
 $\lambda_3$ & $-3 |\mu |^2 T_{d_{li}}
 Y_{d_{kj}} \left(T^*_{d_{lj}}
 Y^*_{d_{ki}}+T^*_{d_{ki}}
 Y^*_{d_{lj}}\right)$ & $-3 |\mu |^2 T_{u_{jl}}
 Y_{u_{ik}} \left(T^*_{u_{jk}}
 Y^*_{u_{il}}+T^*_{u_{il}}
 Y^*_{u_{jk}}\right)$ \\
$\lambda_4$ & $3 |\mu |^2
 T_{d_{li}} T^*_{d_{ki}} Y_{d_{kj}}
 Y^*_{d_{lj}}$ & $3 |\mu |^2 T_{u_{jl}}
 T^*_{u_{il}} Y_{u_{ik}}
 Y^*_{u_{jk}}$ \\
$\lambda_5$ & $-3 \mu ^2
 T_{d_{kj}} T_{d_{li}} Y^*_{d_{ki}}
 Y^*_{d_{lj}}$ & $-3 \mu ^2 T_{u_{ik}}
 T_{u_{jl}} Y^*_{u_{il}}
 Y^*_{u_{jk}}$ \\
$\lambda_6$ & $3 \mu  T_{d_{ki}}
 T_{d_{lj}} T^*_{d_{li}}
 Y^*_{d_{kj}}$ & $3 \mu  |\mu |^2 T_{u_{il}}
 Y_{u_{jk}} Y^*_{u_{ik}}
 Y^*_{u_{jl}}$ \\
$\lambda_7$ & $3 \mu 
 |\mu |^2 T_{d_{lj}} Y_{d_{ki}}
 Y^*_{d_{kj}} Y^*_{d_{li}}$ & $3 \mu 
 T_{u_{il}} T_{u_{jk}} T^*_{u_{jl}}
 Y^*_{u_{ik}}$ \\ \hline
\end{tabular}
\caption{$D_0$ squark contribution to 
         $\lambda_1^{\rm sferm}$--$\lambda_7^{\rm sferm}$ in 
         Eqs.~(\ref{eq:20}), (\ref{eq:lambda14sq}), and (\ref{eq:19-2}).}
\label{tab:d0lambdasquark}
\end{table}
Only $\lambda_4$ receives a contribution from 
$d_4^{ijkl}$: 
\begin{equation}
d_4^{ijkl}= - 3
(T_{d_{ki}} T_{u_{kl}}^* - |\mu|^2 Y_{d_{ki}} Y_{u_{kl}}^*)
(T_{u_{jl}} T_{d_{ji}}^* - |\mu|^2 Y_{u_{jl}} Y_{d_{ji}}^*)
\qquad\qquad 
 \mbox{for~~}\lambda_4,
\end{equation} 
while $d_4^{ijkl}=0$ for $\lambda_i$ with $i\neq 4$.
The contributions to the matching coefficients depend on the Yukawa couplings
$Y_{e,u,d}$ of the charged leptons, the up-type quarks, and the down-type
quarks as well as on the trilinear soft breaking terms $T_{e,u,d}$, defined in
the superflavour basis (see Appendix~A).

We write
$\lambda_{1-4}^{\textrm{sferm}} = \lambda_{1-4}^{\textrm{sl}} + \lambda_{1-4}^{\textrm{sq}}$, and find for the slepton contribution
\begin{equation}
\label{eq:lambda14sl}
\begin{split}
  \lambda_{1-4}^{\textrm{sl}} = 
  & \left( b_1 \delta _{ij} + b_2 Y_{ee_{ii}} \delta _{ij} + 
    b_3 Y_{ee_{ij}} Y_{ee_{ji}} \right) 
  B_0\left(\tilde{m}_{e_i},\tilde{m}_{e_j}\right) + \\
  & \left( b_4 \delta _{ij} + b_5 \bar{Y}_{ee_{ii}} \delta _{ij} +
    b_6 \bar{Y}_{ee_{ij}} \bar{Y}_{ee_{ji}} \right)
  B_0\left(\tilde{m}_{l_i},\tilde{m}_{l_j}\right) + \\
  & \left( c_1 |\mu |^2 Y_{e_{ki}} Y^*_{e_{ki}} \delta _{ij} +
    c_2 T_{e_{ki}} T^*_{e_{ki}} \delta _{ij} +
    c_3 |\mu |^2 Y_{e_{ki}} Y^*_{e_{kj}} Y_{ee_{ij}} + 
\right. \\ & \qquad \qquad \left.
    c_4 T_{e_{ki}} T^*_{e_{kj}} Y_{ee_{ij}} \right) 
    C_0\left(\tilde{m}_{e_i},\tilde{m}_{e_j},\tilde{m}_{l_k}\right)
    + \\
  & \left( c_5 |\mu |^2 Y_{e_{ji}} Y^*_{e_{ji}} \delta _{jk} +
   c_6 T_{e_{ji}} T^*_{e_{ji}} \delta_{jk} +
   c_7 |\mu |^2 Y_{e_{ji}} Y^*_{e_{ki}} \bar{Y}_{ee_{kj}} +
\right. \\ & \qquad \qquad \left.
   c_8 T_{e_{ji}} T^*_{e_{ki}} \bar{Y}_{ee_{kj}} \right)
 C_0\left(\tilde{m}_{e_i},\tilde{m}_{l_j},\tilde{m}_{l_k}\right) \, ,
\end{split}
\end{equation}
while the squark contribution reads
\begin{equation}
\label{eq:lambda14sq}
\begin{split}
  \lambda_{1-4}^{\textrm{sq}} = 
 & \left( b_7 \delta _{ij} + b_8 Y_{dd_{ii}} \delta _{ij} +
   b_9 Y_{dd_{ij}} Y_{dd_{ji}} \right)
 B_0\left(\tilde{m}_{d_i},\tilde{m}_{d_j}\right) +
\\ & \qquad \qquad 
 b_{10} Y_{du_{ij}} Y_{ud_{ji}} 
 B_0\left(\tilde{m}_{d_i},\tilde{m}_{u_j}\right) + \\
 & \left( b_{11} \delta _{ij} + b_{12} Y_{uu_{ii}} \delta _{ij}
   + b_{13} Y_{uu_{ij}} Y_{uu_{ji}} \right)
 B_0\left(\tilde{m}_{u_i},\tilde{m}_{u_j}\right) + \\
 & \left( b_{14} \delta _{ij} + b_{15} \bar{Y}_{dd_{ii}} \delta _{ij} +
   b_{16} \bar{Y}_{uu_{ii}} \delta _{ij} + b_{17} \bar{Y}_{dd_{ij}}
   \bar{Y}_{dd_{ji}} + 
\right. \\ & \qquad \qquad \left.
   b_{18} \bar{Y}_{uu_{ij}} \bar{Y}_{uu_{ji}} +
   b_{19} \bar{Y}_{dd_{ij}} \bar{Y}_{uu_{ji}} \right)
 B_0\left(\tilde{m}_{Q_i},\tilde{m}_{Q_j}\right) + \\
 & \left( c_9 |\mu |^2 Y_{d_{ki}} Y^*_{d_{ki}} \delta _{ij} +
   c_{10} T_{d_{ki}} T^*_{d_{ki}} \delta_{ij} +
   c_{11} |\mu |^2 Y_{d_{ki}} Y^*_{d_{kj}} Y_{dd_{ij}} +
\right. \\ & \qquad \qquad \left.
   c_{12} T_{d_{ki}} T^*_{d_{kj}} Y_{dd_{ij}} \right)
 C_0\left(\tilde{m}_{d_i},\tilde{m}_{d_j},\tilde{m}_{Q_k}\right) + \\
 & \left( 
   c_{13} |\mu |^2 Y_{d_{ji}} Y^*_{d_{ji}} \delta_{jk} +
   c_{14} |\mu |^2 Y_{d_{ji}} Y^*_{d_{ki}} \bar{Y}_{dd_{kj}} +
   c_{15} T_{d_{ji}} T^*_{d_{ji}} \delta _{jk} +
\right. \\ & \qquad \qquad \left.
   c_{16} T_{d_{ji}} T^*_{d_{ki}} \bar{Y}_{dd_{kj}} +
   c_{17} T_{d_{ji}} T^*_{d_{ki}} \bar{Y}_{uu_{kj}} \right)
 C_0\left(\tilde{m}_{d_i},\tilde{m}_{Q_j},\tilde{m}_{Q_k}\right) + \\
 &\left( 
   - c_{18} Y_{d_{ji}} Y^*_{u_{jk}} Y_{du_{ik}} |\mu |^2
   - c_{18} Y_{u_{jk}} Y^*_{d_{ji}}  Y_{{ud}_{ki}} |\mu |^2+
\right. \\ & \qquad \qquad \left.
   + c_{18} T_{u_{jk}} T^*_{d_{ji}} Y_{{ud}_{ki}}
   + c_{18} T_{d_{ji}} T^*_{u_{jk}} Y_{{du}_{ik}}
 \right) 
 C_0\left(\tilde{m}_{d_i},\tilde{m}_{Q_j},\tilde{m}_{u_k}\right)  +\\
 & \left( c_{19} |\mu |^2 Y_{u_{ik}} Y^*_{u_{ik}} \delta_{ij} +
   c_{20} |\mu |^2 Y_{u_{jk}} Y^*_{u_{ik}} \bar{Y}_{uu_{ij}} +
   c_{21} T_{u_{ik}} T^*_{u_{ik}} \delta_{ij} + \right. \\ 
   & \qquad \qquad \left. c_{22} T_{u_{ik}} T^*_{u_{jk}} \bar{Y}_{uu_{ji}} +
   c_{23} T_{u_{ik}} T^*_{u_{jk}} \bar{Y}_{dd_{ji}} \right)
 C_0\left(\tilde{m}_{Q_i},\tilde{m}_{Q_j},\tilde{m}_{u_k}\right) + \\
 & \left( c_{24} |\mu |^2 Y_{u_{ij}} Y^*_{u_{ij}} \delta _{jk} +
   c_{25} |\mu |^2 Y_{u_{ik}} Y^*_{u_{ij}} Y_{uu_{kj}} +
   c_{26} T_{u_{ij}} T^*_{u_{ij}} \delta _{jk} +
\right. \\ & \qquad \qquad \left.
   c_{27} T_{u_{ij}} T^*_{u_{ik}} Y_{uu_{jk}} \right)
 C_0\left(\tilde{m}_{Q_i},\tilde{m}_{u_j},\tilde{m}_{u_k}\right) +
 \\ &
 d_1^{ijkl}
 D_0\left(\tilde{m}_{e_i},\tilde{m}_{e_j},\tilde{m}_{l_k},\tilde{m}_{l_l}\right)
 + d_2^{ijkl}
 D_0\left(\tilde{m}_{d_i},\tilde{m}_{d_j},\tilde{m}_{Q_k},\tilde{m}_{Q_l}\right)
 + 
\\ & \qquad \qquad 
d_3^{ijkl}
 D_0\left(\tilde{m}_{Q_i},\tilde{m}_{Q_j},\tilde{m}_{u_k},\tilde{m}_{u_l}\right)
  + d_4^{ijkl}
 D_0\left(\tilde{m}_{d_i},\tilde{m}_{Q_j},\tilde{m}_{Q_k},\tilde{m}_{u_l}\right)
\, .
\end{split}
\end{equation}
Here we introduced shorthand notations for the products of two Yukawa
coupling matrices:
\begin{equation}
\label{eq:18-2}
Y_{{xy}_{ij}} \equiv Y^\dagger_{x_{il}} Y_{y_{lj}} \, , \quad  
\bar{Y}_{xy_{ij}} \equiv Y_{x_{il}} Y^\dagger_{y_{lj}}  \, ,
\end{equation}
where $x,y=e,u,d$ and we sum over the internal index $l$.  The
coefficients $b_n$, $c_n$, and $d_n^{ijkl}$ for each 
$\lambda_1,\ldots\lambda_4$ are given
in Tables \ref{tab:d0lambdaslepton}, \ref{tab:d0lambdasquark}, 
\ref{tab:slepton1to4}, and \ref{tab:squark1to4}.

We finally give the slepton and squark contributions to $\lambda_{6,7}$:
\begin{equation}
\label{eq:19-2}
\begin{split}
  \lambda_{6,7}^{\rm sferm} =
  & \left( c'_1 \mu  T_{e_{ki}} Y^*_{e_{ki}} \delta _{ij} +
  c'_2 \mu  T_{e_{ki}} Y^*_{e_{kj}} Y_{ee_{ij}} \right)
  C_0\left(\tilde{m}_{e_i},\tilde{m}_{e_j},\tilde{m}_{l_k}\right) + \\
  & \left( c'_3 \mu  T_{e_{ji}} Y^*_{e_{ji}} \delta _{jk}  +
  c'_4 \mu  T_{e_{ji}} Y^*_{e_{ki}} \bar{Y}_{ee_{kj}} \right)
  C_0\left(\tilde{m}_{e_i},\tilde{m}_{l_j},\tilde{m}_{l_k}\right) + \\
  & \left( c'_5 \mu  T_{d_{ki}} Y^*_{d_{ki}} \delta _{ij} +
  c'_6 \mu  T_{d_{ki}} Y^*_{d_{kj}} Y_{dd_{ij}} \right)
  C_0\left(\tilde{m}_{d_i},\tilde{m}_{d_j},\tilde{m}_{Q_k}\right) + \\
  & \left( c'_7 \mu  T_{d_{ji}} Y^*_{d_{ji}} \delta _{jk} +
  c'_8 \mu  T_{d_{ji}} Y^*_{d_{ki}} \bar{Y}_{dd_{kj}} \right)
  C_0\left(\tilde{m}_{d_i},\tilde{m}_{Q_j},\tilde{m}_{Q_k}\right) + \\
  & \left( c'_9 \mu  T_{u_{ik}} Y^*_{u_{ik}} \delta _{ij} +
  c'_{10} \mu  T_{u_{ik}} Y^*_{u_{jk}} \bar{Y}_{uu_{ji}} \right)
  C_0\left(\tilde{m}_{Q_i},\tilde{m}_{Q_j},\tilde{m}_{u_k}\right) + \\
  & \left( c'_{11} \mu  T_{u_{ij}} Y^*_{u_{ij}} \delta _{jk} +
  c'_{12} \mu  T_{u_{ij}} Y^*_{u_{ik}} Y_{uu_{jk}} \right)
  C_0\left(\tilde{m}_{Q_i},\tilde{m}_{u_j},\tilde{m}_{u_k}\right) +
 \\ &
 d_1^{ijkl}
 D_0\left(\tilde{m}_{e_i},\tilde{m}_{e_j},\tilde{m}_{l_k},\tilde{m}_{l_l}\right)
 + d_2^{ijkl}
 D_0\left(\tilde{m}_{d_i},\tilde{m}_{d_j},\tilde{m}_{Q_k},\tilde{m}_{Q_l}\right)
 + \\ & d_3^{ijkl}
 D_0\left(\tilde{m}_{Q_i},\tilde{m}_{Q_j},\tilde{m}_{u_k},\tilde{m}_{u_l}\right)\,
\end{split}
\end{equation}
where the coefficients $c'_n$ and $d_n^{ijkl}$ are given in Tables
\ref{tab:d0lambdaslepton}, \ref{tab:d0lambdasquark}, and
\ref{tab:sfermion_5to6}.
\begin{table}[t]
\centering
\begin{tabular}{|c|c|c|}
  \hline & $\lambda_6$ & $\lambda_7$ \\ \hline
  $c'_1$ & $-\frac{g^{\prime ^2}}{2}$ & $\frac{g^{\prime ^2}}{2}$ \\
  $c'_2$ & $1$ & $0$ \\
  $c'_3$ & $\frac{1}{4} \left(g^{\prime ^2}-g^2\right)$ & $\frac{1}{4} \left(g^2-g^{\prime ^2}\right)$ \\
  $c'_4$ & $1$ & $0$ \\
  $c'_5$ & $-\frac{g^{\prime ^2}}{2}$ & $\frac{g^{\prime ^2}}{2}$ \\
  $c'_6$ & $3$ & $0$ \\
  $c'_7$ & $\frac{1}{4} \left(-g^{\prime ^2}-3 g^2\right)$ & $\frac{1}{4} \left(g^{\prime ^2}+3 g^2\right)$ \\
  $c'_8$ & $3$ & $0$ \\
  $c'_9$ & $\frac{1}{4} \left(3 g^2-g^{\prime ^2}\right)$ &
  $\frac{1}{4} \left(g^{\prime ^2}-3 g^2 \right)$ \\
  $c'_{10}$ & $0$ & $3$ \\
  $c'_{11}$ & $g^{\prime ^2}$ & $-g^{\prime ^2}$ \\
  $c'_{12}$ & $0$ & $3$ \\ \hline
\end{tabular}
\caption{Coefficients of $\lambda_6^{\rm sferm}$ and 
  $\lambda_7^{\rm sferm}$ in \eq{eq:19-2}.}
\label{tab:sfermion_5to6}
\end{table}

\begin{table}[t]
\centering
\begin{tabular}{|c|c|c|c|c|}
  \hline & $\lambda_1$ & $\lambda_2$ & $\lambda_3$ & $\lambda_4$ \\ \hline
  $b_1$ & $-\frac{g^{\prime ^4}}{4}$ & $-\frac{g^{\prime ^4}}{4}$ & $\frac{g^{\prime ^4}}{4}$ & $0$ \\
  $b_2$ & $g^{\prime ^2}$ & $0$ & $-\frac{g^{\prime ^2}}{2}$ & $0$ \\
  $b_3$ & $-1$ & $0$ & $0$ & $0$ \\
  $b_4$ & $\frac{1}{8} \left(-g^4-g^{\prime ^4}\right)$ & $\frac{1}{8} \left(-g^4-g^{\prime ^4}\right)$ &
  $\frac{1}{8} \left(g^4+g^{\prime ^4}\right)$ & $-\frac{g^4}{4}$ \\
  $b_5$ & $\frac{1}{2} \left(g^2-g^{\prime ^2}\right)$ & $0$ & $\frac{1}{4} \left(g^{\prime ^2}-g^2\right)$ &
  $\frac{g^2}{2}$ \\
  $b_6$ & $-1$ & $0$ & $0$ & $0$ \\
  $c_1$ & $0$ & $-g^{\prime ^2}$ & $\frac{g^{\prime ^2}}{2}$ & $0$ \\
  $c_2$ & $g^{\prime ^2}$ & $0$ & $-\frac{g^{\prime ^2}}{2}$ & $0$ \\
  $c_3$ & $0$ & $0$ & $-1$ & $0$ \\
  $c_4$ & $-2$ & $0$ & $0$ & $0$ \\
  $c_5$ & $0$ & $\frac{1}{2} \left(g^{\prime ^2}-g^2\right)$ & $\frac{1}{4} \left(g^2-g^{\prime ^2}\right)$ &
  $-\frac{g^2}{2}$ \\
  $c_6$ & $\frac{1}{2} \left(g^2-g^{\prime ^2}\right)$ & $0$ & $\frac{1}{4} \left(g^{\prime ^2}-g^2\right)$ &
  $\frac{g^2}{2}$ \\
  $c_7$ & $0$ & $0$ & $-1$ & $1$ \\
  $c_8$ & $-2$ & $0$ & $0$ & $0$ \\ \hline
\end{tabular}
\caption{Slepton loop contributions to 
  $\lambda_1^{\rm sferm}\ldots\lambda_4^{\rm sferm}$ in \eq{eq:lambda14sl}.}
\label{tab:slepton1to4}
\end{table}

\begin{table}[tp]
\centering
\begin{tabular}{|c|c|c|c|c|}
  \hline & $\lambda_1$ & $\lambda_2$ & $\lambda_3$ & $\lambda_4$ \\ \hline
  $b_7$ & $-\frac{g^{\prime ^4}}{12}$ & $-\frac{g^{\prime ^4}}{12}$ & 
  $\frac{g^{\prime ^4}}{12}$ & $0$ \\
  $b_8$ & $g^{\prime ^2}$ & $0$ & $-\frac{g^{\prime ^2}}{2}$ & $0$ \\
  $b_9$ & $-3$ & $0$ & $0$ & $0$ \\
  $b_{10}$ & $0$ & $0$ & $0$ & $-3$ \\
  $b_{11}$ & $-\frac{g^{\prime ^4}}{3}$ & $-\frac{g^{\prime ^4}}{3}$ & $\frac{g^{\prime ^4}}{3}$ & $0$ \\
  $b_{12}$ & $0$ & $2 g^{\prime ^2}$ & $-g^{\prime ^2}$ & $0$ \\
  $b_{13}$ & $0$ & $-3$ & $0$ & $0$ \\
  $b_{14}$ & $\frac{1}{24} \left(-9 g^4-g^{\prime ^4}\right)$ & $\frac{1}{24} \left(-9 g^4-g^{\prime ^4}\right)$ &
  $\frac{1}{24} \left(9 g^4+g^{\prime ^4}\right)$ & $-\frac{3}{4} g^4$ \\
  $b_{15}$ & $\frac{1}{2} \left(3 g^2+g^{\prime ^2}\right)$ & $0$ & $\frac{1}{4} \left(-3 g^2-g^{\prime ^2}\right)$ &
  $\frac{3}{2} g^2$ \\
  $b_{16}$ & $0$ & $\frac{1}{2} \left(3 g^2-g^{\prime ^2}\right)$ & $\frac{1}{4} \left(g^{\prime ^2}-3 g^2\right)$ &
  $\frac{3}{2} g^2$ \\
  $b_{17}$ & $-3$ & $0$ & $0$ & $0$ \\
  $b_{18}$ & $0$ & $-3$ & $0$ & $0$ \\
  $b_{19}$ & $0$ & $0$ & $0$ & $-3$ \\
  $c_9$ & $0$ & $-g^{\prime ^2}$ & $\frac{g^{\prime ^2}}{2}$ & $0$ \\
  $c_{10}$ & $g^{\prime ^2}$ & $0$ & $-\frac{g^{\prime ^2}}{2}$ & $0$ \\
  $c_{11}$ & $0$ & $0$ & $-3$ & $0$ \\
  $c_{12}$ & $-6$ & $0$ & $0$ & $0$ \\
  $c_{13}$  & $0$ & $- \frac{1}{2} \left(3 g^2 +g^{\prime ^2}\right)$ & 
  $\frac{1}{4} \left(3 g^2 +g^{\prime ^2}\right)$ & $-\frac{3}{2} g^2$ \\
  $c_{14}$ & $0$ & $0$ & $-3$ & $3$ \\
  $c_{15}$ & $\frac{1}{2} \left(3 g^2+g^{\prime ^2}\right)$ & $0$ & $\frac{1}{4} \left(-3 g^2-g^{\prime ^2}\right)$ &
  $\frac{3}{2} g^2$ \\
  $c_{16}$ & $-6$ & $0$ & $0$ & $0$ \\
  $c_{17}$ & $0$ & $0$ & $0$ & $-3$ \\
  $c_{18}$ & $0$ & $0$ & $0$ & $-3$ \\
  $c_{19}$ & $\frac{1}{2} \left(g^{\prime ^2}-3 g^2\right)$ & $0$ & $\frac{1}{4} \left(3 g^2-g^{\prime ^2}\right)$ &
  $-\frac{3 g^2}{2}$ \\
  $c_{20}$ & $0$ & $0$ & $-3$ & $3$ \\
  $c_{21}$ & $0$ & $\frac{1}{2} \left(3 g^2-g^{\prime ^2}\right)$ & $\frac{1}{4} \left(g^{\prime ^2}-3 g^2\right)$ &
  $\frac{3}{2} g^2$ \\
  $c_{22}$ & $0$ & $-6$ & $0$ & $0$ \\
  $c_{23}$ & $0$ & $0$ & $0$ & $-3$ \\
  $c_{24}$ & $-2 g^{\prime ^2}$ & $0$ & $g^{\prime ^2}$ & $0$ \\
  $c_{25}$ & $0$ & $0$ & $-3$ & $0$ \\
  $c_{26}$ & $0$ & $2 g^{\prime ^2}$ & $-g^{\prime ^2}$ & $0$ \\
  $c_{27}$ & $0$ & $-6$ & $0$ & $0$ \\
  \hline
\end{tabular}
\caption{Squark loop contributions to 
         $\lambda_1^{\rm sferm}\ldots\lambda_4^{\rm sferm}$ in \eq{eq:lambda14sq}.}
\label{tab:squark1to4}
\end{table}

\subsection{Loop Functions}
\label{sec:loop-functions}
In the UV-divergent loop functions we set 
$\epsilon = (4-D)/2$. The loop functions are defined as
\begin{equation}
\begin{split}
  \frac{i}{(4 \pi)^2} A_0 \left( m_1 \right) 
  \left( \frac{4 \pi}{\mu_0^2} e^{-\gamma_E} \right)^\epsilon 
  &=
  \int \frac{d^D q}{(2 \pi)^D}
  \frac{1}{q^2-m_1^2} \\
  \frac{i}{(4 \pi)^2} B_0 \left( m_1, m_2 \right)
  \left( \frac{4 \pi}{\mu_0^2} e^{-\gamma_E} \right)^\epsilon 
  &=
  \int \frac{d^D q}{(2 \pi)^D}
  \frac{1}{q^2-m_1^2} \frac{1}{q^2-m_2^2} \\
  \frac{i}{(4 \pi)^2} C_0 \left( m_1, m_2, m_3 \right)
  \left( \frac{4 \pi}{\mu_0^2} e^{-\gamma_E} \right)^\epsilon 
  &=
  \int \frac{d^D q}{(2 \pi)^D}
  \frac{1}{q^2-m_1^2} \frac{1}{q^2-m_2^2} \frac{1}{q^2-m_3^2} \\
  \frac{i}{(4 \pi)^2} D_0 \left( m_1, m_2, m_3, m_4 \right)
  \left( \frac{4 \pi}{\mu_0^2} e^{-\gamma_E} \right)^\epsilon 
  &=
  \int \frac{d^D q}{(2 \pi)^D}
  \frac{1}{q^2-m_1^2} \frac{1}{q^2-m_2^2} \frac{1}{q^2-m_3^2} 
  \frac{1}{q^2-m_4^2} \\
  \frac{i}{(4 \pi)^2} W \left( m_1, m_2 \right)
  \left( \frac{4 \pi}{\mu_0^2} e^{-\gamma_E} \right)^\epsilon 
  &= 
  \left. \frac{d}{d k^2} \right|_{k^2=0}
  \int \frac{d^D q}{(2 \pi)^D}
  \frac{\textrm{Tr} \left[ 
      (q \hskip-7pt \diagup - k \hskip-7pt \diagup) \, 
      q \hskip-7pt \diagup \right] }
  {\left( (q-k)^2-m_1^2 \right) \left(q^2-m_2^2\right) \, .
}
\end{split}
\label{eq:13}
\end{equation}
These functions read:
\begin{equation}
\label{eq:2}
\begin{split}
  A_0 \left( m_1 \right) &=
  \frac{m_1^2}{\epsilon} + m_1^2 + 
  m_1^2 \log \left(\frac{\mu_0^2}{m_1^2}\right) \\
  B_0 \left( m_1, m_2 \right) &=
  \frac{1}{\epsilon} + 1 + 
  \frac{m_1^2 \log \left(\frac{\mu_0^2}{m_1^2}\right)
    + m_2^2 \log \left(\frac{\mu_0^2}{m_2^2}\right)}{
    m_1^2-m_2^2} \\
  B_0' \left( m_1, m_2 \right) &=
  \frac{m_1^4 -m_2^4+
    2 m_1^2 \, m_2^2 \log \left(\frac{m_2^2}{m_1^2}\right)}{
    2 \left(m_1^2-m_2^2\right)^3} \\
  C_0 \left( m_1, m_2, m_3 \right) &= \frac{
    m_1^2 \, m_2^2 \log \left(\frac{m_2^2}{m_1^2}\right)+
    m_3^2 \, m_2^2 \log \left(\frac{m_3^2}{m_2^2}\right)+
    m_1^2 \, m_3^2 \log \left(\frac{m_1^2}{m_3^2}\right)}{
    \left(m_1^2-m_2^2\right)
    \left(m_1^2-m_3^2\right)
    \left(m_2^2-m_3^2\right)} \\
  D_0 \left( m_1, m_2, m_3, m_4 \right) &= \sum_{\{a,b,c,d\}}^
  {\begin{subarray}{c} \{m_1^2,m_2^2,m_3^2,m_4^2\} \\ + \textrm{cyclic
        permutations}
  \end{subarray}}
  \frac{
    a^2 b \, c \log \left(\frac{b}{c}\right) -
    a \, b^2 c \log \left(\frac{a}{c}\right) +
    b \, c \, d^2 \log \left(\frac{c}{b}\right)}{
    \left(a-b\right) \left(a-c\right) \left(a-d\right) 
    \left(b-c\right) \left(b-d\right) \left(c-d\right) } \\
  W \left( m_1, m_2 \right) &=
  -\frac{2}{\epsilon} -
  2 \log \left(\frac{\mu_0 ^2}{m_1^2}\right)  \\
  & \qquad
  - \log \left(\frac{m_2^2}{m_1^2}\right) 
  \frac{\left(2 m_2^6-6 m_1^2
      m_2^4\right)}{\left(m_1^2-m_2^2\right)^3} 
  -\frac{m_1^4-6 m_2^2 m_1^2+m_2^4}{\left(m_1^2-m_2^2\right)^2}
\end{split}
\end{equation}
\begin{equation}
\begin{split}
\tilde D_2 (m_1,m_2,m_3,m_4) =&
C_0 (m_2,m_3,m_4) + 
m_1^2 D_0(m_1,m_2,m_3,m_4) \\
\tilde D_4 (m_1,m_2,m_3,m_4) =&
B_0(m_3,m_4) + 
\left( m_1^2 +m_2^2 \right) C_0 (m_2,m_3,m_4) + \\
& \qquad m_1^4 D_0(m_1,m_2,m_3,m_4) 
\end{split}
\label{eq:14}
\end{equation}
A further loop function, $H_2$, is defined in \eq{3.20}.

\section{Renormalization group and  bag parameters}\label{sect:rghp}
The standard-model contribution to \bbm\ involves the operator
$Q_1^{\rm VLL}= (\bar b_L\gamma_{\mu}q_L)( \bar b_L\gamma^{\mu} q_L)$
of \eq{3.3}.  The main new supersymmetric contribution
to \bbm\ presented in this paper comes with the four-quark operator
$Q_1^{\rm SLL} = (\bar b_R q_L) (\bar b_R q_L)$ with $q=d$ or $q=s$, see
\eq{defqll}. $Q_1^{\rm SLL}$ mixes under renormalization with
\begin{equation}
\widetilde Q_1^{\rm SLL} = (\bar  b_R^i q_L^j)(\bar b_R^j q_L^i)
\label{defqllt}
\end{equation} 
where $i,j$ are colour indices.
The operators $Q_1^{\rm SLL}$ and $\widetilde Q_1^{\rm SLL}$ are widely studied
in the context of the width difference $\dg$ among the two mass
eigenstates in the \bbm\ system and the CP asymmetry $a_{\rm fs}$ in
flavour-specific decays \cite{bbgln1,bbd}.  

Yet the next-to-leading-order (NLO) anomalous dimensions have been calculated for
an equivalent operator basis in Ref.~\cite{bmu}. These operators,
\begin{equation}
\label{eq:30}
\begin{split}
  \Qbmu_1^\textrm{VLL} &= 
  (\bar b_L \gamma_\mu q_L)(\bar b_L \gamma^\mu q_L)\, , \\[6pt]
  \Qbmu_1^\textrm{LR} &= 
  (\bar b_L \gamma_\mu q_L) (\bar b_R \gamma^\mu q_R)\, , \\
  \Qbmu_2^\textrm{LR} &= 
  (\bar b_R q_L) (\bar b_L q_R)\, , \\[6pt]
  \Qbmu_1^\textrm{SLL} &= 
  (\bar b_R q_L) (\bar b_R q_L)\, , \\
  \Qbmu_2^\textrm{SLL} &= -
  (\bar b_R \sigma_{\mu \nu} q_L) (\bar b_R \sigma^{\mu \nu} q_L)\, , \\[6pt]
  \Qbmu_1^\textrm{VRR} &= 
  (\bar b_R \gamma_\mu q_R)(\bar b_R \gamma^\mu q_R)\, , \\[6pt]
  \Qbmu_1^\textrm{SRR} &= 
  (\bar b_L q_R) (\bar b_L q_R)\, , \\
  \Qbmu_2^\textrm{SRR} &= -
  (\bar b_L \sigma_{\mu \nu} q_R) (\bar b_L \sigma^{\mu \nu} q_R)\, ,
\end{split}
\end{equation}
are split into five sectors (VLL, LR, SLL, VRR, SRR) which separately
mix under renormalization -- note that we define
$\sigma_{\mu\nu}=\frac{i}{2}\left[ \gamma_{\mu},\gamma_{\nu}\right]$.
The anomalous dimensions of the VRR and SRR sectors are the same as those of
the VLL and SLL sectors, respectively. To define the renormalization
scheme for the NLO we first note that we use the $\ov{\rm MS}$ scheme
with anticommuting $\gamma_5$ as in \cite{bmu}. Then we must specify
the definition of the evanescent operators which enter the NLO results
as counterterms. In particular for the SLL sector the evanescent
operators of Ref.~\cite{bmu} read:
\begin{equation}
\label{eq:21}
\begin{split}
  \Ebmu_1^\textrm{SLL} =&
  \left( \bar b_R^i q_L^j \right)
  \left( \bar b_R^j q_L^i \right) + 
  \tfrac{1}{2} \Qbmu_1^{\rm SLL} - \tfrac{1}{8} \Qbmu_2^{\rm SLL}\, , \\
  \Ebmu_2^\textrm{SLL} =& -  
  \left( \bar b_R^i \sigma_{\mu \nu} q_L^j \right)
  \left( \bar b_R^j \sigma^{\mu \nu} q_L^i \right) -
  6 \Qbmu_1^{\rm SLL} - \tfrac{1}{2} \Qbmu_2^{\rm SLL}\, , \\
  \Ebmu_3^\textrm{SLL} =& 
  \left( \bar b_R^i \gamma_\mu \gamma_\nu 
    \gamma_\rho \gamma_\sigma q_L^i \right)
  \left( \bar b_R^j \gamma^\mu \gamma^\nu 
    \gamma^\rho \gamma^\sigma q_L^j \right) +
  (-64 + 96 \epsilon) \Qbmu_1^\textrm{SLL} +
  (-16 + 8 \epsilon) \Qbmu_2^\textrm{SLL}\, , \\
  \Ebmu_4^\textrm{SLL} =&   
  \left( \bar b_R^i \gamma_\mu \gamma_\nu 
    \gamma_\rho \gamma_\sigma q_L^j \right)
  \left( \bar b_R^j \gamma^\mu \gamma^\nu 
    \gamma^\rho \gamma^\sigma q_L^i \right) -
  64 \Qbmu_1^\textrm{SLL} +
  (-16 + 16 \epsilon) \Qbmu_2^\textrm{SLL} \, ,
\end{split}
\end{equation}
where we use $\epsilon\equiv (4-D)/2$.  

The operator basis
\begin{equation}
\label{eq:33}
\begin{split}
  Q_1^\textrm{VLL} &= \Qbmu_1^\textrm{VLL} \, , \\
  Q_1^\textrm{LR} &= \Qbmu_1^\textrm{LR} \, , \\
  Q_2^\textrm{LR} &= \Qbmu_2^\textrm{LR} \, , \\
  Q_1^\textrm{SLL} &= \Qbmu_1^\textrm{SLL} \, , \\
  Q_2^\textrm{SLL} &= \widetilde Q_1^{\rm SLL}\, = (\bar b_R^i q_L^j)(\bar b_R^j q_L^i) \, , \\
  Q_1^\textrm{VRR} &= \Qbmu_1^\textrm{VRR} \, , \\
  Q_1^\textrm{SRR} &= \Qbmu_1^\textrm{SRR} \, , \\
  Q_2^\textrm{SRR} &= \widetilde Q_1^{\rm SRR}\, = (\bar b_L^i q_R^j)(\bar b_L^j q_R^i)  \, ,
\end{split}
\end{equation}
which we adopt in this work agrees with the one of Eq.~(\ref{eq:30}) except
for the SLL sector and the SRR sector. The evanescent operators are defined
as in Refs.~\cite{bbgln1,ln}:
\begin{equation}
\label{eq:31}
\begin{split}
E^{\textrm{SLL}}_1 =& (\bar b^i_R \gamma_\mu \gamma_\nu q_L^i)
(\bar b_R^j \gamma^\nu \gamma^\mu  q_L^j) +
8 (1-\epsilon) \widetilde Q_1^{\rm SLL}\, , \\
E^{\textrm{SLL}}_2 =& 
(\bar b_R^i \gamma_\mu \gamma_\nu q_L^j)
(\bar b_R^j \gamma^\nu \gamma^\mu q_L^i)  + 
8 (1-\epsilon) Q_1^{\rm SLL} \, .
\end{split}
\end{equation}

The hadronic matrix elements in this basis are parametrized in terms of `bag'
parameters $B_1^{\rm VLL}$, $B_1^{\rm SLL\, \prime}$, and $\widetilde B_1^{\rm SLL\, \prime}$ defined as
\begin{equation}
\label{defbag}
\begin{split}
\bra{\Bbar_q} Q_1^{\rm VLL} (\mu) \ket{B_q} &= 
   \frac{2}{3}  M^2_{B_q}\, f^2_{B_q} B_1^{\rm VLL} (\mu), \\ 
\bra{\Bbar_q} Q_1^{\rm SLL} (\mu) \ket{B_q} &= -\frac{5}{12}  M^2_{B_q}\,
                f^2_{B_q} B_1^{\rm SLL\, \prime} (\mu), \\
\bra{\Bbar_q} \widetilde Q_1^{\rm SLL} (\mu) \ket{B_q} &= 
     \frac{1}{12}  M^2_{B_q}\,
        f^2_{B_q} \widetilde B_1^{\rm SLL\, \prime} (\mu). 
\end{split}
\end{equation}
Here $\mu$ is the renormalization scale at which the matrix element is
computed and $f_{B_q}$ is the $B_q$ meson decay constant.  While $f_{B_s}$
exceeds $f_{B_d}$ by 10--30\%, no non-perturbative calculation finds any
dependence of a bag parameter on the flavour of the light valence quark. In the
vacuum insertion approximation the bag parameters equal $B_1^{\rm VLL} (\mu)=1$
and $B_1^{\rm SLL\, \prime} (\mu)= \widetilde B_1^{\rm SLL\, \prime} (\mu)=
M_{B_q}^2/[m_b(\mu)+m_q(\mu)]^2 $.
Lattice computations determine the matrix elements at a low scale around 1 GeV
and results are quoted for $\mu =\ov m_b(\ov m_b)$. In order to use the lattice
results in our calculation we need the renormalization group (RG) evolution of
the bag parameters to the high scale $\mu_h$ which is set by the masses of the
Higgs bosons exchanged in our \bbm\ diagrams. The matrix elements computed on
a finite lattice 
are converted to continuum QCD by a matching calculation.
This lattice-continuum matching is only meaningful beyond the leading order of
perturbative QCD.  Thus 
the dependence of the bag parameters on
the chosen (continuum) renormalization scheme must be addressed: The NLO
anomalous dimension matrices entering the RG
evolution must be defined in the same renormalization scheme as the bag
parameters, so that the scheme dependence properly cancels from physical
observables. The NLO anomalous dimensions have been calculated for $Q_1^{\rm VLL}$
in Ref.~\cite{bjw}.
As said previously, in the case of $(Q_1^{\rm SLL},\widetilde Q_1^{\rm SLL})$ the NLO
anomalous dimensions have been calculated for the equivalent operator basis $(\bar Q_1^{\rm SLL},\bar Q_2^{\rm SLL})$ 
with the evanescent operators of Eq.~(\ref{eq:21}) \cite{bmu}.

The purpose of this section is twofold: First, we present the transformation
of the results of Ref.~\cite{bmu} to the $(Q_1^{\rm SLL},\widetilde Q_1^{\rm SLL})$
basis and the scheme corresponding to the evanescent operators of Eq.~(\ref{eq:31}), for which lattice groups quote their results.
These formulae are useful beyond
the need to evolve the bag parameters given at $\mu=m_b$ up to $\mu=\mu_h\,$:
In particular lattice groups need to evolve $B_1^{\rm SLL\,\prime} (\mu)$
and $\widetilde B_1^{\rm SLL\, \prime}(\mu)$ from a scale around 1 GeV
up to $\mu=m_b$. Second, we exploit a
heavy-quark relation among the bag factors in \eq{defbag} to sharpen the
numerical prediction for $B_1^{\rm SLL\, \prime} (\mu_h)$ entering the SUSY
contribution to \bbm. While constraints from the heavy-quark limit of QCD have
been used to improve the predictions for $\dg$ and $a_{\rm fs}$
\cite{bbgln1,bbd,ln}, they had escaped attention in studies of new physics
contributions to $B$ physics observables so~far.

\subsection{NLO scheme transformation formulae}\label{sect:stf} 
We decompose the anomalous dimension matrix in the usual way as 
\begin{eqnarray}
\gamma &=& \frac{\alpha_s (\mu)}{4 \pi} \, \gamma^{(0)} \;+\;  
         \lt( \frac{\alpha_s (\mu)}{4 \pi}\rt)^2  \, \gamma^{(1)} 
         \; + \; {\cal O} (\alpha_s^3) .
\end{eqnarray}
The NLO correction $\gamma^{(1)}$ has been computed for the basis
$(\Qbmu_1^{\rm SLL}, \Qbmu_2^{\rm SLL})$ in Ref.~\cite{bmu}.
In four dimensions it is related to the basis (\ref{eq:33})
by a simple Fierz identity:
\begin{eqnarray}
\vec Q \;=\; \lt(
\begin{array}{cc}
Q_1^{\rm SLL} \\ 
\widetilde Q_1^{\rm SLL} 
\end{array}
\rt)
&\stackrel{D=4}{=}& 
\hat R \, 
\lt(
\begin{array}{cc}
\Qbmu_1^{\rm SLL} \\ 
\Qbmu_2^{\rm SLL} 
\end{array}
\rt) = \hat R \, \vec \Qbmu,
\end{eqnarray}
where $\hat R$ is given in \eq{eq:23} below. 

Yet in $D$ dimensions our change of basis involves a rotation of the
operator basis -- including the evanescent operators 
$\vec{\Ebmu}=(\bar E^{\rm SLL}_1,\bar E^{\rm SLL}_2)^T$ -- and a change of the renormalization scheme.
We follow Ref.\cite{Gorbahn:2004my} and write the rotation as
\footnote{Two more evanescent
  operators (called $\Ebmu_3^{SLL}$ and $\Ebmu_4^{SLL}$ in
  Ref.~\cite{bmu}) must be specified to fully define the scheme of the
  calculated $\gammabmu^{(1)}$. This information enters the matrix
  $\hat U$ in \eq{eq:23}. We choose to add $\Ebmu_3^{SLL}$ and
  $\Ebmu_4^{SLL}$ also to the evanescent operators of
  Eq.~(\ref{eq:21}), so that we can in practice work with the change
  of basis defined in Eqs.~(\ref{eq:22}) and (\ref{eq:23}).}

\begin{equation}
\label{eq:22}
\vec Q = \hat{R} \left ( \vec{\Qbmu} + \hat{W}
  \vec{\Ebmu} \right ), \quad 
\vec E = \hat{M} \left ( \epsilon \, \hat{U} \vec{\Qbmu} + \left
    [ \hat{1} + \epsilon \, \hat{U} \hat{W} \right ] \vec{\Ebmu} \right )
\, ,
\end{equation}
with
\begin{equation}
\label{eq:23}
\hat R = \left(
  \begin{array}{cc}
    1 & 0 \\ - \frac{1}{2} & \frac{1}{8}
  \end{array}
  \right), \quad
\hat W = \left(
  \begin{array}{cc}
    0 & 0 \\ 8 & 0
  \end{array}
  \right), \quad
\hat U = \left(
  \begin{array}{cc}
    \frac{1}{4} & - \frac{1}{8} \\ 8 & - \frac{1}{4}
  \end{array}
  \right), \quad
\hat M = \left(
  \begin{array}{cc}
    -8 & 0 \\ -4 & 1
  \end{array}
  \right) \, .
\end{equation}
The information on the definition of the evanescent operators
in \eqsand{eq:21}{eq:31} is contained in the matrices $\hat U $ and 
$\hat M $.
Now \eq{eq:22} corresponds to a finite renormalization with
renormalization constants \cite{Gorbahn:2004my}
\begin{equation}
\label{eq:24}
\hat{Z}^{(1, 0)}_{QQ} = 
\hat{R} \left [ \hat{W} \hat{\Zbmu}^{(1,0)}_{EQ} -
  \left ( \hat{\Zbmu}^{(1, 1)}_{QE} + \hat{W} \hat{\Zbmu}^{(1,1)}_{EE} - 
   \frac{1}{2} \gammabmu^{(0)} \hat{W}
\right ) \hat{U}
\right ] \hat{R}^{-1} .
\end{equation}
While the one-loop anomalous dimension matrix is just rotated, the
two-loop anomalous dimension matrix undergoes an additional scheme
transformation: 
\begin{equation}
\label{eq:25}
\begin{split}
  \gamma^{(0)} & = \hat{R} \hspace{0.2mm} \gammabmu^{(0)}
  \hspace{0.2mm} \hat{R}^{-1} \, , \\[2mm]
  \gamma^{(1)} & = \hat{R} \hspace{0.2mm} \gammabmu^{(1)}
  \hspace{0.2mm} \hat{R}^{-1} - \left [ \hat{Z}^{(1, 0)}_{QQ},
    \gamma^{(0)} \right ] - 
   2 \beta^{(0)} \hat{Z}^{(1,0)}_{QQ} \, ,
\end{split}
\end{equation}
with the one-loop operator renormalization constants
\begin{equation}
\label{eq:26}
\hat{\Zbmu}_{QE}^{(1,1)} =
\left(
  \begin{array}{cc}
    0 & \frac{1}{2}  \\
    -8 & -8 
  \end{array}
\right), \quad 
\hat{\Zbmu}_{EE}^{(1,1)} =
\left(
  \begin{array}{cc}
    2 & \frac{11}{6}  \\
    -\frac{16}{3} & -\frac{44}{3}
  \end{array}
\right), \quad
\hat{\Zbmu}_{EQ}^{(1,0)} =
\left(
  \begin{array}{cc}
    \frac{37}{12} & \frac{29}{48} \\
    -\frac{73}{3} & -\frac{1}{12} \\
  \end{array}
\right)\, .
\end{equation}

We can now calculate the new two-loop anomalous dimension matrix $\gamma$ from 
the NLO anomalous dimension matrix $\bar \gamma$ of Ref.~\cite{bmu},
\begin{equation}
\label{eq:27}
\gammabmu^{(0)} =
\lt(
\begin{array}{cc}
\begin{array}{ll}
\ds -10 &\ds  \frac{1}{6} \\[12pt]
\ds -40 &\ds  \frac{34}{3}
\end{array}
\end{array}
\rt), \qquad
\gammabmu_{\mbox{\scriptsize\cite{bmu}}}^{(1)} =
\lt(
\begin{array}{c@{~~~}c}
\ds -\frac{1459}{9}+ \frac{74}{9}f & \ds -\frac{35}{36} -\frac{1}{54}f\\[12pt]
\ds -\frac{6332}{9}+ \frac{584}{9}f & \ds \frac{2065}{9}-\frac{394}{27}f
\end{array}
\rt).
\end{equation}
We obtain
\begin{equation}
\label{eq:28}
\gamma^{(0)} = \left(
  \begin{array}{rr}
   \ds -\frac{28}{3} &\ds \frac{4}{3} \\[12pt]
   \ds  \frac{16}{3} &\ds \frac{32}{3}
  \end{array}
\right), \qquad
\gamma^{(1)} = 
\lt(
\begin{array}{rr}
\ds  -\frac{260}{3}+\frac{88}{27}f &\ds -\frac{44}{3}+\frac{8}{27}f \\[12pt]
\ds   \frac{242}{3}-\frac{76}{27}f &\ds 198-\frac{332}{27}f
\end{array}
\rt).
\end{equation}
Here $f$ denotes the number of active flavours and $\gamma^{(0)}$ coincides
with the result in \cite{bbgln1}.  As a check we have calculated the result of
\eq{eq:28} also in a different way: It is possible to define evanescent
operators such that the Fierz identity holds for the one-loop matrix elements.
This choice fixes the definitions of both $E^{\rm SLL}_1$ and $E^{\rm SLL}_2$ in \eq{eq:31} and of
the evanescent operators on the $(\bar Q_1^{\rm SLL}, \bar Q_2^{\rm SLL})$ basis.  (One
of the latter operators equals $\epsilon$ times a physical operator. Its
impact is equivalent to a finite multiplicative renormalization of
$\bar Q_1^{SLL}$.)  In this approach one can simply rotate $\gamma^{(1)}$ in the
same way as $\gamma^{(0)}$ in \eq{eq:25}.  Finally the result is transformed
to the scheme of Ref.~\cite{bmu} using the scheme transformation formula of
Ref.~\cite{hn}.

Next we calculate the matrices governing the RG evolution in the
$(Q_1^{\rm SLL},\widetilde Q_1^{\rm SLL})$ basis.  The bag factors at the scale
$\mu_h$ are obtained from those at the low scale $\mu_b={\cal O} (m_b)$
via
\begin{eqnarray}
\lt(
\begin{array}{c}
\ds  - 5 B_1^{\rm SLL\, \prime } (\mu_h)\\[12pt]
\ds  \widetilde B_1^{\rm SLL\, \prime} (\mu_h)
\end{array}
\rt) &=& 
U( \mu_b, \mu_h)^T \,
\lt(
\begin{array}{c}
\ds  - 5 B_1^{\rm SLL\, \prime} (\mu_b)\\[12pt]
\ds  \widetilde B_1^{\rm SLL\, \prime} (\mu_b)
\end{array}
\rt) \label{evo}
\end{eqnarray}
In the spirit of \cite{Buras:2001ra} we write the evolution matrix as
\begin{eqnarray}
U( \mu_b, \mu_h)  
&=&  U^{(0)} \Big(\frac{\alpha_s(\mu_h)}{\alpha_s(\mu_b)} \Big) \; + \; 
   \frac{\alpha_s (\mu_b) }{4\pi}\,
   \Delta U \Big( \frac{\alpha_s(\mu_h)}{\alpha_s(\mu_b)} \Big),
\label{defdeu} 
\end{eqnarray}
where $U^{(0)}$ is the LO evolution matrix and the NLO correction reads
\begin{eqnarray}
   \Delta U ( \eta) 
&=&  J_f \, U^{(0)} (\eta) \,-\, \eta \, U^{(0)} (\eta) \, J_f . \label{deu}
\end{eqnarray}
The $2\times 2$ matrix $J_f$ is calculated from the anomalous dimension matrix
$\gamma$ \cite{bjl}. We only need $J_5$, since we run with 5 active
flavours to the scale $\mu_h$. For applications in kaon physics one
also involves $J_4$ and $J_3$. We quote all three matrices here, so that
the formulae of Ref.~\cite{Buras:2001ra} can be easily extended to the
$(Q_1^{\rm SLL},\widetilde Q_1^{\rm SLL})$ basis: 
\begin{eqnarray}
\!\!\!\!\!\!
J_5\! &=&\! 
\left(
\begin{array}{rr}
1.474 & 0.707 \\
0.306 & -5.350
\end{array}
\right), \;\;
J_4 =  
\left(
\begin{array}{rr}
0.964 & 1.452 \\
0.375 & -4.982
\end{array}
\right), \;\;
J_3 =  
\left(
\begin{array}{rr}
0.652 & 2.597 \\
0.421 & -4.804
\end{array}
\right)\, .
\end{eqnarray}

We quote handy formulae for the five-flavour evolution matrix,
similarly to Ref.~\cite{Buras:2001ra}:
\begin{eqnarray}
\!\! U^{(0)}_{f=5} ( \eta ) \!\!
&=& \!\!
 \left(\!
 \begin{array}{ll} 
  \phantom{-} 
    0.9831  &
  - 0.2577  \\
   - 0.0644  &
  \phantom{-} 
  0.0169 
 \end{array} \! \right) 
 \, \eta^{-0.6315} 
 \; +\; 
 \left(\!
 \begin{array}{ll} 
    0.0169 & 
   0.2577 \\
   0.0644   &
   0.9831 
 \end{array} \! \right) \,\eta^{0.7184} \, .
\label{u0}
\end{eqnarray}
The NLO correction reads:
\begin{eqnarray}
\Delta U_{f=5} ( \eta )
&=& \phantom{\; +\;}
 \left(
 \begin{array}{ll}
  1.4040 - 1.3707 \,\eta\
  &
  - 0.3680 - 2.0731 \,\eta\
   \\
 0.6454 + 0.0898 \,\eta\
  &
  -0.1692 + 0.1358 \,\eta\
 \end{array} \right)
 \, \eta^{-0.6315} \no\\[2mm]
&&   \; + \;
 \left(
 \begin{array}{ll}
  \phantom{-}
   0.0704 -  0.1037  \,\eta\  &
  \phantom{-}
    1.0746 + 1.3665 \,\eta\  \\
  - 0.3395  - 0.3958   \,\eta\  &
   - 5.1807 + 5.2141 \,\eta\
 \end{array} \right) \, \eta^{0.7184}\, .
\label{deltau}
\end{eqnarray}
In our numerical analysis we drop the terms which are linear in $\eta$
in the two matrices in \eq{deltau}, because they are scheme-dependent.
The scheme dependence of these terms cancels with that of the NLO QCD
corrections to the \bbm\ diagrams with SUSY Higgs exchange. Yet these
QCD corrections are unknown.

\subsection{Hadronic matrix elements  and heavy-quark relations} 
The three bag factors $B_1^{\rm VLL}$, $B_1^{\rm SLL\, \prime} (\mu_b)$,
and $\widetilde B_1^{\rm SLL\, \prime} (\mu_b)$ obey a heavy quark 
relation \cite{bbd}:
\begin{eqnarray}
B_1^{\rm SLL\, \prime} (\mu_b) &=& 
   \frac{4}{5} \alpha_2(\mu_b) B_1^{\rm VLL} \; +\;
   \frac{1}{5} \alpha_1(\mu_b) \widetilde B_1^{\rm SLL\, \prime} \; +\;
   {\cal O} \lt( \frac{\lqcd}{m_b} \rt)\, . \label{hqer}
\end{eqnarray}
Here $\alpha_1(\mu)$  and $\alpha_2(\mu)$ comprise NLO QCD 
corrections \cite{bbgln1,ln}:
\begin{eqnarray}
\!  \!   \!  \!  
\alpha_1 (\mu_b) = 1+\frac{\alpha_s(\mu_b)}{4\pi}
          \lt( 16 \log \frac{\mu_b}{m_b} + 8 \rt),
     &&\!\!\!
\alpha_2 (\mu_b) = 1+\frac{\alpha_s(\mu_b)}{4\pi}
         \lt( 8 \log \frac{\mu_b}{m_b} + \frac{26}{3} \rt).
\label{defr12}
\end{eqnarray}
These values are specific to the definition of the evanescent operators
as in \eq{eq:31}. As mentioned in Sect.~\ref{sect:stf}, this definition
allows to maintain the validity of Fierz identities at the loop level.
Such a definition is preferred, if the bag factors are meant to
parametrize the deviation of matrix elements from the vacuum insertion
approximation (VIA), because the calculation of matrix elements in VIA
approximation involves a Fierz transformation. In particular the choice
in \eq{eq:31} is crucial for \eq{hqer} to hold in the limit of a large
number $N_C$ of colours \cite{bbgln1}.

The bag factor $B_1^{\rm VLL}$ is very well studied in lattice QCD,  
so that it is worthwile to study the constraint on the other bag factors
when \eq{hqer} is combined with lattice results for $B_1^{\rm VLL}$. 
Indeed, one can use \eq{hqer} to pinpoint the ratio 
\begin{eqnarray}
\frac{B_1^{\rm SLL\, \prime}(m_b)}{B_1^{\rm VLL}(m_b)} &=& 
0.93 \; +\; 0.23 \, \frac{\widetilde B_1^{\rm SLL\, \prime}(m_b)}{B_1^{\rm VLL}(m_b)}
\;+\; (0.23 \pm 0.05) \frac{1}{B_1^{\rm VLL}(m_b)} \label{hqer2}   
\end{eqnarray}
quite precisely, 
even if $\widetilde B_1^{\rm SLL\, \prime}$ is only poorly known, because
its coefficient in \eq{hqer2} is small. The last term in \eq{hqer2}
quantifies the $\lqcd/m_b$ corrections, see \cite{ln} for details.  
The lattice results of \cite{lattice} have been combined in Ref.~\cite{ln}
to 
\begin{eqnarray}
B_1^{\rm VLL}(m_b)    & = & 0.85 \pm 0.06 
\qquad \mbox{and} \qquad
\widetilde B_1^{\rm SLL\, \prime} (m_b) \; =\; 1.41 \pm 0.12. 
\label{setone}
\end{eqnarray}
Inserting these values into \eq{hqer2} yields
\begin{eqnarray}
\frac{B_1^{\rm SLL\, \prime} (m_b)}{B_1^{\rm VLL}(m_b)} &=& 
1.57 \pm 0.08 ,\label{hqenum} 
\end{eqnarray}
which is consistent with the direct determination
\begin{eqnarray}
B_1^{\rm SLL\,\prime} (m_b) \; =\; 1.34 \pm 0.12
\end{eqnarray}
from the lattice \cite{lattice}. 

We are now in the position to accurately predict the bag factors at the
high scale $\mu_h$. Choosing $\mu_h=\ov m_t(\ov m_t)=164\gev$, 
$\alpha_s(M_Z)=0.1189$ and $\ov m_b (\ov m_b)=4.2\gev$ and using 
\eqsand{hqenum}{setone} we find 
\begin{eqnarray}
B_1^{\rm SLL\, \prime} (m_t) &=& 1.62 \, B_1^{\rm SLL\, \prime} (m_b) + 
            0.01 \, \widetilde B_1^{\rm SLL\, \prime} (m_b) \nn 
&=&           (2.54 \pm 0.13) \, B_1^{\rm VLL} (m_b) + 0.01  
\nn
\widetilde B_1^{\rm SLL\, \prime} (m_t) &=& 
            1.29 \, B_1^{\rm SLL\, \prime}(m_b) + 
            0.54 \, \widetilde B_1^{\rm SLL\, \prime}(m_b) \nn
& =&  (2.03 \pm 0.10)  \, B_1^{\rm VLL} (m_b) + 0.77 \pm 0.07 
            \label{bagnum}
\end{eqnarray}
Here we have omitted the scheme-dependent terms proportional to $\eta$
in \eq{deltau}. The small $(2,1)$ element of $U^{(0)}_{f=5}$ in \eq{u0}
ensures that $\widetilde B_1^{\rm SLL\, \prime} (m_b)$ is inessential for $B_1^{\rm SLL\, \prime}
(m_t)$. One realises from \eq{bagnum} that the uncertainty of the 
high-scale bag factors stems almost completely from the error of
the lattice result for $ B_1^{\rm VLL} (m_b)$. 

Switching finally to the $P_i$'s defined in \eq{3.6} we get
\begin{eqnarray}
P_1^{\rm SLL} &=& -\frac{5}{8} B_1^{\rm SLL\, \prime} (m_t) \; =\; 
            - (1.59  \pm 0.08) \, B_1^{\rm VLL} (m_b) - 0.01 \;=\; 
            - 1.36 \pm 0.12 \nn
P_1^{\rm VLL} &=& B_1^{\rm VLL} (m_t) \; = \; 0.83 \, B_1^{\rm VLL} (m_b) \; =\; 
            0.71 \pm 0.05 .\label{pnum}      
\end{eqnarray}
In the last line the full NLO result of \cite{bjw} has been used.  We
don't need $\widetilde P_1^{\rm SLL} = \widetilde B_1^{\rm SLL\,\prime}(m_t)/8$
for our analysis. Parity ensures that $Q_1^{\rm SLL}$ and the
chirality-flipped operator $Q_1^{\rm SRR}$ defined in \eq{defqrr} have the
same matrix element, i.e.\ $P_1^{\rm SRR}=P_1^{\rm SLL}$.

Finally we compute $P_2^{\rm LR}$ using the formulae of
Ref.~\cite{Buras:2001ra} with the bag factors of Be\'cirevi\'c et al.
\cite{lattice}. This time the conversion between the bases of
Ref.~\cite{Buras:2001ra} and Ref.~\cite{lattice} is straightforward,
since the renormalization scheme used in Refs.~\cite{bmu,Buras:2001ra}
respects the Fierz symmetry and lattice results are already quoted for
this scheme.  The result is
\begin{eqnarray}
P_2^{\rm LR}= 3.2 \pm 0.2 . \label{p2lrnum}
\end{eqnarray}
The number in \eq{p2lrnum} is significantly larger than $P_2^{\rm LR}=2.46$
quoted in Ref.~\cite{Buras:2001ra}, because our value for $m_b$ is smaller and 
the lattice bag factors are larger than one. The error in \eq{p2lrnum} 
does not include the systematic error from the quenching approximation. 

\section{Trilinear Higgs couplings}
\label{app:Vltb3}
The trilinear terms of the effective Lagrangian
at $\tan\beta=\infty$ introduced in Sect.\ \ref{sec:ltbeft} 
read
\begin{equation}     \label{eq:V3ltb}
\begin{split}
  V^{(3)}_{\rm ltb} &= \frac{v}{\sqrt{2}}\, \Bigg\{
      \frac{\lambda_2}{\sqrt{2}} r_u (r_u^2 +  (G^0)^2 + 2 |h_u^+|^2)
      + \sqrt{2} \lambda_3 r_u H_d^\dagger H_d
 \\
& \qquad \qquad
      + \lambda_4 \Big( \sqrt{2}\, r_u |h_d^-|^2
                         + \big[ h_d^0 h_d^- h_u^+ + \mbox{h.c.} \big]
                  \Big)
 \\
& \qquad 
      + \Bigg[
        \lambda_5 h_d^{0*} \Big(\frac{1}{\sqrt{2}} h_d^{0*} (r_u +i\, G^0)
                           - h_d^- h_u^+ \Big)
               + \lambda_6 (H_d^\dagger H_d) h_d^{0*}
 \\
&     \qquad \qquad
         + \lambda_7 \Big(h_d^{0*} \big[ \frac{3}{2} r_u^2 +
                    + i r_u G^0 + \frac{1}{2} (G^0)^2 + |h_u^+|^2 \big]
                  - \sqrt{2}\, r_u h_u^+ h_d^- \Big)  + \mbox{h.c.} \Bigg]
     \Bigg\}\, .
\end{split}
\end{equation}
Again, the first two lines respect the $U(1)$ symmetry introduced
in Sect.\ \ref{sec:ltbeft},
while the last two lines break it, and the breaking is
proportional to loop-induced couplings.  Finally, the quartic
Lagrangian is obtained from the quartic terms in Eq.~(\ref{2.1}) by
substituting $H_u \to (h_u^+, \frac{1}{\sqrt{2}} \phi_u^0)$ 
and $H_d \to (h_d^{0*}, -h_d^-)$.
Also there, only $\lambda_5$, $\lambda_6$ and $\lambda_7$ break the
symmetry.

\end{appendix}



\end{document}